%% file: paper.tex
\documentclass[12pt]{report}
\usepackage{epsfig,psfig,float}
\usepackage{a4wide,graphicx}
\usepackage{amssymb,amsbsy}
\usepackage{fancyhdr}
\usepackage{indentfirst}
\usepackage[notcite,notref]{} 
\usepackage{cite}
\usepackage[spanish,english]{babel}
\newcommand{\NP}[1]{ Nucl.\ Phys.\ {#1}}
\newcommand{\ZP}[1]{ Z.\ Phys.\ {#1}}

\newcommand{\PL}[1]{ Phys.\ Lett.\ {#1}}
\newcommand{\NC}[1]{Nuovo Cimento {#1}}
\newcommand{\AN}[1]{Ann. Phys. NY {#1}}

\newcommand{\PR}[1]{Phys.\ Rev.\ {#1}}
\newcommand{\PRL}[1]{ Phys.\ Rev.\ Lett.\ {#1}}

\newcommand{\bi}{\bibitem}
\newcommand{\vs}{\vspace{-0.2cm}}
\newcommand{\La}{{\cal L}}

\newcommand{\Opd}{\mathcal{O}(p^2)}
\newcommand{\Opc}{\mathcal{O}(p^4)}
\newcommand{\Ima}{{\rm Im}\,}
\newcommand{\Rea}{{\rm Re}\,}

\newcommand{\be}{\begin{equation}}
\newcommand{\ee}{\end{equation}}
\newcommand{\ba}{\begin{eqnarray}}
\newcommand{\ea}{\end{eqnarray}}

\newcommand{\nn}{\nonumber}

\begin{document}
\include{first}
\topmargin -2.5cm
\textheight 26cm
\tableofcontents
\include{intro}
\include{cl}
\include{iam}

\include{fsi}
\include{isifsi}
\include{nuclear}

\include{conclu}
\addcontentsline{toc}{part}{Bibliography}

\include{biblio}
\end{document}

%% file: first.tex
\title{
\vspace*{-5cm} 
\hfill {\small  FZJ-IKP(TH)-1999-37, FTUV-IFIC-99-1215}\\
~\\
Chiral Unitary approach to meson-meson and 
meson-baryon interactions and nuclear applications}

\author{
J. A. OLLER \\
{\small\it Forschungzentrum J\"ulich,
Institut f\"ur Kernphysik (Theorie)}\\ 
{\small\it D-52425 J\"ulich, Germany} \\
~\\
E. OSET \\
{\small\it Departamento de F\'{\i}sica Te\'orica and
IFIC,}
{\small \it Centro Mixto Universidad de Valencia-CSIC,}\\
{\small\it 46100 Burjassot (Valencia), Spain}  \\
~\\
and \\
~\\
A. RAMOS \\
{\small\it Departament d'Estructura i Constituents
de la Mat\`eria,
Universitat de
Barcelona,}\\
{\small\it 08028-Barcelona, Spain} \\
~\\
~\\
{\bf\small Abstract}\\
~\\
\begin{minipage}[t]{0.8\linewidth}
{\small{
We report on recent nonperturbative techniques that combine the
information of  chiral Lagrangians (with and without resonances) with
unitarity in coupled channels
and other requirements of the S-matrix theory of the strong interactions. As
a result, the region of applicability of such techniques is much larger than
the one of Chiral Perturbation Theory allowing one to study also resonance
physics. Applications to meson-meson and meson-baryon scattering, as
well as to problems where pairs of mesons or a meson and a baryon appear in the
initial or final state are shown. Implications in several nuclear
problems are also discussed.
}}
~\\
~\\
{\small 
PACS:~11.15.Tk,~11.80.Gw,~12.38.Lg,~12.39.Fe,~13.25.Jx,~13.40.Hq,~13.60.Le,
13.60.Rj, 13.75.Gx, 13.75.Jz, 13.75.Lb, 25.80.Hp, 25.80.Nv
} \hfill \\ 
{\small Keywords: Chiral symmetry, coupled channels, unitarity,
nonperturbative methods, meson-meson scattering, 
meson-baryon scattering, final state interactions} \hfill 
\vspace*{-5cm}
\end{minipage}
}
\date{}
\maketitle

%% file: intro.tex
\chapter{Introduction}

\vspace*{3.5cm}
Nowadays it is believed that QCD is the theory for the strong interactions.
However, while in the high energy regime, due to the asymptotic freedom, the
theory has been successfully tested by the experiment, this is not the case for
low energies. In this case, one is in the confinement regime of QCD
and perturbative methods cannot be applied.

	On the other hand, in the low energy region the spectrum of QCD 
presents an interesting fact which is the appearance of the isospin pion 
triplet with a mass much smaller than the rest of the QCD states. This can be 
extended to SU(3) by considering the lowest octet of pseudoscalar states 
$(\pi,K,\eta)$. This fact
can be understood by the presence in the QCD Lagrangian of a chiral symmetry for
the light quark sector $(u,d,s)$ which spontaneously breaks down giving rise, by
the Goldstone theorem, to the aforementioned states. The presence of this
symmetry pattern constrains also tremendously the interactions between these
Goldstone bosons. As a consequence, a successful theory has emerged, Chiral
Perturbation Theory ($\chi PT$), which exploits these facts giving rise to a
series of Lagrangians in a power momentum expansion treating the quark masses in
a perturbative way. This power series is valid up to some high energy scale,
$\Lambda_{\chi PT}$, which is of the order of 1 GeV. Hence, the $\chi PT$ 
expansion is valid for momenta $p<<\Lambda_{\chi PT}\approx 1 $ GeV.

Although $\chi PT$ is a powerful tool for the low energy region, its convergence
is limited to a narrow interval (for example, for energies below 0.5 GeV
in meson-meson scattering or close to the threshold region in
meson-baryon scattering). As a consequence, one of
the most representative and interesting facts of strong interacting phenomena,
resonances and their properties, cannot be studied. Furthermore, when one tries
to increase the energy region of applicability of $\chi PT$ just by including
higher orders, the predictive power of the theory is rapidly lost since the
number of free parameters increases tremendously with the order. For instance, 
the leading order $\chi PT$ Lagrangian has essentially no free parameters, in 
the next-to-leading one there are 12 and in the next-next-to-leading order 
there are more than 100.

Hence, the development of nonperturbative techniques which can
extend the energy region of applicability of the theory, without loosing
predictive power, is an important issue. This is the topic to which this
report is devoted. We shall indeed see that this can be done
successfully and that a good description of the meson-meson and
meson-baryon interactions, even for energies above 1 GeV,
arises from $\chi PT$, supplied with
exact unitarity with or without explicit resonance exchanges.

The manuscript is organized as follows. In {\it{chapter 2}} an overall
introduction to the emerging chiral Lagrangians is given. 
{\it{Chapter 3}} is devoted
to deducing the various nonperturbative methods reported here, as well as
to discussing their
applications to strong meson-meson and meson-baryon scattering processes. In 
{\it{chapters 4}} and {\it{5}} it is discussed how the previous strong
amplitudes are implemented to account for final and initial state
interactions, which turn
out to be crucial
to understand the physics of many reactions. Applications of the 
former nonperturbative methods in nuclear physics are reported in {\it{chapter
6}}. Finally, conclusions and remarks are collected in the last section. Every
chapter has its own introduction where one can find a more detailed, although
still general, overview of its contents.

%% file: cl.tex
\chapter{Effective chiral Lagrangians}

In this chapter we want to review the chiral Lagrangians that are going to be used 
in the following. After giving a brief account of chiral symmetry from 
the QCD Lagrangian we will report on the lowest and next to leading order 
$\chi PT$ Lagrangians without baryons.  We will also discuss the 
inclusion of explicit meson resonance fields. Then, we will consider the 
meson-baryon system where the lowest order $\chi PT$ Lagrangian will be given. 
There are many good reviews about Chiral Perturbation Theory 
\cite{Pich,Meissner,Eckerep} to which the interested reader is referred for
further details.

\section{Chiral symmetry}

The QCD Lagrangian with massless $u$, $d$ and $s$ quarks coupled to several
external sources reads:

\begin{equation}
\label{LQCD}
{\La_{QCD}}=\La^0_{QCD}+i\bar{q} D^\mu \gamma_\mu q+\bar{q}\gamma^{\mu}
(v_\mu+\gamma_5 a_\mu)q-\bar{q}(s-i\gamma_5 p)q
\end{equation}
where $\La^0_{QCD}$ is the part of the QCD Lagrangian for the heavier quarks
$c$, $b$ and $t$ and gluons, $v_\mu$, $a_\mu$, $s$ and $p$ are the vector, axial,
scalar and pseudoscalar external sources, $D_\mu$ is the covariant derivative
for the $SU(3)$-colour gauge symmetry and 

\begin{equation}
q=\left(
\begin{tabular}{c}
$u$ \\ $d$ \\ $s$
\end{tabular}
\right)
\end{equation}
is a vector in the three dimensional flavour space.

The Lagrangian eq. (\ref{LQCD}) exhibits a local $SU(3)_L\otimes SU(3)_R$
flavour symmetry under the following transformation rules

\begin{eqnarray}
\label{trans}
q&\rightarrow & g_R \frac{1}{2}(1+\gamma_5)q+g_L\frac{1}{2}(1-\gamma_5)q \\ \nn
v_\mu\pm a_\mu &\rightarrow & g_{R,L}(v_\mu\pm a_\mu)g_{R,L}^\dagger+i\, g_{R,L}
\partial_\mu \, g_{R,L}^\dagger \\ \nn
s+ip &\rightarrow& g_R(s+ip)g_L^\dagger \\ \nn
g_{R,L}&\in& SU(3)_{R,L}
\end{eqnarray}

This chiral symmetry, which should be rather good in the light quark sector, is
not seen in the hadronic spectrum. Although hadrons can be classified in
$SU(3)_V \equiv SU(3)_{R+L}$ representations, degenerate multiplets with 
opposite parity are not observed. Moreover, the octet of the lightest 
pseudoscalar mesons ($\pi$, $K$, 
$\eta$) can be understood if the chiral $SU(3)_L\otimes SU(3)_R$ symmetry
spontaneously breaks down to $SU(3)_V$. Then, according to the
Goldstone theorem \cite{1961TR}, an octet of pseudoscalar massless bosons
appears
in the theory, and these mesons will have the same quantum numbers as the broken
generators of $SU(3)_A \equiv SU(3)_{R-L}$. One thus expects that ($\pi$, K, 
$\eta$)
are to be identified with this octet of Goldstone bosons.

Furthermore, Chiral Symmetry is also explicitly broken by a mass term in eq.
(\ref{LQCD}) 

\begin{equation}
\label{massterm}
-\bar{q} M_q q
\end{equation}
when fixing the scalar source $s(x)=M=diag(m_u,m_d,m_s)$, where $m_u$, $m_d$ and
$m_s$ are the masses of the respective quarks. Because of this term,
the Goldstone bosons acquire a small mass giving rise to the masses of the
($\pi$, K, $\eta$). On the other hand $SU(3)_V$ in the
hadronic spectrum is only an approximate symmetry because $m_s$ is much larger than
$m_u$ or $m_d$.

\section{Chiral Perturbation Theory}

Taking advantage of the mass gap separating the lightest
pseudoscalar octet 
from the 
rest of the hadronic spectrum one can build an effective field theory
containing only
the Goldstone modes. These Goldstone fields, $\vec{\phi}$, can be
collected in a 
traceless $SU(3)$ matrix

\begin{equation}
\label{Fi}
\Phi=\frac{\vec{\lambda}}{\sqrt{2}}\, \vec{\phi}=\left(
\matrix{
\frac{1}{\sqrt{2}} \pi^0+\frac{1}{\sqrt{6}}\eta_8 & \pi^+ &
K^+ \cr
\pi^- & -\frac{1}{\sqrt{2}}\pi^0+\frac{1}{\sqrt{6}}\eta_8 & K^0 \cr
K^- & \bar{K}^0 & -\frac{2}{\sqrt{6}}\eta_8}
\right)
\end{equation}
where $\lambda_i$ are the Gell-Mann's matrices with Tr$(\lambda_i
\lambda_j)=2\delta_{ij}$. From the field $\Phi$ one builds the matrix 
$U(\phi)=e^{i \sqrt{2} \Phi/f}$, with $f$ a constant.
This matrix transforms linearly under $SU(3)_L \otimes SU(3)_R$ as:

\begin{equation}
\label{Utrans}
U(\phi)\rightarrow g_R U(\phi)g_L^\dagger
\end{equation}
and, hence, the Goldstone boson fields $\vec{\phi}$ transform in a
non-linear
form.

The effective Lagrangian will be constructed as a power expansion series in
terms of the external Goldstone momenta and the quark mass matrix.

\section{Lowest and next to leading order {\boldmath $\chi PT$} Lagrangian}

The lowest
order chiral Lagrangian invariant under Lorentz transformations, parity and 
charge conjugation with only two derivatives and linear in the quark masses 
is \cite{xpt}

\begin{equation}
\label{L2}
\La_2=\frac{f^2}{4}< D_\mu U^\dagger \, D^\mu U+U^\dagger \,\mathcal{M}+
\mathcal{M}^\dagger \,U> \ ,
\end{equation}
where $<>$ means $SU(3)$-flavour trace and 
\begin{equation}
\label{xi}
\mathcal{M}=2 B_0(s+ip) \ ,
\end{equation}
with $B_0$ a constant and the covariant derivative is defined as
\begin{equation}
\label{covariant}
D_\mu U=\partial_\mu U-i r_\mu U+i U \ell_\mu \ ,
\end{equation}
where $r_\mu(\ell_\mu)=v_\mu+(-)a_\mu$.

The external sources can also be used to incorporate the electromagnetic and
semileptonic weak interactions through the following relations:

\begin{eqnarray}
\label{electro}
r_\mu&=&e \mathcal{Q} A_\mu+... \nn \\
\ell_\mu&=&e \mathcal{Q} A_\mu+ \frac{e}{\sqrt{2} \sin \theta_W}(W_\mu^\dagger
T_+ + HC)+...
\end{eqnarray}
where $\displaystyle{\mathcal{Q}=\frac{1}{3}diag(2,-1,-1)}$ is the quark-charge matrix and
$T_+$ is a $3\times 3$ matrix containg the relevant
Cabibbo-Kobayashi-Maskawa factors

\be
T_+=\left(
\matrix{
0&V_{ud}&V_{us} \cr
0& 0& 0 \cr
0&0&0}
 \right)  \ .
\ee

On the other hand, the former Lagrangian given in eq. (\ref{L2}) is quoted as 
$\Opd$ because it contains 
at most masses squared, see eq.
(\ref{mass2}), and two 
derivatives. In general, when no baryons are 
present, the Lagrangian 
will have an even number of power of masses and derivatives. In this way, we 
will have $\La=\La_2+\La_4+\La_6+...$, where the subindex indicates the power 
in the momenta.

Fixing $s(x)= M$ and $p(x)=0$, the $\mathcal{M}$ term in eq. (\ref{L2}) gives 
rise to a quadratic pseudoscalar mass term plus additional interactions 
proportional to
the quark masses. This is the reason why in $\chi PT$ the quark masses are 
considered as $\Opd$.

 In the isospin limit ($m_u=m_d$) with 
 ${\displaystyle{\hat{m}=\frac{m_u+m_d}{2}}}$, the following relations arise
 from the lowest order $\chi PT$ Lagrangian, $\mathcal{L}_2$:

\begin{eqnarray}
\label{mass2}
m_\pi^2&=&2 \hat{m} B_0 \\ \nn
m_K^2&=&(\hat{m}+m_s) B_0 \\ \nn
m_{\eta_8}^2&=&\frac{2}{3}(\hat{m}+2 m_s)B_0
\end{eqnarray}
satisfying the Gell-Mann\cite{Mann1962TP}-Okubo\cite{Okubo1962TP} mass relation.

\begin{equation}
\label{GMO}
3\, m_{\eta_8}^2=4\, m_K^2-m_\pi^2 \ .
\end{equation}
From eqs. (\ref{mass2}), valid in the isospin limit, we can write the mass 
matrix $\mathcal{M}$ ($s(x)=M$ and $p=0$) present in the lowest order 
Lagrangian, $\mathcal{L}_2$, as:

\be
\label{mmatrix}
\left(
\matrix{
m_\pi^2 & 0 & 0\cr
0& m_\pi^2 & 0 \cr
0&0& 2 m_K^2 - m_\pi^2} 
\right) \ .
\ee

The meaning of the constant $f$ can be appreciated when calculating from
the
lowest order Lagrangian, eq. (\ref{L2}), the axial current. Then $f$
becomes the pion decay constant in the chiral limit, that is, 

\begin{equation}
\label{f}
f=f_\pi+\mathcal{O}(m_q) \ .
\end{equation}

The next to leading order Lagrangian, $\La_4$, is constructed with the same building 
blocks than $\La_2$, namely, eqs. (\ref{trans}), (\ref{Fi}), (\ref{Utrans}) and
(\ref{covariant}). Preserving Lorentz invariance, parity and charge conjugation
one has \cite{xpt}:

\begin{eqnarray}
\label{L4}
{\cal L}_4&=&L_1 \left<D_{\mu} U^{\dagger} D^{\mu} U 
\right>^2+
L_2 \left<D_{\mu} U^{\dagger} D_{\nu} U 
\right>\left<D^{\mu} U^{\dagger} D^{\nu} U \right>
\nonumber \\
&+&L_3 \left<D_{\mu} U^{\dagger} D^{\mu} U D_{\nu} 
U^{\dagger} D^{\nu} U \right>+L_4 \left<D_{\mu} 
U^{\dagger} 
D^{\mu} U \right> \left<U^{\dagger} {\cal M}+{\cal M}^
{\dagger} U \right> \nonumber \\
&+&L_5 \left<D_{\mu} U^{\dagger} D^{\mu} U 
\left(U^{\dagger} {\cal 
M}+{\cal M}^{\dagger} U \right)\right>+ L_6 \left<U^{\dagger} {\cal M}
+{\cal M}^{\dagger} U \right>^2
\nonumber \\
&+&L_7 \left<U^{\dagger} {\cal M}-{\cal M}^{\dagger} U \right>^2+L_8 
\left<{\cal M}^{\dagger} U {\cal M}^{\dagger}U +U^{\dagger} {\cal M} 
U^{\dagger} {\cal M} \right> \nonumber \\
&-&i L_9<F_R^{\mu\nu} D_\mu U D_\nu U^\dagger+F_L^{\mu\nu} D_\mu U^\dagger D_\nu
U >+L_{10} <U^\dagger F_R^{\mu\nu} U F_{L,\,\mu\nu}>
\nonumber \\
&+&H_1 \left< F_{R\,\mu\nu}\,F_R^{\mu\nu}+F_{L\,\mu\nu}\,F_L^{\mu\nu}
\right>+H_2\left< \mathcal{M}^\dagger \mathcal{M}\right>
\end{eqnarray}
In $\La_4$ there is also the anomalous term \cite{Wess1971TR,Witten1983TR}
although we will not consider it.  
In the former equation we have also included the strength tensor:
\begin{eqnarray}
\label{strength}
F_L^{\mu\,\nu}&=&\partial^\mu \, \ell^\nu-\partial^\nu \, \ell^\mu-i[\ell^\mu,\ell^\nu] \\
\nn
F_R^{\mu\,\nu}&=&\partial^\mu \, r^\nu-\partial^\nu \, r^\mu-i[r^\mu,r^\nu] \\
\nn
\end{eqnarray}

In the $\La _4$ Lagrangian the terms proportional to $H_1$ and $H_2$ do not 
contain the pseudoscalar fields and are therefore not directly measurable. 
Thus, at $\Opc$
we need ten additional coupling constants $L_i$ to determine the low-energy
behaviour of the Green functions. These couplings have an infinite plus a finite
part. The infinite part cancels with the infinites from loops, so that at the
end only the finite parts, $L_i^r$, remain. In $\chi PT$ the 
$\overline{MS}$-1 scheme
is the usual renormalization scheme. At the present time these $L^r_i$ constants have
to be fitted to the phenomenology. In general, the number of free parameters
increases drastically with the order of the chiral expansion so that for 
$\La_6$ there are more than one hundred free couplings. This implies that the 
predictive power of the theory is rapidly lost with higher orders.

On the other hand, the convergence of the $\chi PT$ series is 
restricted to low energies, typically for $\sqrt{s}<500$ MeV, although this
upper limit depends strongly on the process. Note that the
lightest well established resonance, the $\rho$(770) has a mass of 770 MeV. This
resonance introduces a pole in the $T$-matrix which cannot be reproduced by a
power expansion. Thus, the masses of the heavier states not
included in eq. (\ref{Fi}), put a clear upper limit to the $\chi PT$ series and
also give us an upper limit of the 
scale $\Lambda_{\chi PT}$ over which $\chi PT$ is constructed

\begin{equation}
\label{Lam}
\frac{\Opc}{\Opd} \sim \frac{p^2}{\Lambda_{\chi PT}^2}
\end{equation}
with $\Lambda_{\chi PT}\approx M_\rho\approx 1$ GeV.

One can also obtain an estimation of $\Lambda_{\chi PT}$ by taking into 
account those
contributions coming from loops when allowing a change in the regularization 
scale by a factor of $\mathcal{O}(1)$ \cite{Georgi}. The result is that

\begin{equation}
\label{Lam2}
\Lambda_{\chi PT} \leq 4 \pi f_\pi\approx 1.2 \hbox{ GeV}
\end{equation}
which is of the same order of magnitude than $M_\rho$.

\section{Chiral Lagrangians with meson resonances}

Following ref. \cite{EPR} we include hadron states heavier than the lightest
pseudoscalar mesons ($\pi$, $K$, $\eta$). The former states will include 
vector
($V$), axial ($A$), scalar ($S$) and pseudoscalar ($P$) octets and scalar
($S_1$) and pseudoscalar ($P_1$) singlets. The exchange of these resonances
between the Goldstone bosons contains the resonance propagators which for
$p^2<<M_R^2$ can be expanded as

\begin{equation}
\label{bareprop}
\frac{1}{p^2-M_R^2}=\frac{-1}{M_R^2}\left( 1+\frac{p^2}{M_R^2}+
\left( \frac{p^2}{M_R^2} \right)^2+...\right)
\end{equation}
giving rise to contributions which should be embodied in the $\chi PT$
counterterms. However, from the equation above it is obvious that a resummation
to all orders of such local contributions is obtained when including explicit 
resonance fields.

The lowest order Lagrangian with resonance fields, conserving parity and 
charge conjugation is given in ref. \cite{EPR}. Its kinetic part, expressing the
vector and axial vector octets in terms of antisymmetric tensor fields 
$V_{\mu\nu}$ and $A_{\mu\nu}$ (see below), is:

\begin{eqnarray}
\La_{Kin}(R=V,A)&=&-\frac{1}{2}<\bigtriangledown^\lambda R_{\lambda \mu} \bigtriangledown_\nu
R^{\nu\mu}-\frac{1}{2}M_R^2 R_{\mu \nu}R^{\mu \nu}> \nonumber \\
&-&\frac{1}{2}\partial^\lambda R_{1,\lambda\mu}\partial_\nu
R_{1}^{\nu\mu}+\frac{1}{4}M^2_{R_1} R_{1,\mu\nu} R_1^{\mu\nu}\nonumber \\
\La_{Kin}(R=S,P)&=&\frac{1}{2}<\bigtriangledown^\mu R \bigtriangledown_\mu R-M_R^2
R^2>+\frac{1}{2}\partial^\mu R_1 \partial_\mu R_1-\frac{1}{2}M_{R_1}^2 R_1^2
\end{eqnarray}
where the covariant derivative $\bigtriangledown_\mu R$ is defined as 
\begin{equation}
\label{cov}
\bigtriangledown_\mu R=\partial_\mu R+[\Gamma_\mu,R]
\end{equation}
with
\begin{equation}
\label{Gammamu}
\Gamma_\mu=\frac{1}{2}\left\{ u^\dagger(\partial_\mu-i r_\mu)u+
u(\partial_\mu-i l_\mu)u^\dagger \right\}
\ee
with $u$ such that $u^2=U$.

The interaction Lagrangians of the octets and singlets of resonances with
spin$\leq$1 to lowest order in the chiral expansion are given in ref. \cite{EPR}:

\vspace{0.8cm}
\begin{minipage}[t]{0.5\textwidth}
{\bf{Vector Octet, J$^{\hbox{PC}}=1^{--}$}}
\begin{equation}
\label{V}
\frac{F_V}{2\sqrt{2}}<V_{\mu\nu}f^{\mu\nu}_+>+\frac{iG_V}{\sqrt{2}}<V_{\mu\nu}u^\mu
u^\nu>
\ee
\end{minipage}
\hfill
\begin{minipage}[t]{0.45\textwidth}
{\bf{Axial Octet, J$^{\hbox{PC}}=1^{++}$}}
\begin{equation}
\label{A}
\frac{F_A}{2\sqrt{2}}<A_{\mu\nu}f^{\mu\nu}_->
\end{equation}
\end{minipage}

\begin{minipage}[t]{0.5\textwidth}
{\bf{Scalar Octet, J$^{\hbox{PC}}=0^{++}$}}
\begin{equation}
\label{S}
c_d \,<S\, u_\mu u^\mu>+c_m \,<S\, \chi_+>
\ee
\end{minipage}
\hfill
\begin{minipage}{0.5\textwidth}
{\bf{Scalar Singlet, J$^{\hbox{PC}}=0^{++}$}}
\begin{equation}
\label{S1}
\widetilde{c}_d\,S_1\,<u_\mu u^\mu>+\widetilde{c}_m\,S_1\,<\chi_+>
\ee
\end{minipage}

\begin{minipage}{0.5\textwidth}
{\bf{Pseudosalar Octet, J$^{\hbox{PC}}=0^{-+}$}}
\begin{equation}
\label{P}
i\,d_m\,<P\,\chi_->
\ee
\end{minipage}
\begin{minipage}{0.5\textwidth}
{\bf{Pseudoscalar Singlet, J$^{\hbox{PC}}=0^{-+}$}}
\be
\label{P1}
i\,\widetilde{d}_m \, P_1 \,< \chi_{-}>
\ee
\end{minipage}
where $V_{\mu\nu}$ is

\begin{equation}
\label{Vmunu}
V_{\mu\nu}=\left(
\matrix{\frac{1}{\sqrt{2}}\rho^0+\frac{1}{\sqrt{6}}w_8
& \rho^+
& K^{* \ +} \cr \rho^- &
-\frac{1}{\sqrt{2}}\rho^0+\frac{1}{\sqrt{6}}w_8 &
K^{*\ 0} \cr K^{* \ -} & \bar{K}^{* \, 0} & -\frac{2}{\sqrt{6}}w_8}
\right)_{\mu\nu}
\end{equation}
and similarly for the rest of the octets. From eq. (\ref{V}) and (\ref{A}) the
$V$ and $A$ resonances only couple at lowest order as octets. In the former
equations the vector and axial octets are included as antisymmetric tensor
fields, such that if $|W,p>$ represents a vector or axial resonance with
momentum $p$ and mass $M$, then

\begin{equation}
\label{antiva}
<0|W_{\mu\nu}|W,p>=i\,M^{-1}\left[ p_\mu\,\epsilon_\nu(p)-p_\nu\, \epsilon_\mu(p)\right]
\end{equation}
with $\epsilon_\mu(p)$ the polarization (axial)vector of the resonance state. 
The propagator is given by:

\begin{equation}
\label{propva}
\frac{M^{-2}}{M^2-p^2-i\epsilon}\left[g_{\mu\rho}\, g_{\nu\sigma}\,(M^2-p^2)
+g_{\mu\rho}\,
p_\nu \, p_\sigma-g_{\mu\sigma}\,p_\nu \, p_\rho-(\mu \leftrightarrow \nu) \right]
\ee

In the Lagrangians given above we have also used the following objects which 
transform as $SU(3)_V$ octets:

\ba
u_\mu&=&i u^\dagger D_\mu U u^\dagger=u^\dagger_\mu \\ \nn
\chi_\pm&=&u^\dagger \mathcal{M} u^\dagger\pm u\mathcal{M}^\dagger u \\ \nn
f^{\mu \nu}_{\pm}&=&u F_{L}^{\mu \nu} u^{\dagger}\pm u^\dagger F_R^{\mu \nu} u
\ea

In ref. \cite{EPR} the $\Opc$ contributions resulting from 
the exchange of the above resonances are also studied. Note that this is the 
first order to which
resonance exchange contributes to the $\chi PT$ series since their couplings 
to the Goldstone fields are $\Opd$.

At $\Opc$ the resonance exchange gives contribution to all the terms of $\La_4$
in eq. (\ref{L4}). In fact, it is seen in that reference that,
under
certain assumptions for the mass and couplings of the scalar resonances,
 the numerical
values obtained for the  couplings $L_i$ present in $\La_4$ are saturated 
by 
the resonance contributions to them.

\section{Chiral Lagrangians with baryons}

The inclusion of baryons in the chiral formalism is done in a similar way than
the one used for the meson resonances, that is, exploiting their well
defined linear transformation laws under $SU(3)_V$. We consider here the octet of
baryons

\begin{equation}
B=\left(
\matrix{\frac{1}{\sqrt{2}}\Sigma^0 +\frac{1}{\sqrt{6}}\Lambda^0
& \Sigma^+
& p \cr \Sigma^- &
-\frac{1}{\sqrt{2}}\Sigma^0 +\frac{1}{\sqrt{6}}\Lambda^0 & 
n \cr \Xi^- & \Xi^0 & -\frac{2}{\sqrt{6}}\Lambda^0}
\right)
\end{equation} 

The lowest order baryon-meson Lagrangian with at most two baryons can
be written as:

\begin{eqnarray}
\label{BaryonL}
\La_1=<\bar{B}i\gamma^\mu \bigtriangledown_\mu B >-M_B
<\bar{B}B>&+&\frac{1}{2}D<\bar{B}\gamma^\mu \gamma_5
\left\{u_\mu,B \right\}>\\
&+&\frac{1}{2}F<\bar{B}\gamma^\mu \gamma_5 [u_\mu,B] >\nonumber
\end{eqnarray}
where
\begin{equation}
\label{CovDer}
\bigtriangledown_\mu B=\partial_\mu B+[\Gamma_\mu,B] \ ,
\end{equation}
$\Gamma_\mu$ is already defined in Eq. (\ref{Gammamu}) and 
$D+F=g_A=1.257$ and $D-F=0.33$ \cite{Pich}.

Note that, while in the Lagrangians without baryons the number of derivatives is
always even, in the case with baryons an odd number of derivatives also appear,
and in fact, the former Lagrangian is $\mathcal{O}(p)$. On the other hand in eq.
(\ref{BaryonL}) $M_B$ is the baryon octet mass in the chiral limit. The baryon
mass splitting begins to appear at $\Opd$ satisfying the 
Gell-Mann\cite{Mann1962TP}-Okubo\cite{Okubo1962TP} mass relation for baryons
\cite{Pich}.

	Finally, from eq. (\ref{BaryonL}) one can easily derive the well known
Goldberger-Treiman \cite{GT} relation and the Kroll-Ruderman term \cite{KR}.

%% file: iam.tex
\chapter{Nonperturbative models from chiral symmetry for
meson-meson and 
meson-baryon interactions}

The effective chiral Lagrangian techniques have become a widespread tool to
address the problem of the low energy interactions of Goldstone bosons
\cite{xpt,Weinberg}. We have presented in {\it{chapter 2}}, the $\chi PT$ formalism
\cite{xpt} which is the low energy effective theory of the strong interactions
(QCD). Another example is the standard model strongly interacting symmetry
breaking sector (SISBS) \cite{appe} or the effective chiral Lagrangians in
solid-state physics for high-$T_c$ superconductors \cite{solid}. In all the
cases, the chiral symmetry constraints are a powerful tool to determine
the low energy matrix elements in a systematic way. 

	These Lagrangians consist of an expansion on the powers of the external 
momenta of the Goldstone bosons over some typical scale $\Lambda$, which is 
smaller than the masses of the heavier particles. For instance in QCD, 
resonances typically appear for $\sqrt{s}\gtrsim 0.8$ GeV, so that 
$\Lambda_{\chi PT}\lesssim 1$ GeV. 
Of course, when a resonance appears, there is no way to
reproduce it from the perturbative expansion since it is associated to a pole in
the scattering amplitude. 
Furthermore, as explained in {\it{section 2.3}}, there are also higher order 
corrections, as chiral loops, which make that 
$\Lambda_{\chi PT}<4\,\pi\,f_\pi\approx 1.2$ 
GeV \cite{Georgi}. As a result, the $\chi PT$ expansion is typically 
valid up to
energies around 500 MeV with $\Lambda_{\chi PT} \approx 1$ GeV. Nevertheless, 
the constraints imposed by chiral
symmetry breaking are rather powerful and not restricted to the region where
$\chi PT$ is meant to converge \cite{Steele}.

	Another drawback of the effective chiral theories, is the appearance of
a fast increasing number of free parameters (not fixed by the symmetry) as one
increases the order of the calculation. At ${\mathcal{O}(p^2)}$
the $\chi PT$ Lagrangian without baryons only contains the masses of pions, 
kaons and etas and
$f_\pi$. At $\mathcal{O}(p^4)$ several new free parameters appear: for 
instance in $\chi PT$ \cite{xpt} there are 12 parameters and in the SISBS
\cite{Longi} one 
needs 13. At
$\mathcal{O}(p^6)$ in $\chi PT$ there are more than 100 new parameters. That is,
the predictive power of the theory is lost as we go higher in order.

	Because of the former reasons, nonperturbative schemes become necessary
in order to go to higher energies and maintain the predictive power of the
theory.

\section{The inverse amplitude method with coupled channels}

An attempt to extend the constraints of chiral symmetry to higher energies, 
constructing a unitary $T$-matrix, is the Inverse 
Amplitude Method (IAM) \cite{IAM}. This approach proved efficient 
in reproducing low energy 
data and produced poles in the amplitudes associated to the $\rho$ and 
$K^*$ in the vector channel as well as the $\sigma$ in the scalar one. 
It has also been applied to study the SISBS resonances that could appear 
at LHC \cite{LHCres}.
Since only elastic unitarity
was imposed in the IAM, multichannel problems could not be
addressed. In fact, the treatment of coupled channels
has proved to be crucial in order to reproduce the basic features 
of the $f_0$ and $a_0$ resonances \cite{Weinis,Jansen,npa} and in general for 
all the scalar sector with $I=0,1,1/2$ \cite{Tornqvist,OOnd}.
As a consequence, neither the $f_0$ nor $a_0$ resonances could be obtained by
the elastic IAM \cite{IAM}.

We now proceed to the extension of the IAM with coupled channels which was 
given for first time in ref. \cite{prl}.

Let $T_L^I$ be a meson-meson partial wave amplitude with definite isospin $I$ and
angular momentum $L$. If $T$ is the scattering matrix, $S=I-i\,T$ with $S$ the
$S$-matrix, then

\begin{eqnarray}
\label{TL2}
\frac{T^I}{\sqrt{2}^\alpha}&=&\sum_{L=0}^\infty (2L+1) T^I_L(s) P_L(\cos \theta) \nn\\
T^I_L(s)&=&\frac{1}{2\, (\sqrt{2})^\alpha} \int_{-1}^1 d\cos \theta \,
P_L(\cos \theta)\, T^I(s,\cos \theta)
\end{eqnarray}
where $(\sqrt{2})^\alpha$ is a symmetry factor to take care of the presence of
identical particle states as $\eta\eta$ or $\pi\pi$ in the isospin
limit. The index $\alpha$ can be 0,1 or 2 depending on the number of times these
identical particle states appear in the corresponding partial wave amplitude.
For instance, $\alpha=2$ for $\pi\pi\rightarrow \pi\pi$, $\alpha=1$ for
$\eta\eta \rightarrow K\bar{K}$, $\alpha=0$ for $K\pi\rightarrow K\pi$ and so
on. $P_L(\cos \theta)$ is the Legendre polynomial of $L_{th}$ degree. In the
following we will omit the indexes $L$ and $I$ in a partial wave, although it
should be kept in mind that we are considering partial wave amplitudes with
definite $L$ and $I$, unless the contrary is said.

 In our normalization unitarity in coupled channels reads:
\begin{equation}
\Ima T_{if} = -T_{in} \, \rho_{nn} \, T^*_{nf}
\label{Tunit}
\end{equation} 
where $\rho$ is a real diagonal matrix whose elements account 
for the phase space of the two meson intermediate states $n$ which are
physically accessible. With our normalization, $\rho$ 
is given by

\begin{equation}
\label{1.0}
\rho_{nn}(s)=\frac{k_n}{8 \pi \sqrt{s}} \theta (s-(m_{1n}+m_{2n})^2)
\end{equation}
where $k_n$ is the on shell center mass (CM) momentum of the meson in the 
intermediate state $n$ and $m_{1 n}, m_{2 n}$ are the masses of the two mesons 
in this state.

Isolating $\rho$ from eq. (\ref{Tunit}) one has:

\begin{eqnarray}
\rho &=& -T^{- 1} \cdot \Ima T \cdot T^{* - 1}\nn \\
&=& -\frac{1}{2 i} T^{- 1}\cdot (T - T^*)\cdot T^{* - 1} \nn\\
&=& -\frac{1}{2 i} (T^{- 1 *} - T^{- 1}) = \Ima T^{- 1}
\label{ImG}
\end{eqnarray}

The former result is in fact the basis for the $K$-matrix formalism 
\cite{Martinspearman}. From eq. (\ref{ImG}) we can write:

\begin{equation}
\label{1.1}
T^{-1}=\Rea T^{-1}+i \rho\equiv K^{-1}+i\rho
\end{equation}
where $K$ is the $K$-matrix which from the former equation 
is given by

\begin{equation}
\label{1.2}
K^{-1}=\Rea T^{-1}
\end{equation} 

Once the $K$-matrix ($\Rea T^{-1}$) is given, the $T$-matrix follows by inverting eq. (\ref{1.1}):
 
\begin{equation}
T = [K^{-1}+i \rho]^{-1}=[\Rea T^{- 1} + i \, \rho]^{- 1}
\label{Tinverse}
\end{equation}

We will  
approach the $K$-matrix by expanding $\Rea T^{-1}$
from the ${\mathcal{O}}(p^4)$ $\chi PT$ expansion of the
$T$-matrix.
In this way, unitarity will be
fulfilled to all orders since 
we know exactly the imaginary part of $T^{-1}$ 
from eq. (\ref{ImG}). Another advantage of considering the expansion of $T^{-1}$ 
is clear when $T$ has a pole. In this case, $T^{-1}$ will have just a zero and hence 
its expansion will not be affected by this pole of $T$.

Thus, expanding $T^{-1}$ in powers of $p^2$ from the expansion of $T$, one has:

$$
T \simeq T_2 + T_4 + ...
$$
where $T_2$ is the lowest order $\chi PT$ amplitude and $T_4$ the $\Opc$
contribution,

\begin{equation}
T^{- 1} = T_2^{- 1}\cdot [1 + T_4 \cdot T_2^{-1}+ ...]^{- 1} 
= T_2^{- 1}\cdot [1 - T_4 \cdot T_2^{- 1}+...]\nn\\
=T_2^{-1}\cdot \left[T_2-T_4 +...\right]\cdot T_2^{-1}
\label{TIChPT}
\end{equation}

Inverting the former equation we finally have:

\be
\label{iamcc}
T=T_2\cdot\left[T_2-T_4\right]^{-1}\cdot T_2
\ee

Note that eq. (\ref{TIChPT}) fulfills the unitarity requirements given in eq.
(\ref{ImG}) because $\Ima T_4=-T_2\cdot \rho \cdot T_2$ above the physical
thresholds. Taking into account eq. (\ref{1.2}), the $K$-matrix resulting from 
eq. (\ref{TIChPT}) is given by:

\be
\label{kmat}
K=T_2\cdot\left[T_2-\Rea T_4\right]^{-1}\cdot T_2 
\ee

Eq. (\ref{iamcc}) is the extension of the IAM to coupled channels. In the next
two sections we describe the application of the former formalism to the study
of the meson-meson interactions up to $\sqrt{s}\lesssim 1.2$ GeV. For higher 
energies more than two meson states are needed since multipion states become
increasingly important.

\subsection{{\boldmath $\pi\pi$} and {\boldmath $K\bar{K}$} amplitudes}

In this section we are going to present the results of applying the IAM with
coupled channels to the
study of the $\pi\pi$ partial wave amplitudes with $(I,L)$=(0,0),
(1,1) and (2,0). The pions couple with the $K\bar{K}$ channel in the waves
(0,0) and (1,1), although, as we will see below, this coupling is negligible in
the (1,1) case. This study was developed in ref. \cite{NPB}.

In order to apply eq. (\ref{iamcc}) we need the $\chi PT$ amplitudes up to 
$\Opc$.
For $\pi \pi\rightarrow \pi\pi$ this calculation was done in ref. \cite{xpt} for the
$SU(2)$ case and extended to $SU(3)$ in ref. \cite{UMNR}. The $\pi\pi \rightarrow
K\bar{K}$ amplitude can be obtained by crossing from the $K\pi\rightarrow K\pi$
one calculated in ref. \cite{UMNR}. The $K\bar{K}\rightarrow K\bar{K}$ 
was first calculated in ref. \cite{NPB}. 

The amplitudes we are considering in this section depend on the $L_i$ couplings
of the $\Opc$ $\chi PT$
Lagrangian, eq. (\ref{L4}), that enter in their calculations. These constants
are: $L_1$, $L_2$, $L_3$, $L_4$, $L_5$ and $2 L_6+L_8$. They are fitted to the 
elastic
$\pi\pi$ phase shifts in the partial waves $(I,L)=(0,0)$ and (1,1) as shown in
Fig. 3.1 and 3.2. The fit was done using MINUIT.
In the energy region $\sqrt{s} =$ 500--950 MeV the data 
from different experiments for S-wave $\pi\pi$ phase shifts are incompatible.
Given that situation, the central value for each energy is taken as the mean 
value between the different experimental results 
\cite{Kaminski,Hyams,Estabrooks,Grayer,Protopopescu}. For $\sqrt{s}=$ 
0.95--1 GeV, the mean value comes from \cite{Hyams,Grayer}. In both cases 
the error is the maximum between the experimental errors and the largest 
distance between the experimental points and the average value. The fit is good
with a $\chi^2=1.3$ per degree of freedom. The values of the $L_i$ at the
$M_\rho$ scale are shown in Table 3.1. In the second
column the values obtained
from $\chi PT$ fits at $\Opc$ to the low energy data are also shown. 
We can see that the values obtained by applying the IAM \cite{NPB}, when the 
errors are taken into account, are compatible with those from $\chi PT$.

\begin{figure}
\begin{minipage}[t]{0.48\textwidth}
\includegraphics[width=0.75\textwidth,angle=-90]{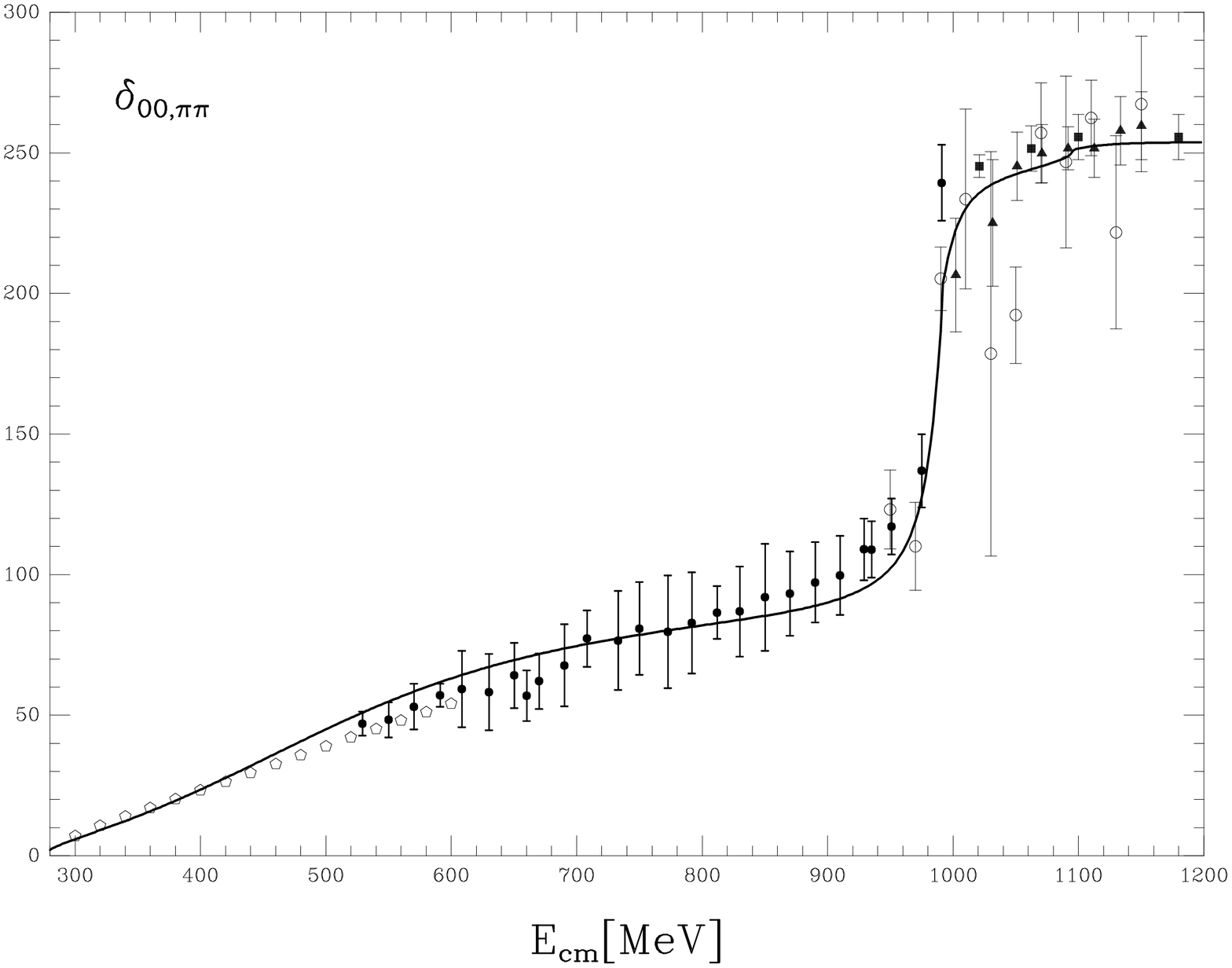}
\caption{Phase shift for $\pi\pi \rightarrow \pi\pi$ in $I=L=0$. Data: Empty
pentagon \cite{Frogatt}; empty
circle \cite{Kaminski}; full square \cite{Grayer}; full triangle \cite{Hyams};
full circle represents the average explained in the text.}
\end{minipage}
\hfill
\begin{minipage}[t]{0.48\textwidth}
\includegraphics[width=0.75\textwidth,angle=-90]{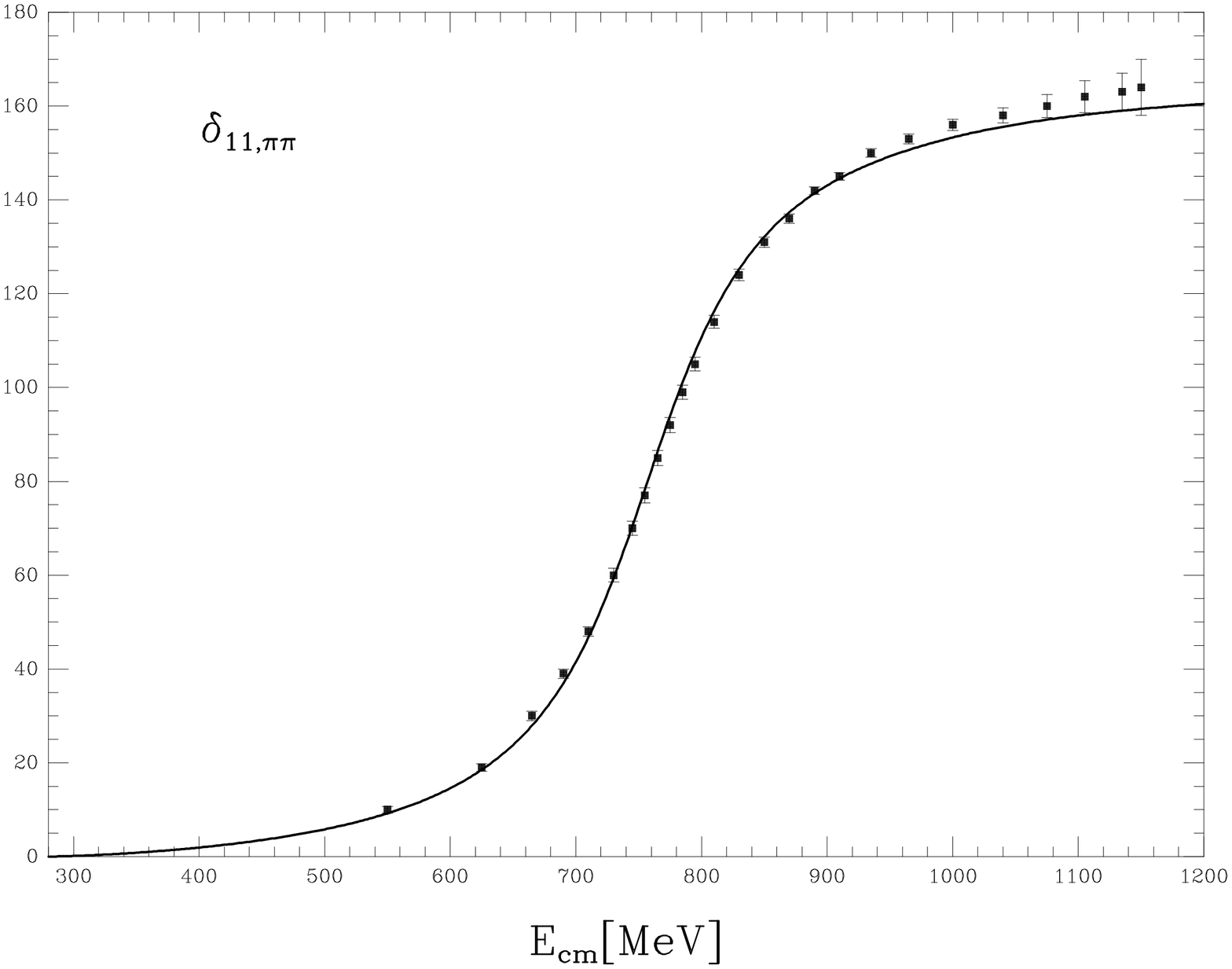}
\caption{Phase shift for $\pi\pi \rightarrow \pi\pi$ in $I=L=1$. Data:
\cite{Protopopescu}}
\end{minipage}
\end{figure}

\begin{table}[ht]
\begin{center}
\caption{$L_i\,10^3$ coefficients.}
\begin{tabular}{ccc}
\hline
& Fit&$\chi PT$ \\
\hline
$L_1$& $0.72^{+0.03}_{-0.02}$ & $0.4\pm 0.3$ \\
$L_2$& $1.36^{+0.02}_{-0.05}$ & $1.4\pm 0.3$ \\
$L_3$& $-3.24\pm 0.04$ & $-3.5 \pm 1.1$ \\
$L_4$&$0.20\pm 0.10$ & $-0.3\pm 0.5$\\
$L_5$&$0.0^{+0.8}_{-0.4}$&$1.4\pm 0.5$\\
$2 L_6+L_8 $&$0.00 ^{+0.26}_{-0.20}$&$0.5\pm 0.7$\\
\hline
\end{tabular}
\end{center}
\end{table}

With the former values for the $L_i$ coefficients, the 
inelastic S-wave phase shifts for $K\bar{K}\rightarrow \pi\pi$, Fig. 3.3,
$\displaystyle{\frac{1-\eta_{00}^2}{4}}$ where $\eta_{00}$ is the inelasticity
in the $I=L=0$ channel, Fig. 3.4, and the elastic $\pi\pi$ S-wave phase shifts 
with $I=2$, Fig. 3.5, are also calculated. In Fig. 3.3 one sees clearly the 
$\eta\eta$ threshold.
Although this channel is not included in the unitarization process, it
appears as an intermediate state in the loops of $T_4$. If we
remove from $T_4$ the imaginary part coming from the intermediate $\eta\eta$
state above its threshold the dashed line results. It is clear then that the
inclusion of the $\eta\eta$ channel in the unitarization procedure for the (0,0)
channels is important. This point will be further considered in 
{\it{section 3.2.3}} and at the end of {\it{section 3.3.1}}.

\begin{figure}[H]
\begin{minipage}[t]{0.48\textwidth}
\includegraphics[width=.75\textwidth,angle=-90]{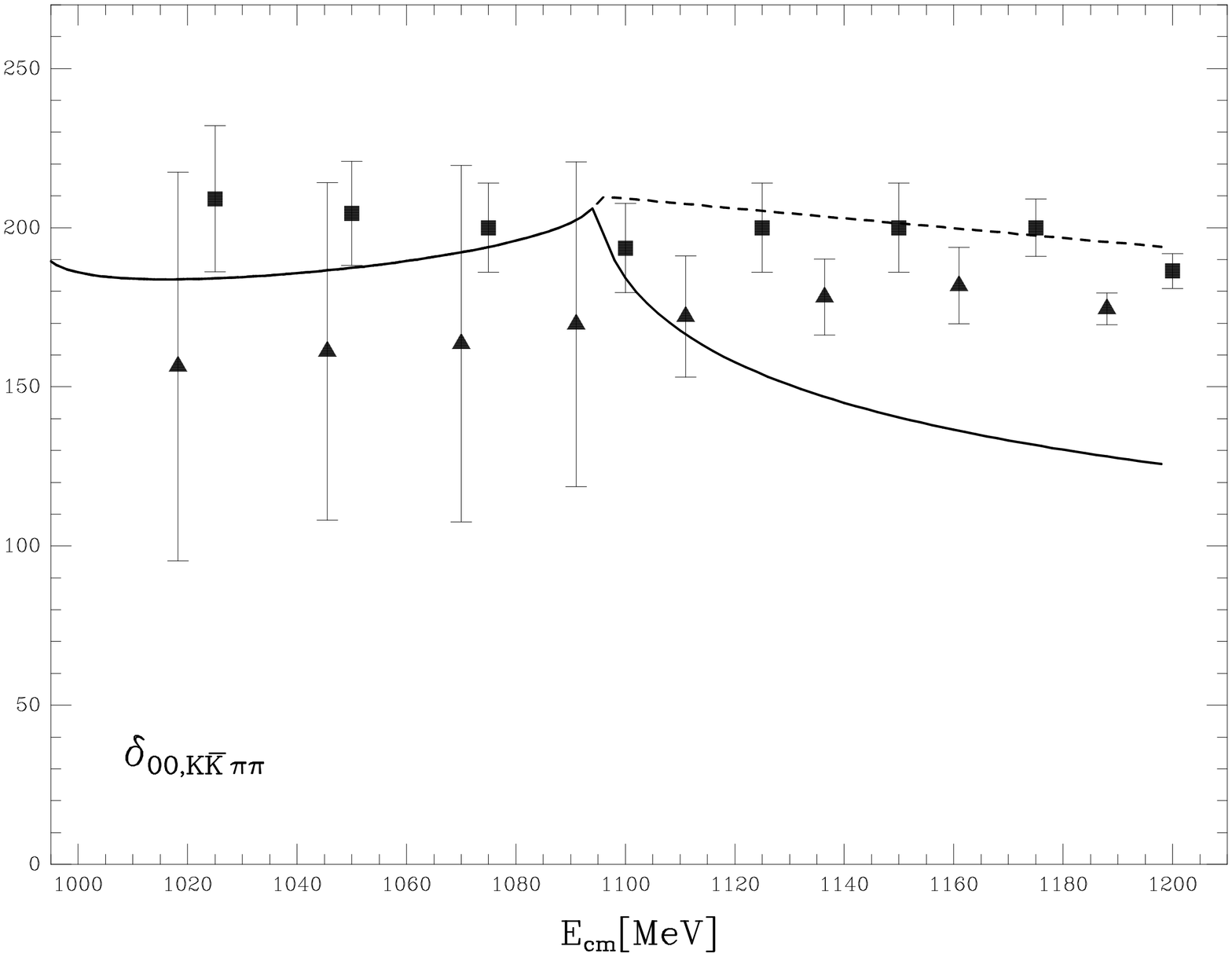}
\caption{Phase shift for $\pi\pi \rightarrow K {\bar K}$ in $I=L=0$. Data:
full square \cite{Cohen}, full triangle \cite{Martin}.}
\end{minipage}
\hfill
\begin{minipage}[t]{0.48\textwidth}
\includegraphics[width=0.75\textwidth,angle=-90]{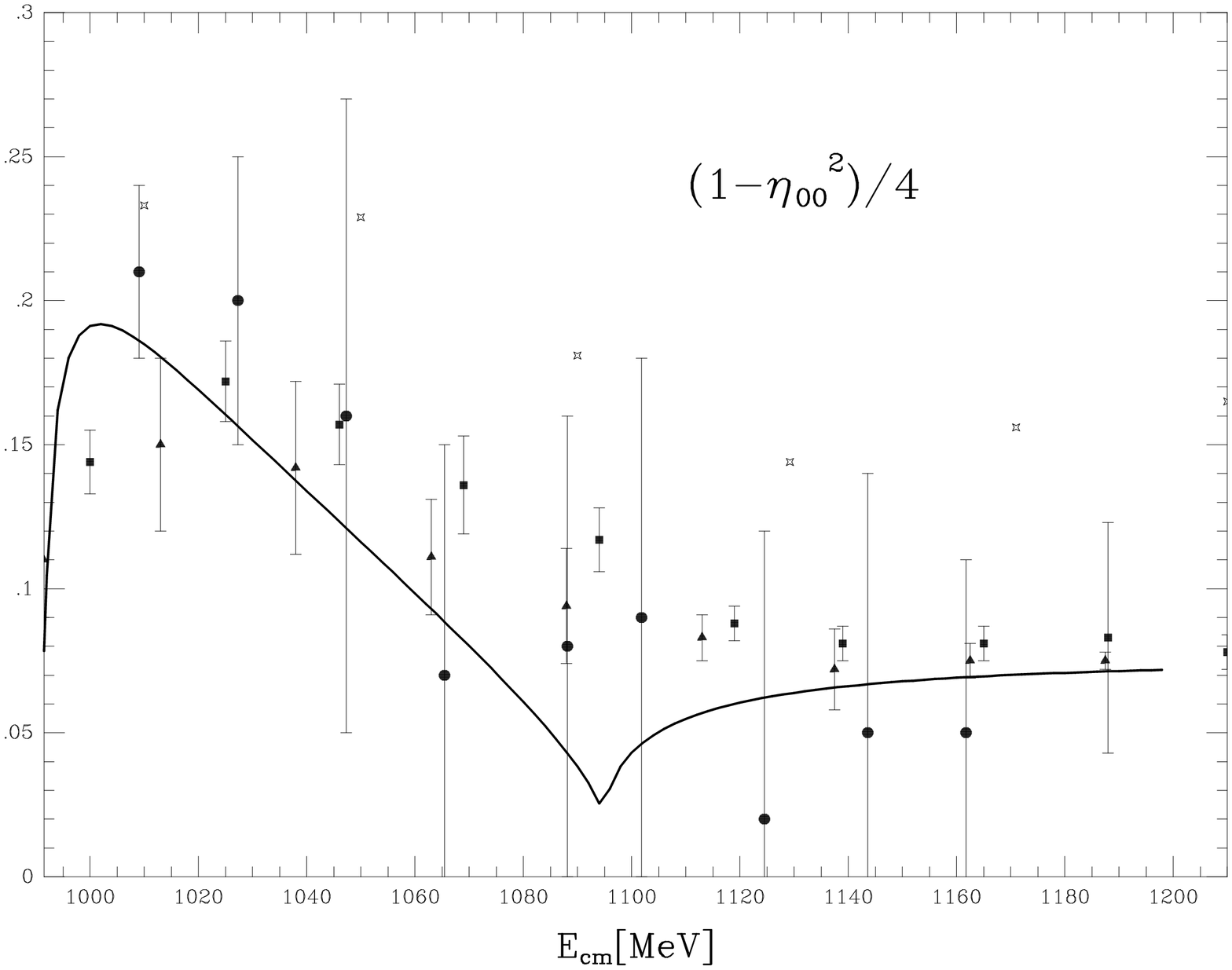}
\caption{($1-(\eta_{00})^2)/4$, where $\eta_{00}$ is the inelasticity in
 $I=L=0$. Data: starred square \cite{Frogatt}, full square \cite{Cohen}, full 
triangle \cite{Martin}, full circle \cite{Ochs}.}
\end{minipage}
\begin{minipage}[t]{0.48\textwidth}
\includegraphics[width=0.75\textwidth,angle=-90]{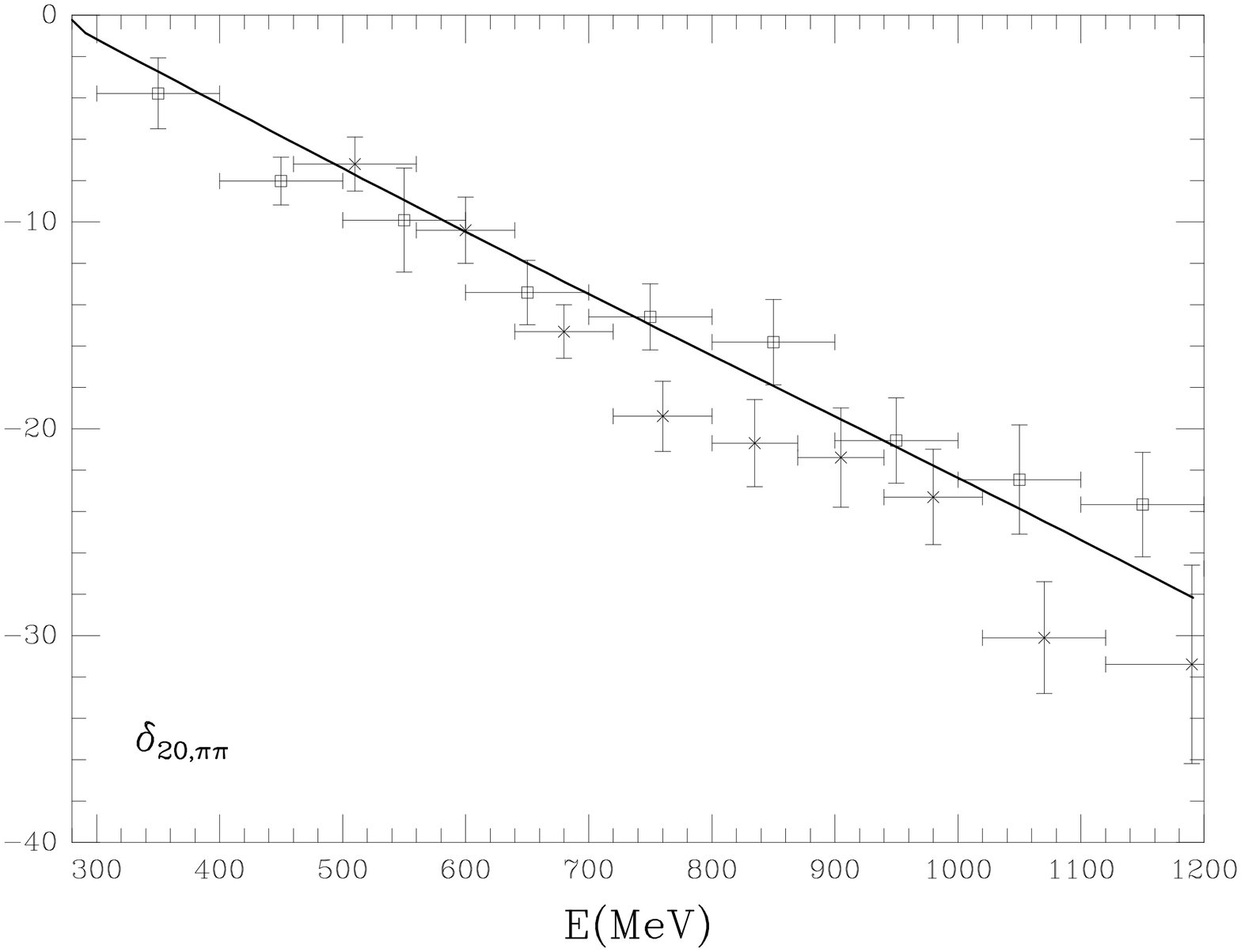}
\caption{Phase shift for $\pi\pi \rightarrow \pi\pi$ in $I=2,L=0$. Data:
cross \cite{Rosselet}, empty square \cite{Schenk}.}
\end{minipage}
\end{figure}

The scattering lengths of the channels $(I,L)$=(0,0), (1,1) and (2,0) were also
calculated in ref. \cite{NPB}. These scattering lengths are denoted by $a^I_L$. In 
Table 3.2 we show the values obtained for $a^I_L$ in ref. \cite{NPB} 
together with the 
experimental ones and the $\chi PT$ values up to ${\cal O} (p^4)$. 
We see in this table that a good agreement 
with experiment is accomplished. The values of ref. \cite{NPB} are also close 
to the ones from $\chi PT$ as one should expect because at low energies
the
IAM recovers the chiral expansion up to $\Opc$.

\begin{table}[ht]
\begin{center}
\caption{Comparison of scattering lengths in different channels.}
\begin{tabular}{cccc}
\hline
$a^I_L$ & $\chi PT$ & Results from ref. \cite{NPB} & Experiment\\
\hline
$a^0_0$ & $0.20 \pm 0.01$ & $0.210 \pm 0.002$ & $0.26 \pm 0.05$\\
$a^1_1$ & $0.037 \pm 0.002$ & $0.0356 \pm 0.0008$ & $0.038 \pm 0.002$\\
$a^2_0$ & $-0.041 \pm 0.004$ & $-0.040 \pm 0.001$ & $-0.028 \pm 0.012$\\
\hline
\end{tabular}
\end{center}
\end{table}

\subsection{Two-meson scattering below 1.2 GeV}

In the previous section the IAM with coupled channels was applied with the 
full $\Opc$ $\chi PT$ amplitudes. The calculation of the latter amplitudes is 
rather involved and it is not
done in all two-meson channels, for instance, in those channels with the $\eta$
meson. In this section, following ref. \cite{PRD}, an approximation to 
calculate the $\Opc$ amplitude, which 
turns out to be technically much simpler and rather accurate at the 
phenomenological level, is presented. Then, eq. (\ref{iamcc}) will be applied 
to study the
partial waves $(I,L)$=(0,0), (1,0), (2,0), (1/2,0), (3/2,0), (1,1), (1/2,1) and
(0,1). One considers the channels shown in Table 3.3, which are
supposed to be the dominant ones up to $\sqrt{s}\lesssim 1.2$ GeV.

\begin{table}[ht]
\label{channels}
\begin{center}
\caption{Channels used in the different $I,L$ channels}
\begin{tabular}{cccccc}
\hline
& $I=0$ & $I=1/2$ & $I=1$ & $I=3/2$ & $I=2$ \\
\hline       
L=0 & $\begin{array}{c} \pi\pi \\ K \bar{K}\\ \end{array} $&
$\begin{array}{c} K \pi \\ K \eta \\ \end{array}$ & $\begin{array}{c} 
\pi \eta \\ K \bar{K} \\ \end{array}$ & $K \pi$ & $\pi\pi$ \\
\hline 
L=1 & $K \bar{K}$ & $\begin{array}{c} K \pi \\ K \eta \\ \end{array}$ & $
\begin{array}{c} \pi \pi \\ K \bar{K} \\ \end{array}$ & & \\
\hline
\end{tabular}
\end{center}
\end{table}

The $T_4$ amplitudes are then approximated in ref. \cite{PRD} by

\begin{equation}
\label{t4ap}
T_4\approx T_4^P+T_2\cdot g(s) \cdot T_2 
\end{equation}
where $T_4^P$ represents the polynomial tree level amplitudes from the $\Opc$
Lagrangian, eq. (\ref{L4}), and are given in the Appendix A of ref. \cite{PRD}. On the
other hand, $g(s)$ is a diagonal matrix
corresponding to the loop integral with two meson propagators, given by:

\begin{equation}
g_{nn}(s) = i \int \frac{d^4 q}{(2 \pi)^4} \;
\frac{1}{q^2 - m^2_{1 n} + i \epsilon} \;
\frac{1}{ (P - q)^2 - m^2_{2 n} + i \epsilon}
\label{5.0}
\end{equation} 
where $P$ is the total initial four-momentum of the two meson system. This $g$ 
matrix has the property

\begin{equation}
\label{Img}
\Ima \, g_{nn}(s)=-\rho_{nn}(s)
\end{equation}
as can be easily checked. 

The real part of $g(s)$ is divergent and requires regularization. 
In ref. \cite{PRD} it is evaluated by a cut off regularization with a maximum 
value, 
$q_{max}$, for the modulus of the three-momentum in the integral. 
Because the
divergence in eq. (\ref{5.0}) is only logarithmic its calculation making use of 
dimensional regularization is numerically equivalent when, depending on
the considered renormalization scheme, the dimensional 
regularization scale $\mu$ is properly chosen as a function of $q_{max}$.
For instance,
in the $\overline{MS}-1$ renormalization scheme, the one used in $\chi PT$,
$\mu\approx 1.2 \,q_{max}$ \cite{PRD}.  In {\it{section 3.2.2}} the
expression of $g(s)$ in dimensional regularization is given.

When approximating $T_4$ by eq. (\ref{t4ap}) one is taking into account the
close relationship between $T_4^P$ and the vector mesons \cite{EPR} and the 
dominant role that the unitarization with coupled channels of the lowest order 
$\chi PT$ amplitudes has in the scalar sector \cite{npa,kaiser}. In fact, the
approach of ref. \cite{npa} follows in the limit $T_4^P=0$. With respect to a full 
$\Opc$ $\chi PT$ calculation, neither tadpoles nor loops in crossed
channels are included. These are soft contributions which will be reabsorbed 
in the $L_i$
coefficients, which are now denoted by $\hat{L}_i$ since differences begin to 
rise even at $\Opc$ with respect to the $L_i$ of $\chi PT$.

\begin{figure}[H]
\begin{center}
    \hbox{ \psfig{file=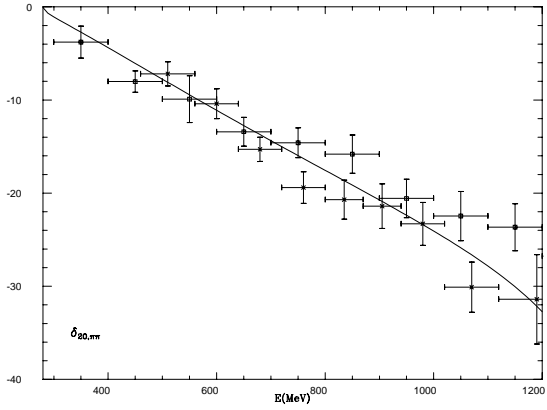,width=.35\textwidth,angle=-90}}
  \caption{Phase shifts for $\pi\pi\rightarrow\pi\pi$ in the 
  $I=2$, $L=0$ channel. Data: cross \cite{Rosselet}, empty square 
  \cite{Schenk}.}
\end{center}
\end{figure}

\begin{figure}[H]
\begin{center}
    \hbox{ \psfig{file=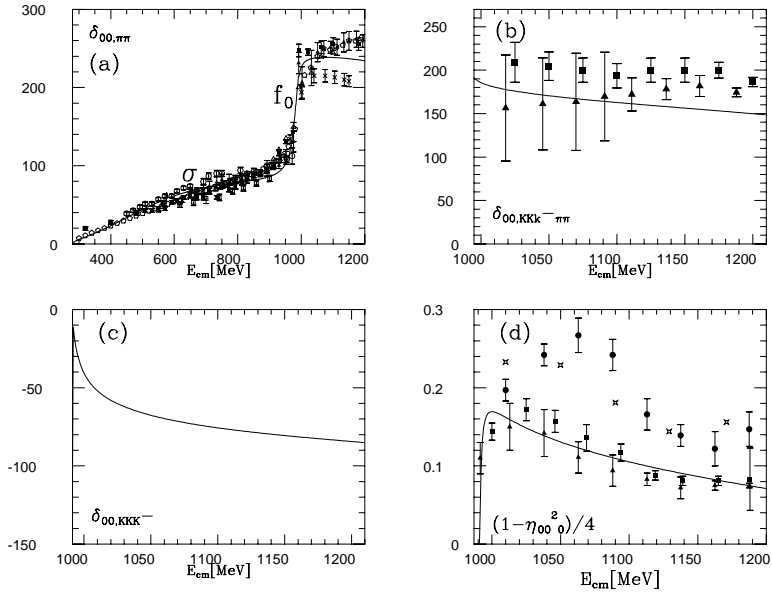,width=.5\textwidth,angle=-90}}
  \caption{Results in the $I=L=0$ channel. (a) phase shifts for
  $\pi\pi\rightarrow\pi\pi$ as a fraction of the c.m. energy of the
  meson pair: full triangle \cite{Hyams}, open circle \cite{Estabrooks}, 
  full square \cite{Grayer}, open triangle \cite{Grayer2}, open square 
  \cite{Grayer3} 
  (all these are analysis of the same experiment \cite{Grayer4}), 
  cross \cite{Protopopescu}, full circle \cite{Manner}, empty pentagon
  \cite{Frogatt}. 
  (b) phase shifts for $K\bar{K}\rightarrow \pi\pi$: full square \cite{Cohen}
  , full triangle \cite{Martin}. (c) Phase shifts for $K \bar{K} \rightarrow 
  K \bar{K}$. (d)
  Inelasticity: results and data for $(1-\eta^2)/4$: starred square 
  \cite{Frogatt}, full square \cite{Cohen}
  , full triangle \cite{Martin}, full circle \cite{Linden}.}
\end{center}
\end{figure}

From Fig. 3.6 to 3.13 we show the fit of ref. \cite{newPRD} to the meson-meson 
S and P-wave experimental data, phase shifts and inelasticities. In this
reference an error was detected in the $T_4^P$ amplitude 
$K^+K^- \rightarrow K^0 \bar{K}^0$ calculated in ref. \cite{PRD} (the corrected
expression  is given in the Appendix of ref. \cite{newPRD}). 
As a result, 
the fit presented in ref.
\cite{PRD} was redone making use of MINUIT and the results are the ones 
displayed in the former
figures. In ref. \cite{newPRD} several sets of values of the $\hat{L}_i$
coefficients were found giving rise to fits of similar quality. The main
difference between the different sets of values is in the value of $\hat{L}_7$ 
which can even change sign. We report here the fit with the $\hat{L}_i$ 
coefficients closer to the $L_i$ of $\Opc$ $\chi PT$. In Table 3.4 
the corresponding values of the couplings are given and compared with the ones 
of $\Opc$ $\chi PT$. As we see from the figures, the results are 
in rather good agreement with a vast amount of experimental data. The resonances that 
appear are indicated with their corresponding names.

\begin{figure}[ht]
    \hbox{ \psfig{file=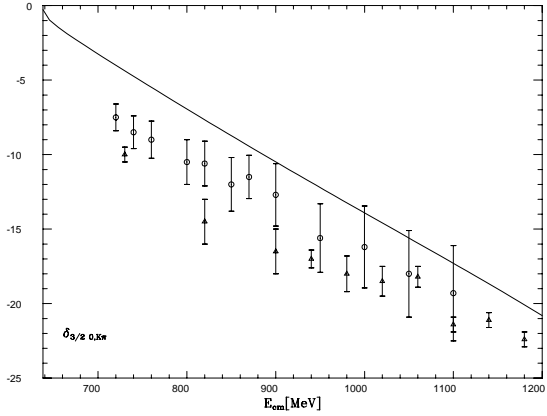,width=0.35\textwidth,angle=-90}}
  \caption{Phase shifts for $K\pi\rightarrow K\pi$ in the $I=3/2$,
  $L=0$ channel. Data: open triangle \cite{Estabrooks3}, open circle 
  \cite{Linglin}.}
\end{figure}

\begin{figure}[H]
\hbox{
\psfig{file=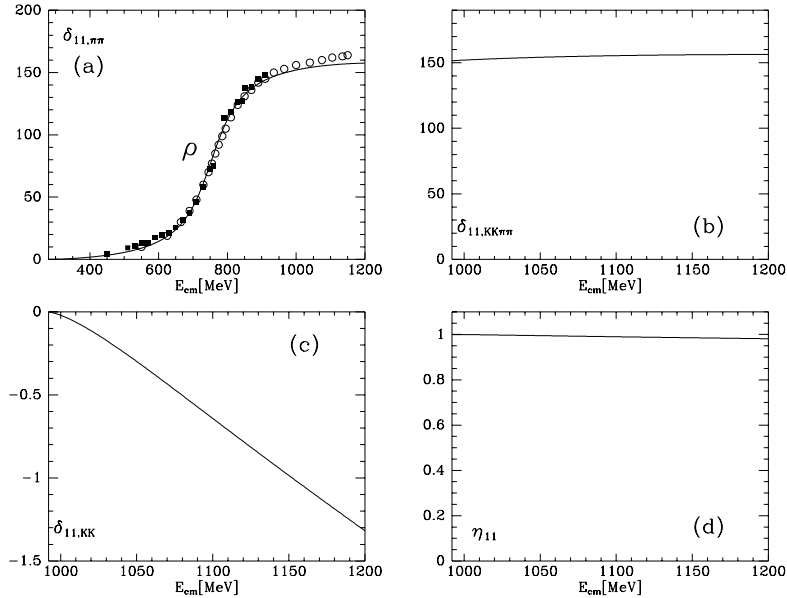,width=0.5\textwidth,angle=-90}}
  \caption{Results in the $I=L=1$ channel. (a) phase shifts for
  $\pi\pi\rightarrow\pi\pi$. Data: open circle \cite{Protopopescu}, black square
  \cite{Estabrooks}. (b), (c) same as in Fig. 3.7. (d) inelasticity.}
\end{figure}

\begin{figure}[ht!]
\begin{center}
\hbox{
\psfig{file=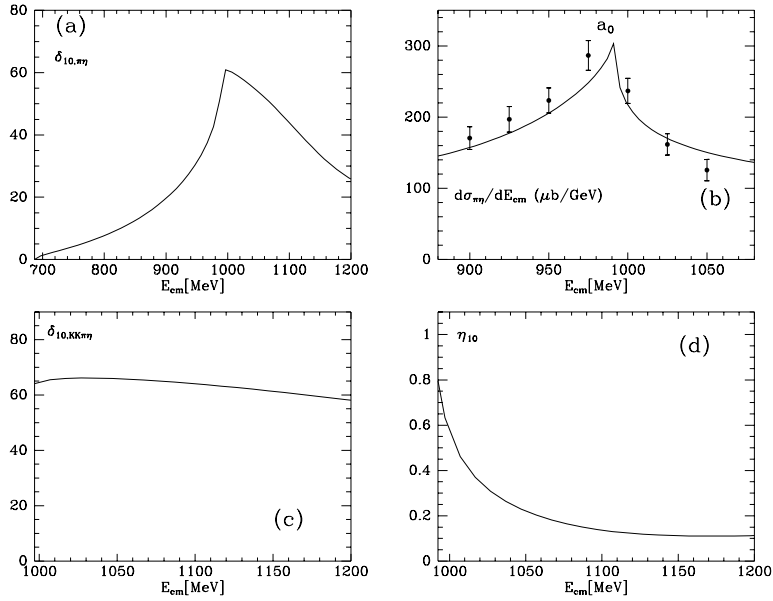,width=0.5\textwidth,angle=-90}}
  \caption{Results in the $I=1$, $L=0$ channel. (a) phase shifts 
  for $\pi\eta\rightarrow \pi\eta$. (b) Invariant mass distribution for
  $\pi\eta$ data from \cite{Amsterdam}. (c) Phase shifts for $K \bar{K}
  \rightarrow \pi \eta$. (d) inelasticity.}
\end{center}
\end{figure}

Some comment is needed with respect to the (0,1) channel, Fig. 3.13. In this channel
a pole appears with a mass around 910 MeV. Below 1.2 GeV there are two 
resonances with such quantum numbers. They are the
$\omega$ and the $\phi$, which fit well within the $q\bar{q}$ 
scheme, with practically ideal mixing, as 
$\frac{1}{\sqrt{2}}(u\bar{u}+d\bar{d})$ and $s\bar{s}$, respectively. In the 
limit of exact SU(3) symmetry these resonances manifest as one antisymmetric 
octet state and a symmetric singlet state. Since the spatial function of the 
$K\bar{K}$ state is antisymmetric its SU(3) wave function has to be also
antisymmetric and therefore it only couples to the antisymmetric octet
resonance. Of course, the amplitudes given by eq. (\ref{t4ap}), do contain 
some SU(3) breaking, but,
in this channel only the $K\bar{K}$ state is considered,
neglecting states with other mesons (like the three pion channel) and, 
hence, the formulae for this process do not contain any SU(3) symmetry breaking term.
Thus, one just sees one pole, $\omega_8$, corresponding to the antisymmetric 
octet state of the exact SU(3) limit. The $\omega_8$ meson will be studied in further 
detail in {\it{section 4.2}}, in connection with the $\phi$
resonance and its decays.

\begin{figure}[ht!]
\hbox{
\psfig{file=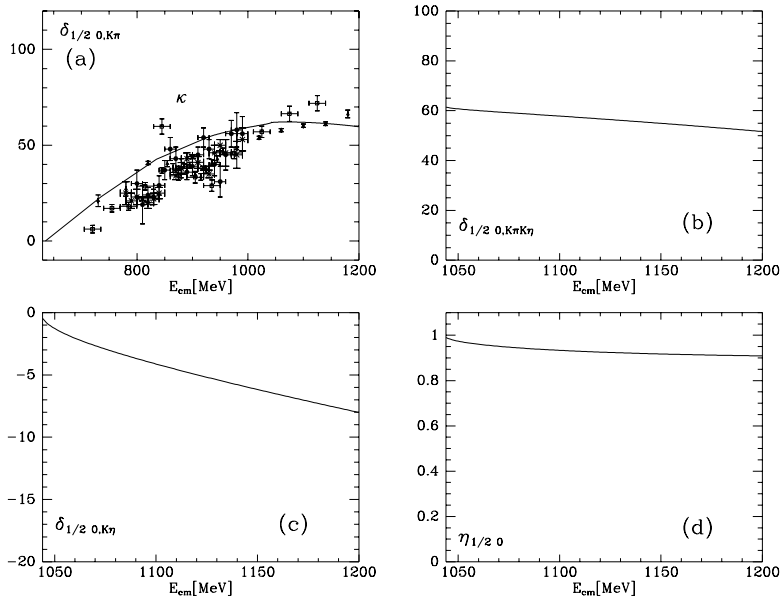,width=.5\textwidth,angle=-90}}
  \caption{Results in the $I=1/2$, $L=0$ channel. (a) phase shifts for
  $K\pi\rightarrow K\pi$. Data: full circle \cite{Mercer}, cross \cite{Bingham}
, open square \cite{Baker}, full triangle \cite{Estabrooks3}, open circle 
\cite{Aston}. (b) phase shifts for $K\pi\rightarrow K\eta$. (c) phase shifts 
for $K\eta \rightarrow  K\eta$. (d) inelasticity.}
\end{figure}

\begin{figure}[ht!]
\hbox{
\psfig{file=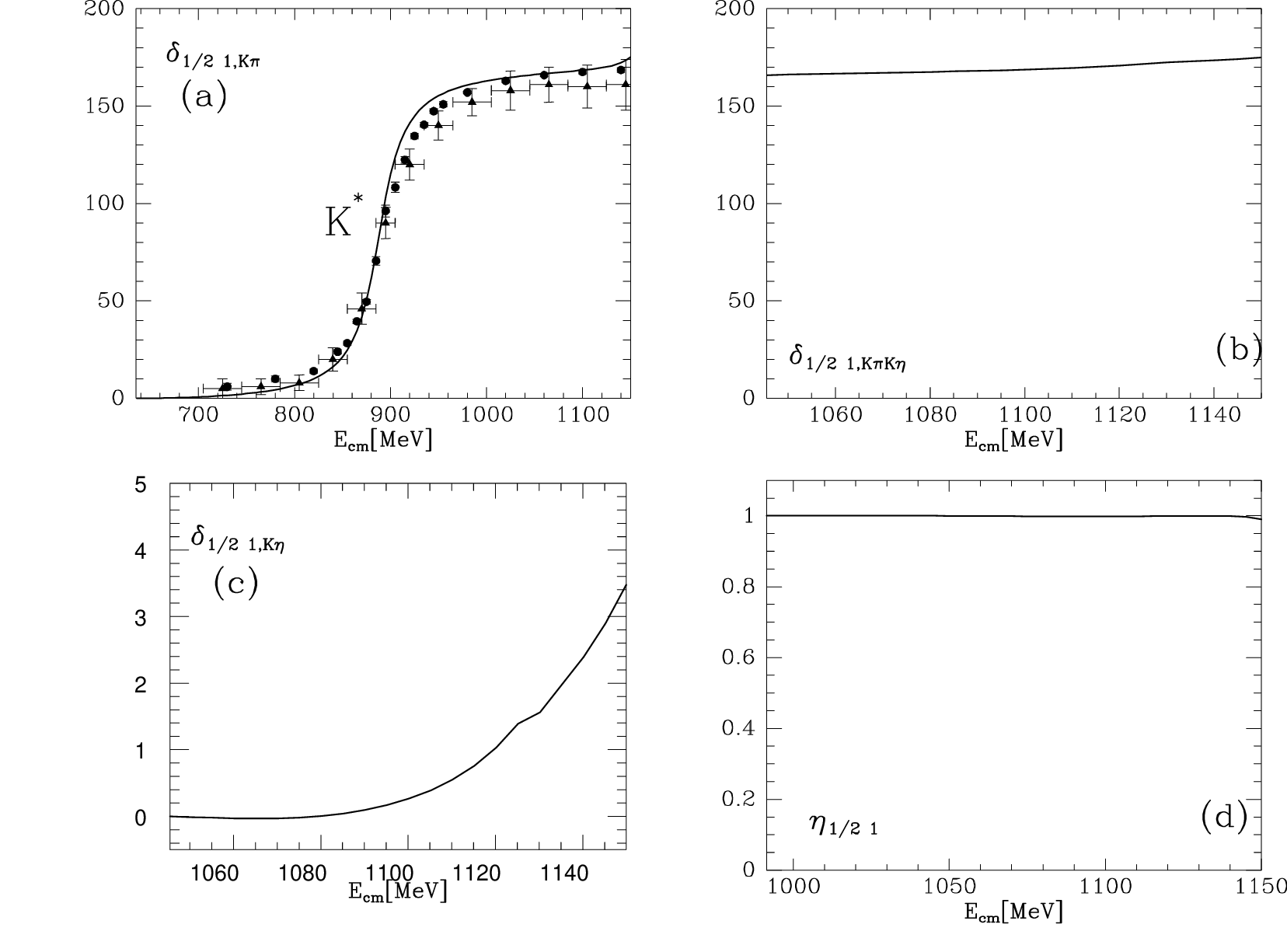,width=.5\textwidth,angle=-90}}
  \caption{Results in the $I=1/2$, $L=1$ channel. (a) phase 
  shifts for $K\pi\rightarrow K\pi$. Data: full triangle \cite{Mercer}, 
  open circle \cite{Estabrooks3}.  (b) phase shifts for
  $K\pi\rightarrow K\eta$. (c) phase shifts for $K\eta\rightarrow
  K\eta$. (d) inelasticity.}
\end{figure}

\begin{figure}[ht!]
    \hbox{ \psfig{file=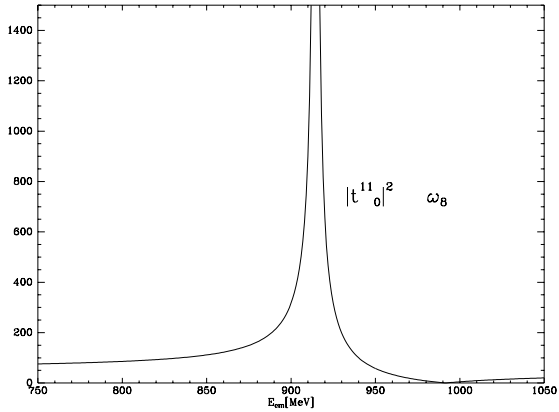,width=0.35\textwidth,angle=-90}}
\caption{ $\vert T_{IJ=01} \vert^2$ for $K \bar{K} \rightarrow K
  \bar{K}$ showing a singularity corresponding to a resonance belonging to the
  antisymmetric vector octet.}
\end{figure}

\begin{table}[H]
\begin{center}
\label{libru}
\caption{Fit parameters $\hat L_i \cdot 10^{3}$ \cite{newPRD} and comparison 
with the $L_i^r \cdot 10^{3}$ of $\chi PT$ }
\begin{tabular}{c|cccccccc}
\hline
 &
$q_{max}$ (MeV)  
& $ \hat L_1$ & $\hat L_2$& $\hat L_3$ & $\hat L_4$ & $\hat L_5$ & $2 \hat L_6 + \hat L_8$& 
$ \hat L_7 $ \\ 
\hline     
Fit  & 647 & 0.94& 1.60 & -3.74& -.019& 0.94& 0.67& -.051 \\
\hline 
\hline

$\chi PT$ & DR Scale & $L_1$ & $L_2$ & $L_3$ & $L_4$ & $L_5$ & $2 L_6+L_8$&
$ L_7 $\\ 
\hline

 &$\mu= M_\rho$&$ \begin{array}{cc} 0.4 \\ \pm 0.3 \\ \end{array}$&$ 
\begin{array}{c} 1.4 \\ \pm 0.3\\ \end{array}$ 
& $ \begin{array}{cc} -3.5 \\ \pm 1.1 \\ \end{array} $ & $ 
\begin{array}{cc} -0.3 \\ \pm 0.5\\ \end{array}$ & $ \begin{array}{cc} 1.4 
\\ \pm 0.5 \\ \end{array}$ 
& $ \begin{array}{cc} 0.5 \\ \pm 0.7 \\ \end{array}$& $\begin{array}{cc} -0.4
\\ \pm 0.2 \\ \end{array}$ \\   
\hline
\end{tabular}
\end{center}
\end{table}

\subsubsection{Pole positions, widths and partial decay widths}

In ref. \cite{PRD} the poles of the $T$-matrix in the complex
plane, which appear in the unphysical Riemann sheets, are also studied and we
refer to this reference for further details. These poles correspond to the 
$f_0(980)$,
$a_0(980)$, $\sigma$, $\rho(770)$, $K^*(890)$, $\kappa$ and the $\omega_8$ resonance. 
The changes in the values of the masses, partial and total decay widths of the
different resonances \cite{PRD}, due to the error detected in 
ref. \cite{newPRD}, are rather small except in the case of the $a_0(980)$ 
resonance. For the values of the $\hat{L}_i$ coefficients given in Table 3.4, the pole of 
the $a_0(980)$ in the second sheet ($k_{\pi^0 \eta}<0$, 
$k_{K\bar{K}}>0$) appears at $(1195-i\,350)$ MeV. However, a new one appears in
the fourth sheet 
($k_{\pi^0 \eta}>0$, $k_{K\bar{K}}<0$) at: $(1111-i\,118)$ MeV. When doing a fit, 
the $I=1$, $L=0$ channel has very little statistical weight due to the lack
of experimental data so that an improvement of this experimental situation would be very 
welcome. For instance, we will see in the next section a good reproduction of the
experimental scalar data showing an $a_0(980)$ resonance with just a pole also in the second 
sheet.

\section{Inclusion of Explicit Resonance Fields}
	
	$\chi PT$ \cite{xpt} can be supplied with the exchange of explicit 
resonance fields \cite{EPR}. In doing this, a resummation up to an infinite
order in the chiral expansion can be achieved from the
expansion of the bare propagator of a resonance, as we have already seen in eq.
(\ref{bareprop}). 
However, the amplitudes that can be
built directly from $\chi PT$ at $\mathcal{O}(p^4)$ plus resonance exchanges as
in ref. \cite{UMR} need a unitarization procedure, in order to compare directly
with experimental data (phase shifts, inelasticities...) for the different
energy regions, in particular, around the resonance masses. This is one of the
aims of the present section.

	On the other hand, it is well known that the scalar sector is much more
controversial than the vector one. In the latter case, the
properties of the associated spectroscopy can be understood in terms of first 
principles coming directly from QCD as Chiral
Symmetry and Large $N_c$ plus unitarity, once we admit VMD as dictated by
phenomenology \cite{FP,NewP}. In ref. \cite{OOnd} the same 
basic principles than before, that is, Chiral Symmetry, Large $N_c$ and 
unitarity in coupled channels, were applied in order to study the scalar 
resonant channels. We will 
also pay special attention to the issue of the nature of the scalar resonances 
that we will find in the amplitudes. As it is well known, the low energy scalar 
resonances have been ascribed \cite{Montanet} to different models as: 
conventional $q\bar{q}$ mesons \cite{Tornqvist,Morgan}, 
$q^2\bar{q}^2$ states \cite{Jaffe,Achasov}, $K\bar{K}$ molecules 
\cite{Weinis,Jansen,npa}, glueballs \cite{5deMP93} 
  and/or hybrids \cite{6deMP93}. The question about the nature of the resonances
  is specially important in order to determine their contributions to the $L_i$ 
counterterms. For instance, the resonances considered in ref. \cite{EPR}
are supposed to be preexisting ones 
(their masses are $\mathcal{O}(1)$ in Large $N_c$ counting rules) and their
contributions to the previous couplings arise dominantly from their bare
propagators. However, as it is shown below, there are also resonances, like 
the $\sigma$ or the $a_0(980)$, that are 
originated from the interactions between the pseudoscalars \cite{OOnd} and 
hence their contributions to the $L_i$ come just from the loops of 
the pseudoscalars.

\subsection{Formalism}

We present here a formalism based on the N/D method \cite{17nd} in order
to 
provide physical amplitudes from $\chi PT$ supplied with the exchange of 
explicit resonance fields as given in ref. \cite{EPR}. This formalism was derived 
in ref. \cite{OOnd}.

A $T(s)$ partial wave amplitude has two kinds of cuts. The right 
hand cut 
required by unitarity and the unphysical cuts from crossing 
symmetry. In our chosen normalization, the right hand cut leads to 
eq. (\ref{ImG}): 

\begin{eqnarray}
\label{rhc}
\Ima T^{-1}&=&\rho (s)  
\end{eqnarray}
for $s>s_{threshold}\equiv s_{th}$. In the case of two particle scattering, 
the one we are concerned about, $s_{th}=(m_1+m_2)^2$ and $\rho(s)$ is 
given in eq. (\ref{1.0}).

The unphysical cuts comprise two types of cuts in the complex s-plane. For 
processes of the type $a+a\rightarrow a+a$ with $m_1=m_2=m_a$, there is only a 
left hand cut for $s<s_{Left}$. However for those of the type 
$a+b\rightarrow a+b$ with $m_1=m_a$ and $m_2=m_b$, apart from a left hand cut 
there is also a circular cut in the complex s-plane for $|s|=m_2^2-m_1^2$, 
where we have taken $m_2>m_1$. In the rest of this section, for simplicity 
in the formalism, we will just refer to the left hand cut as if it were the 
full set of unphysical cuts. This will
be enough for our purposes in this section. In any case, if one works in
the
complex $p^2$-plane all the cuts will be linear cuts and then only the left hand
cut will appear in this variable.

The left hand cut, for $s<s_{Left}$, reads:

\begin{equation}
\label{lhc}
T(s+i\epsilon)-T(s-i\epsilon)=2 i \Ima T_{Left}(s)
\end{equation}

The standard way of taking into account eqs. (\ref{rhc}) and (\ref{lhc})
is 
the N/D 
method \cite{17nd}. In this method a $T(s)$ partial wave is 
expressed as a ratio of two functions,

\begin{equation}
\label{n/d}
T(s)=\frac{N(s)}{D(s)}
\end{equation}
with the denominator function, $D(s)$, bearing the right hand cut and
the numerator function, $N(s)$, the unphysical cuts.
 
 In order to take explicitly into account the behavior of a partial wave 
 amplitude near threshold, which vanishes as $p^{2L}\equiv \nu^L$, we
 consider the new quantity, $T'$, given by:

\begin{equation}
\label{T'}
T'(s)=\frac{T(s)}{\nu^L}
\end{equation}
which also satisfies relations of the type of eqs. (\ref{rhc}) and 
(\ref{lhc}). Hence, we can write:

\begin{equation}
\label{n/d'}
T'(s)=\frac{N'(s)}{D'(s)}
\end{equation}

From eqs. (\ref{rhc}), (\ref{lhc}) and (\ref{T'}), $\hbox{N}'(s)$ and 
$\hbox{D}'(s)$ will obey the following equations:

\begin{equation}
\label{eqs1}
\begin{array}{ll}
\Ima  D'=\Ima T'^{-1} \; N'=\rho(s)
N' \nu^L,&  s>s_{th} \\
\Ima D'=0, &  s<s_{th}
\end{array}
\end{equation}
\begin{equation}
\label{eqs2}
\begin{array}{ll}
\Ima  N'=\Ima T'_{Left} \; D', & s<s_{Left}  \\
\Ima N'=0, & s>s_{Left}
\end{array}
\end{equation}
where $\displaystyle{\Ima T'_{Left}=\frac{\Ima T_{left}}{\nu^L}}$.

It is important to notice that $N'$ and $D'$ can be simultaneously multiplied 
by any arbitrary real analytic function without changing its ratio, 
$\hbox{T}'$, nor eqs. (\ref{eqs1}) and (\ref{eqs2}). In this way, we can 
always consider $N'$ free of poles.

Thus, using dispersion relations for $N'_L(s)$, we 
write from eq. (\ref{eqs2}):

\begin{equation}
\label{n2}
N'(s)=\frac{(s-s_0)^{n+1}}{\pi}\int_{-\infty}^{s_{Left}} ds' 
\frac{\Ima  T'_{Left}(s') D'(s')}{\nu(s')^L (s'-s_0)^{n+1} (s'-s)}+
\sum_{m=0}^{n} \overline a '_m s^m
\end{equation}      
with $n$ such that

\begin{equation}
\label{qen}
\lim_{s\rightarrow \infty}\frac{N'(s)}{s^{n+1}}=0
\end{equation}

In the following, the unphysical cuts will be considered in a 
perturbative 
way. In fact, this was the case in the former section where we 
reported the IAM
results from refs. \cite{prl,NPB,PRD,newPRD}. In this method the unphysical 
cuts are only considered up to $\Opc$ 
as in the $\chi PT$ calculations. That is, from the resummation done by the IAM 
one obtains fully unitarized partial wave amplitudes but satisfying crossing
symmetry perturbatively up to $\Opc$. As a matter of fact, we have 
seen that one can
reproduce rather accurately the meson-meson interactions. Hence, 
the approach of considering the unphysical cuts in a perturbative way seems 
to be a
realistic one for the physical region. First, we study the zeroth order approach, 
that is, no unphysical
cuts at all, obtaining the most general structure that a
partial wave amplitude has in such case. Note that any unitarization method 
without unphysical cuts must then implement this structure (as an example
see
the amplitudes of refs. \cite{npa,nieves}). 
Later on, we will also consider the inclusion of the unphysical cuts up to 
one loop calculated at $\Opc$. In this case, the $\chi PT$ expansion 
up to $\Opc$ will be recovered for the low energy region and the IAM, 
eq. (\ref{iamcc}), will also appear as a special case.

\subsubsection{No unphysical cuts}

In this case, we have $\Ima T_{Left}=0$ and then from eq. (\ref{n2}):

\begin{equation}
\label{nlhc}
N'(s)=\sum_{m=0}^{n} a'_m s^m
\ee
that is, $N'$ is just a polynomial. Then, after dividing $N'$ and $D'$ by
this polynomial, one has:

\be
\label{n'1}
N'=1
\ee
and the dispersion relation for $D'$ will read from eq. (\ref{eqs1})

\be
\label{fin/d}
D'(s)=\frac{(s-s_0)^{L+1}}{\pi}\int_{s_{th}}^\infty ds' \frac
{\nu(s')^L \rho (s')}{(s'-s)(s'-s_0)^{L+1}}+\sum_{m=0}^L a_m s^m+
\sum_i^{M_L} \frac{R_i}{s-s_i}
\ee
where the last sum takes into account the possible presence of poles of $D'$
(zeros of $T'$) inside and along the integration contour, which is given by a
circle in the infinity deformed to engulf the real axis along the right hand
cut. Each term of this sum is referred to as a Castillejo-Dalitz-Dyson (CDD) 
pole after ref. \cite{CDD}. Note that since $N'=1$ from eq. (\ref{n/d'})
$\displaystyle{T'=\frac{1}{D'}}$.

Let us come back to QCD and split the subtraction constants $a_m$ of 
eq. (\ref{fin/d}) in two pieces

\begin{equation}
\label{split}
a_m=a^L_m+a^{SL}_m(s_0)
\end{equation}
The term $a_m^L$ will go as $N_c$, because in the $N_c \rightarrow \infty$ limit, 
the meson-meson amplitudes go as $N_c^{-1}$ \cite{Witten}. Since 
the integral in eq. (\ref{fin/d}) is $\mathcal{O}(1)$ in this counting, the 
subleading term $a_m^{SL}(s_0)$ is of the same order and depends on the 
subtraction point $s_0$. This implies that eq.
 (\ref{fin/d}), when $N_c \rightarrow 
\infty$, will become 

\begin{equation}
\label{limitd'}
D'(s)\equiv D'^{\infty}(s)=\sum_{m=0}^L 
a^L_m s^m+\sum_i^{M^\infty_L}\frac{R^\infty_i}{s-s_i}
\end{equation}    
where $R^\infty_i$ is the leading part of $R_i$ and $M^\infty_L$ counts the
number of leading CDD poles.

Clearly eq. (\ref{limitd'}) represents tree level structures, contact and pole 
terms,  which have nothing to do with any kind of potential scattering, 
which in large $N_c$ QCD is suppressed. 

In order to determine eq. (\ref{limitd'}) one can make use of $\chi PT$ 
\cite{xpt} and of the paper \cite{EPR}. In this latter reference the way to include 
resonances with spin $\leq 1$, consistently with chiral 
symmetry at lowest order in the chiral power counting, is shown. It is also 
seen 
that, when integrating out the resonance fields, the contributions of the 
exchange of these resonances essentially saturate the next to leading 
$\chi PT$ Lagrangian. We will make use of this result in order to state 
that in the inverse of eq. (\ref{limitd'}) the contact terms come just from 
the lowest order $\chi PT$ Lagrangian and the pole terms from the exchange 
of resonances in the s-channel in the way given by ref. \cite{EPR} (consistently 
with our approximation of neglecting the left hand cut the exchange 
of resonances in crossed channels is not considered). In the latter statement 
it is assumed that the result 
of ref. \cite{EPR} at ${\mathcal{O}}(p^4)$ is also applicable to higher orders. 
That is, 
local terms appearing in $\chi PT$ and from eq. (\ref{limitd'}) of order higher than 
${\mathcal{O}}(p^4)$ are also saturated from the exchange of resonances for 
$N_c>>1$, where loops are suppressed.

In ref. \cite{OOnd} it is proved that eq. (\ref{limitd'}) can accommodate the
tree level amplitudes coming from lowest order $\chi PT$ \cite{xpt} and the
Lagrangian given in ref. \cite{EPR} for the coupling of resonances (with spin$\leq$1)
with the lightest pseudoscalars ($\pi$, $K$, $ \eta$).

Thus, if we denote by $T_2$ the $\Opd$ $\chi PT$ amplitudes and by $T^R$ the
contribution from the s-channel exchange of resonances according to ref. \cite{EPR},
we can write:

\be
\label{input1}
T^\infty\equiv T_2+T^R=\nu^L \left[D'^\infty \right]^{-1}
\ee

On the other hand, we define the function $g_L(s)$ by

\begin{equation}
\label{g(s)}
g_L(s) \nu^L=-\sum_{m=0}^L a^{SL}_m(s_0) s^m-\frac{(s-s_0)^{L+1}}{\pi}
\int_{s_{th}}^\infty ds' \frac{\nu(s')^L \rho(s')}{(s'-s)(s'-s_0)^{L+1}}
\end{equation}

After these definitions, we can write our final formula for $T(s)$, in the case
that unphysical cuts are not considered, as:

\be
\label{lastndf}
T(s)=\left[(T^\infty)^{-1}-g_L(s)\right]^{-1}
\ee

The physical meaning of eq. (\ref{lastndf}) is clear. The $T^\infty$ amplitudes
correspond to the tree level structures present before unitarization. The
unitarization is then accomplished through the function $g_L(s)$.

From the former comments the generalization of eq.
(\ref{lastndf}) to coupled channels should be obvious. In this case, 
$T^\infty (s)$ is a matrix determined by the 
tree level partial wave amplitudes given by the lowest order 
$\chi PT$ Lagrangian \cite{xpt} and the exchange of resonances \cite{EPR}. 
For instance, $[T^\infty(s)]_{11}=(T_2)_{11}+T^R_{11}$, 
$[T^\infty_L(s)]_{12}=(T_2)_{12}+T^R_{12}$ and so on. Once we have 
$T^\infty(s)$ its inverse is the one 
which enters in eq. (\ref{lastndf}). Because $N'(s)$ is proportional to 
the  identity, $g_L(s)$ will be a diagonal matrix, accounting for the right
 hand cut, as in the elastic case. In this way, unitarity, which 
 in coupled channels reads (above the thresholds of the channels $i$ and $j$) 

\begin{equation}
\label{ccu}
[\Ima  T^{-1}]_{ij}=\rho_{ii}(s) \delta_{ij}
=-\Ima  g_L(s)_{ii} \delta_{ij}
\end{equation}
is fulfilled. The matrix element $g_L(s)_{ii}$ obeys eq. (\ref{g(s)}) with the 
right masses corresponding to the channel $i$ and its own subtraction constants 
$a^{SL}_i(s_0)$.

In ref. \cite{OOnd} the coupling constants and resonance masses contained in 
$\hbox{T}^\infty_L(s)$ are fitted to the experiment. The same 
happens with the $a^{SL}_i$ although, as we will discuss below, they are 
related by $SU(3)$ considerations.

In Appendix A of ref. \cite{OOnd} the already mentioned coupled channel
version 
of eq. (\ref{lastndf}) is deduced directly from the N/D method in coupled 
channels \cite{Bjorken}. 

\subsubsection{The unphysical cuts at one loop at $\Opc$}
   
In a full calculation for a meson-meson partial wave amplitude combining 
$\chi PT$ at $\Opc$ and the exchange of resonances as done in ref. \cite{UMR} one
has to include the diagrams depicted in Fig. 3.14. The lowest order 
$\chi PT$ amplitudes,
$T_2$, plus the exchange of resonances in the s-channel, $T^R$, were already
taken into account in the previous section. The sum of both contributions was
denoted by $T^\infty$.  One also generates
through the $g_L(s)$ function the loops in the s-channel responsible for
unitarity. Thus, in matrix notation at $\Opc$ we have from eq. (\ref{lastndf}): 
$T_2\cdot g_L(s) \cdot T_2$. Note that the difference between the loops 
in the s-channel calculated in $\chi PT$ at $\Opc$ and the ones we have 
obtained can be at most of a polynomial of second order in $s$. We denote the 
rest of 
contributions coming from the exchange of resonances and loops in the crossed 
channels, tadpole-like contributions and the previous difference
between our loops in the s-channel and the ones from $\chi PT$ by $T_{Left}$,
since they only have unphysical cuts. Then, in the present notation, a 
partial wave amplitude calculated as in ref. \cite{UMR} can be written as:

\be
\label{ournotation}
T^\infty+T_{Left}+T_2\cdot g_L(s) \cdot T_2
\ee

\begin{figure}
\label{figUMNR}
\includegraphics[width=9cm,angle=-90]{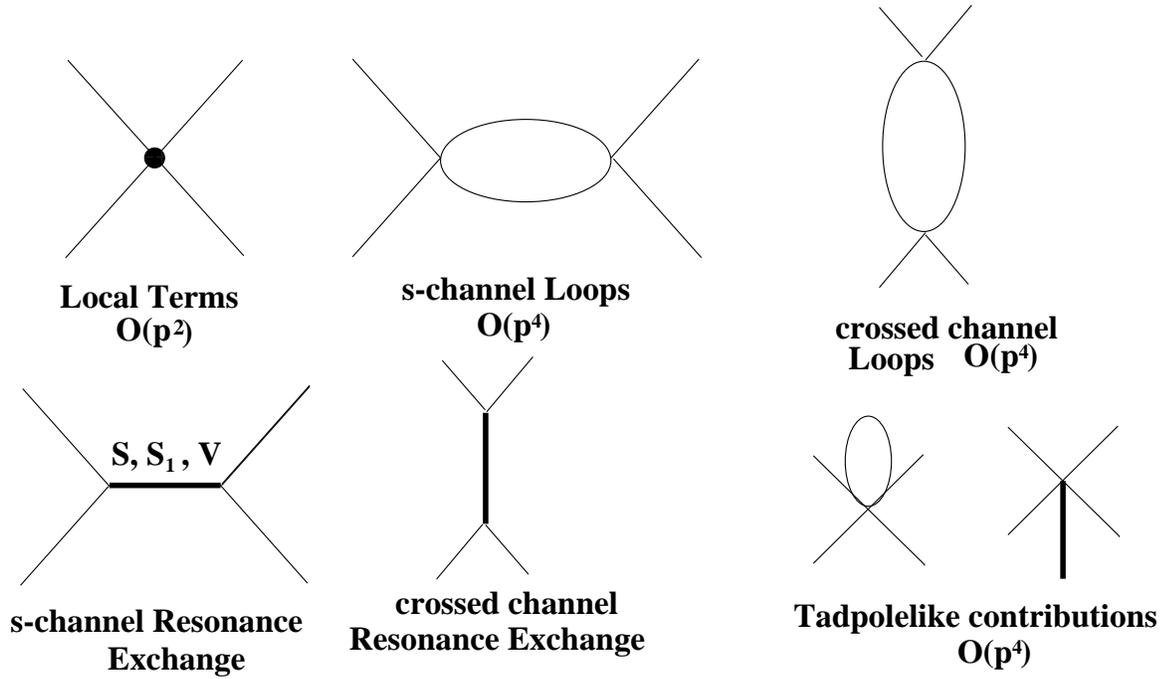}
\caption{Diagrams contained in eq. (\ref{ournotation}), from $\chi PT$ up to
$\Opc$ and from the exchange of resonances \cite{EPR}. Wave function
renormalization is not depicted.}
\end{figure}

An N/D representation of the former amplitude can be done as follows. This
representation contains the unphysical cuts up to the order considered in
eq. (\ref{ournotation}), that is, up to one loop calculated at $\Opc$. In matrix
formalism:

\begin{eqnarray}
\label{ndrep}
N&=&T^\infty+T_{Left} \\ \nn
D&=&I-N \cdot g_L \\ \nn
T&=&D^{-1}\cdot N
\end{eqnarray} 
Making a chiral expansion of the former result, one has:

\begin{equation}
T=N+N\cdot g \cdot N + \mathcal{O}(p^6 \hbar,\hbar^2)=T^\infty_0+T_{Left}+T_2
 \cdot g_L \cdot T_2+ \mathcal{O}(p^6 \hbar,\hbar^2)
\end{equation}
reproducing eq. (\ref{ournotation}). In the former equation, 
$\mathcal{O}(p^6 \hbar,\hbar^2)$ indicates that the result is valid up to one
loop calculated at $\Opc$. One 
can also check that, up to the same order in the unphysical cuts, the $N$ and 
$D$ functions satisfy eqs. (\ref{eqs1}) and (\ref{eqs2}):

\begin{eqnarray}
\Ima \, D&=&N\cdot \Ima \, g_L=N \cdot \rho(s) \hspace{2.7cm} 
s>s_{th} \\ \nn
\Ima \, D&=&\mathcal{O}(p^6 \hbar,\hbar^2)  
\phantom{D \cdot \Ima \, T= \Ima \, T_{Left}+}
\ \ \ \
s\hbox{ in unphysical cuts}
\\ \nn
\Ima \, N&=&D \cdot \Ima \, T= \Ima \, T_{Left}+
\mathcal{O}(p^6 \hbar,\hbar^2) \ \ \ 
s\hbox{ in unphysical cuts} 
\end{eqnarray}

One can reabsorb $T^\infty_0$ in $D$ just by multiplying $N$ and $D$ at the same
time by $(T^\infty_0)^{-1}$. In this way, neither their ratio nor their cut structure 
are modified since $T^\infty_0$ is just a matrix of real rational 
functions. Then one has

\begin{eqnarray}
\label{ndrep2}
N&=&I+(T^\infty)^{-1}\cdot T_{Left} \\ \nn
D&=&(T^\infty)^{-1}-(I+(T^\infty)^{-1}\cdot T_{Left})\cdot g_L(s) 
\end{eqnarray}

In any case eq. (\ref{ndrep}) can also be written as:

\begin{equation}
\label{fin/d6}
T=\left[ (T^\infty+T_{Left})^{-1}-g_L(s) \right]^{-1}
\end{equation}
setting $T_{Left}$ to zero we recover once again the limit case of 
eq. (\ref{lastndf}), where no unphysical cuts are included.

The former equation has being used in ref. \cite{Jamin} to describe the coupled 
channel scattering of $K\pi$ and $K\eta'$ in order to obtain the 
scalar $K\pi$ form factor. It has also being used in ref. \cite{ohiggs} to describe
the strong $WW$ scattering for a heavy Higgs boson. The main conclusion of this
work is that for a general scenario with heavy particles with a mass much larger
than $4\pi v\approx 3$ TeV, a isoscalar-scalar WW resonance would appear with a 
mass $\lesssim$1 TeV. As a consequence, this resonance, which would not be 
responsible for the spontaneous breaking of the electroweak symmetry 
$SU(2)_L\times U(1)_Y$, could be confused with a true Higgs particle with a mass 
around 1 TeV at LHC.

From eq. (\ref{fin/d6}) one can also reobtain the IAM with coupled channels, 
eq. (\ref{iamcc}). In order to do this, let us expand the inverted matrix in eq.
(\ref{fin/d6}) analogously as in eq. (\ref{TIChPT}):

\ba
\label{iamred}
(T^\infty+T_{Left})^{-1}-g_L(s)&=&(T_2+\mathcal{T}_4+...)^{-1}-g_L(s)
\nonumber\\
&=&T_2^{-1}\cdot (1+\mathcal{T}_4\cdot T_2^{-1}+...)^{-1}-g_L(s)
\nonumber \\ &=& T_2^{-1}
\cdot (1-\mathcal{T}_4\cdot T_2^{-1}+...)-g_L(s)\nonumber \\
&=&T_2^{-1}\cdot (T_2-\mathcal{T}_4 -T_2 \cdot g_L(s) \cdot T_2+...) \cdot
T_2^{-1}\nonumber \\
&\approx&T_2^{-1}\cdot (T_2-T_4) \cdot T_2^{-1}
\ea

 In the previous equation $\mathcal{T}_4$ is the $\Opc$ contribution of 
$T^\infty+T_{Left}$. Inverting the former result and assuming the saturation 
of the $L_i$ coefficients by the exchange of resonances \cite{EPR}, one 
recovers eq. (\ref{iamcc}).

\subsection{Results I: The vector sector}

In ref. \cite{OOnd} the $\pi\pi$ and $K\pi$ scattering 
with $I$=$L$=1 and  $I$=1/2, $L$=1, respectively, were studied. As we will see 
below, one reproduces the well known
features associated with the vectors: VMD and the KSFR relation \cite{KSFR} for
the couplings of these resonances to the pseudoscalars.

For the special case of the P-waves, since the zero at threshold is a simple
one, instead of eq. (\ref{fin/d}), we consider the slightly modified formula:

\begin{equation}
\label{d1}
\hbox{D}(s)=\sum_{i}\frac{\gamma_i}{s-s_i}+a-\frac{s-s_0}{\pi} 
\int_{s_{th}}^\infty ds' \frac{\rho(s')}{(s'-s)(s'-s_0)}
\end{equation}
where the threshold zero has passed to poles in the denominator function, $D$.
This equation is derived analogously to eq. (\ref{fin/d}) but working 
directly
with $T$ rather than with $T'$. Another advantage of 
using eq. (\ref{d1}) instead of eq. (\ref{fin/d}) is that the comparison with 
the scalar sector will be more straightforward, because the dispersive integral 
will be the same.

The integral in eq. (\ref{d1}) will be evaluated making use of dimensional 
regularization. It can be identified up to a constant with eq. (\ref{5.0}). 
This identification is a consequence of the fact that both the integral
in eq. (\ref{d1}) and the one in eq. (\ref{5.0}) have the same cut and 
the same imaginary part along this cut, as it can be easily checked. Thus, one
has:

\begin{eqnarray}
\label{g(s)dr}
g_0(s)&=&-a^{SL}(s_0)-\frac{s-s_0}{\pi}\int_{s_{th}}^\infty ds' 
\frac{\rho(s')}{(s'-s)(s'-s_0)}\nonumber\\
&=&\frac{1}{(4\pi)^2}\Bigg[\tilde{a}^{SL}(\mu)+\ln \frac{m_2^2}{\mu^2} 
-\frac{m_1^2-m_2^2+s}{2s}\ln\frac{m_2^2}{m_1^2}-\frac{\lambda^{1/2}(s,m_1^2,
m_2^2)}{2s} \cdot \nonumber \\
&& \cdot  \ln\Big( \frac
{m_1^2+m_2^2-s+\lambda^{1/2}(s,m_1^2,m_2^2)}
{m_1^2+m_2^2-s-\lambda^{1/2}(s,m_1^2,m_2^2)}\bigg)\Bigg]
\end{eqnarray}  
for $s\geq s_{th}$. For $s<s_{th}$ or $s$ complex one has the analytic 
continuation of eq. (\ref{g(s)dr}). The function $\lambda^{1/2}(s,m_1^2,
m_2^2)$ is given by $\sqrt{(s-(m_1+m_2)^2)(s-(m_1-m_2)^2)}$. 
The regularization scale $\mu$, appearing in the last equality of eq.
(\ref{g(s)dr}), plays a similar role than the arbitrary subtraction point
$s_0$
present in the first one. In fact, both of them are arbitrary although the
resulting
$g_0(s)$ function is well defined because any change in $\mu$ or $s_0$ is
reabsorbed in $\tilde{a}^{SL}(\mu)$ or $a^{SL}(s_0)$, respectively.
The $\tilde{a}^{SL}(\mu)$ `constant' will change under a 
variation of the scale $\mu$ to another one $\mu'$ as 

\begin{equation}
\label{scale}
\tilde{a}^{SL}(\mu')=\tilde{a}^{SL}(\mu)+\ln\frac{\mu'^2}
{\mu^2}
\end{equation}
in order to have $g_0(s)$ invariant under changes of the regularization scale. 
We will take $\mu=M_{\rho}=770$ MeV \cite{PDG}. The function $g_0(s)$ is 
also symmetric under the exchange $m_1\leftrightarrow m_2$ and for the 
equal mass limit it reduces to 

\begin{equation}
\label{g(s)dr2}
g_0(s)=\frac{1}{(4\pi)^2}\bigg[\tilde{a}^{SL}(\mu)+\ln\frac{m_1^2}{\mu^2}
+\sigma(s)\ln\frac{\sigma(s)+1}{\sigma(s)-1}\bigg]
\end{equation}
with 
\begin{equation}
\label{sigma2}
\sigma(s)=\sqrt{1-\frac{4m_1^2}{s}}
\end{equation}

Let us consider first the case of the P-wave $\pi\pi$ scattering \cite{OOnd}.
Taking into account the zero at threshold, from eq. (\ref{d1}) we have:

\begin{equation}
\label{Rho}
\hbox{T}^{\pi\pi}(s)=\bigg[\frac{\gamma^{\pi\pi}_1}{s-4m_{\pi}^2}+
\tilde{a}^L_{\pi\pi}-g^{\pi\pi}_0(s)+\sum_{i=2}\frac{\gamma_i}{s-s_i}\bigg]^{-1}
\end{equation}

On the other hand, from $\chi PT$ and the exchange of the $\rho$ \cite{EPR}, one 
has:

\begin{equation}
\label{t1n2}
\hbox{T}^{\pi\pi \, \infty}(s)=-\frac{2}{3}\frac{p_{\pi\pi}^2}
{f^2}+g_v^2 \frac{2}{3}\frac{p_{\pi\pi}^2}{f^2}\frac{s}{s-M_\rho^2}
\end{equation} 
with $p_{\pi\pi}^2$ the three-momentum squared of the pions in the c.m., 
$f=87.3$ MeV the pion decay constant in the chiral limit \cite{xpt}. 
The deviation of $g_v^2$ with respect to unity measures the variation of the 
value of the $\rho$ coupling to two pions with respect to the KSFR relation 
\cite{KSFR}, $g_v^2=1$. In ref. \cite{EP2} this KSFR relation is justified 
making use of large $N_c$ QCD (neglecting loop contributions) and an 
unsubtracted dispersion relation for the pion electromagnetic form factor 
(a QCD inspired high-energy behavior).

Comparing eqs. (\ref{Rho}) and (\ref{t1n2}), one needs only one additional CDD
pole apart from the one at threshold and we obtain

\begin{eqnarray}
\label{aLpi}
\tilde{a}^L&=& 0 \nonumber \\
\gamma^{\pi\pi}_1&=&\frac{6 f^2 (4m_\pi^2-M_\rho^2)}
{(M_\rho^2-4m_\pi^2(1-g_v^2))}\nonumber\\
\gamma^{\pi\pi}_2&=&\frac{6 f^2}{g_v^2-1}\; \frac{g_v^2 M_\rho^2}{M_\rho^2-
(1-g_v^2)4 m_\pi^2} \nonumber \\
s_2&=&\frac{M_\rho^2}{g_v^2-1}
\end{eqnarray}

The former equation is an example of the matching between both representations:
the one given by the N/D method \cite{17nd}, eq. (\ref{d1}), and the one derived from
chiral symmetry \cite{xpt,EPR}. In the following, we will not consider more this
matching and we will take directly $T^\infty$ as given by the lowest order $
\chi PT$ amplitudes \cite{xpt} and the exchange of resonances in the s-channel
\cite{EPR}, as discussed above.

For the P-wave $I=1/2$ $K\pi$ elastic amplitude the tree level amplitude, 
$T^{K\pi\,\infty}$, is the same than for pions but multiplying it by 3/4 and
substituting $p_{\pi\pi}$ by $p_{K\pi}$ and $M_\rho$ by $M_{K^*}$, the mass of
the $K^*(890)$ resonance. 

The subleading constant $\tilde{a}^{SL}$ present in $g_0(s)$, eq. 
(\ref{g(s)dr}), should be the same for the $\pi\pi$ and $K\pi$ states 
because the dependence of the loop in eq. (\ref{5.0}) 
on the masses of the intermediate particles is given by eq. (\ref{g(s)dr}). 
This point can be used in the opposite sense. That is, if it is not possible 
to obtain a reasonable good fit after setting $\tilde{a}^{SL}$ to be the same 
in both channels, some kind of SU(3) breaking is missing.

Making use of the minimization program MINUIT, in ref. \cite{OOnd} a 
simultaneous fit to the
elastic $\pi\pi$ and $K\bar{K}$ phase shifts from threshold up to
$\sqrt{s}\lesssim 1.2 $ GeV was given. As a result one has:

\begin{eqnarray}
\label{fit}
g_v^2&=&0.879\pm0.016\nonumber\\
\tilde{a}^{SL}&=&0.341\pm0.042
\end{eqnarray}    
the errors are just statistical and are obtained by increasing in one unit the 
$\chi^2$ per degree of freedom, $\chi^2_{d.o.f.}$. The $\chi^2_{d.o.f.}$ obtained
is around 0.8. 

The fact that $g_v$ deviates from 1 just by a 6$\%$ states clearly that the KSFR
result is phenomenologically successful.

\subsection{Results II: The scalar sector}

We know consider the S-wave $I$=0,1 and 1/2 amplitudes \cite{OOnd}. 
For the partial wave amplitudes with $L$=0 and $I$=0 and 1, coupled 
channels are fundamental in order to get an appropriate description of the 
physics involved up to $\sqrt{s}\leq1.3$ GeV. This is an important difference 
with respect to the former vector channels, essentially elastic in the considered
energy region. Up to $\sqrt{s}=1.3$ GeV the most important channels are:

\begin{equation}
\label{channelsnd}
\begin{array}{ll}
$I=0$ & \pi\pi(1), \; K\bar{K}(2), \; \eta\eta(3)\\
$I=1$ & \pi\eta(1), \;  K\bar{K}(2)\\
$I=1/2$& K\pi(1), \; K\eta(2) 
\end{array}
\end{equation}
where the number between brackets indicates the index associated to the 
corresponding channel when using a matrix notation.

For the $I$=0 S-wave, the $4\pi$ state becomes increasingly important at energies 
above $1.2-1.3$ GeV, so that, in this channel, one is at the limit of 
applicability of only two meson states when $\sqrt{s}$ is close to 1.4 GeV. 
In the $I$=1/2 channel, the threshold of the important $K\eta'$ state is also 
close to 1.4 GeV. Thus, one cannot go higher in energies in a realistic 
description of the scalar sector without including the $K\eta'$ and 
$4 \pi$ states.

In order to fix $T^\infty$ one needs to include explicit resonance fields. From
ref. \cite{PDG}, two sets of resonances appear in the $L=0$ partial wave
amplitudes. A first one, with a mass around 1 GeV, contains the $I$=0 
$f_0(400-1200)$ and $f_0(980)$ and the $I$=1 $a_0(980)$. A second set appears 
with a mass around $1.4$ GeV as the $I$=0 $f_0(1370)$ 
and the $f_0(1500)$, the $I$=1 $a_0(1450)$ or the $I$=1/2 $K^*_0(1430)$. As a 
consequence, we first include the exchange of two scalar 
nonets, with masses around 1 and 1.4 GeV. In ref. \cite{OOnd} the expressions
for the $T^\infty_{ij}$ partial waves are collected. Once again, the 
minimization program MINUIT was used in order to fit the SU(3) related 
experimental data represented in Figs. 3.15--3.19. It happens that the
couplings of the octet around 1 GeV and the singlet around 1.4 GeV are
compatible with zero, giving rise to very narrow peaks that are not seen in
experiment. In fact, one obtains an equally good fit by including one singlet
around 1 GeV and an octet around 1.4 GeV. The values of the parameters given 
by the fit are \cite{OOnd}:

\begin{equation}
\label{fit2}
\begin{array}{ll}
\hbox{Nonet (MeV)} & \\
c_d=19.1^{+2.4}_{-2.1} & a^{SL}=-.75\pm 0.2\\
c_m=15\pm 30& {\mathcal{N}}=( 9.4 \pm 4.5)\, 10^{-5}\;\,\hbox{MeV}^{-2}\\
M_8=1390\pm 20  &\\
\tilde{c}_d=20.9^{+1.6}_{-1.0} &   \chi^2_{d.o.f.}=1.07 \\
\tilde{c}_m=10.6^{+4.5}_{-3.5} & 188 \;\, \hbox{points}\\
M_1=1021^{+40}_{-20}&
\end{array}
\end{equation}

\subsubsection{Resonances} 

We consider now the resonance content of the former fit. The octet around 1.4 
GeV gives rise to eight resonances which
appear with masses very close to the physical ones, $f_0(1500)$, $a_0(1450)$ 
and $K^*_0(1430)$ \cite{PDG}. However, a detailed
study of the former resonances is not given because one has not 
included channels which become increasingly important for energies above 
$\gtrsim$ 1.3 GeV as $4\pi$ in $I$=0 or $K\eta'$ for $I$=1/2. This makes that the 
widths obtained from the pole position of the former resonances are 
systematically smaller than
the experimental ones \cite{PDG}. Thus, a more detailed study, which
included all the relevant channels for energies above 1.3 GeV, should be done 
in order to obtain a better determination of the parameters for this octet 
around 1.4 GeV. 

On the other hand, from Figs. 3.15 and 3.18, one can easily see two resonances with 
masses around 1 GeV, the well known $f_0(980)$ and $a_0(980)$ resonances. 
The first one is related to the singlet bare state with $M_1=1020$ MeV, 
but for the second there are no bare resonances to associate with, because the 
tree level resonance
was included with a mass around 1.4 GeV and has evolved to the physical
$a_0(1450)$. The situation is even more complex, because we also find in the 
amplitudes other
poles corresponding to the $f_0(400-1200)\equiv \sigma$ and to the
$K^*_0\equiv \kappa$. In Table 3.5 the pole positions of the resonances in
the second sheet\footnote{I sheet: Im $p_1>$0, Im $p_2>$0, Im $p_3>$0; II 
sheet: Im $p_1<$0, Im $p_2>$0, Im $p_3>$0} are given and also the modulus of 
the residues
corresponding to the resonance $R$ and channel $i$, $\zeta^R_i$, given by

\begin{equation}
\label{residue}
|\zeta^R_i \zeta^R_j|=\lim_{s \to s_R} |(s-s_R) \hbox{T}_{ij}|
\end{equation}
where $s_R$ is the complex pole for the resonance $R$.

\begin{figure}[ht!]
\begin{minipage}[t]{0.48\textwidth}
\includegraphics[width=0.75\textwidth,angle=-90]{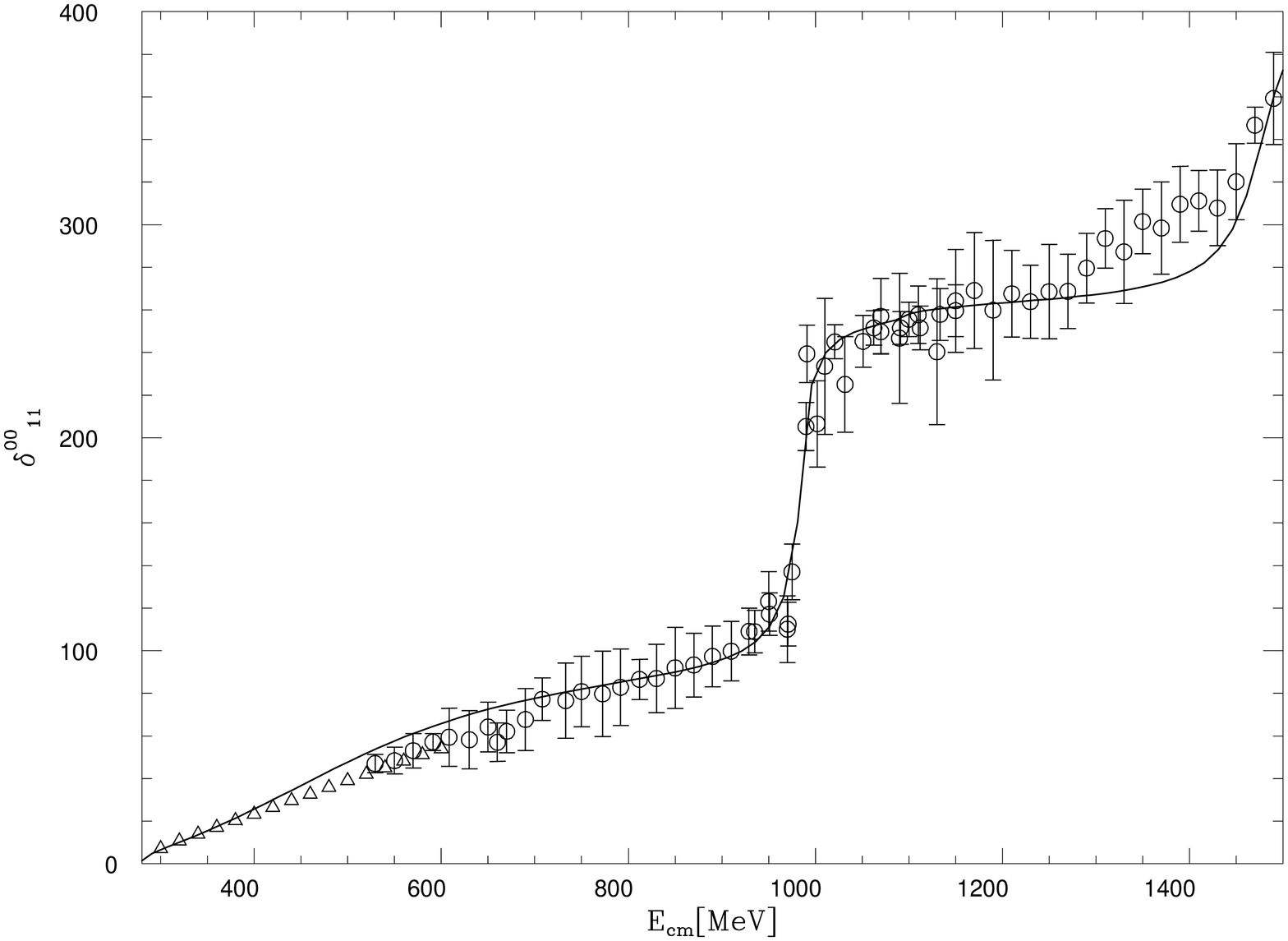}
\caption{Elastic isoscalar $\pi\pi$ phase shifts, $\delta^{00}_{11}$. The
circles correspond to the average of [45--48] and \cite{Kaminski,Ochs}, 
as discussed in {\it{section 3.1.1}}. We have also included the 
triangle points form \cite{Frogatt} to have some data close to threshold, 
although these
points have not been included in the fit because they are given without errors.}
\end{minipage}
\hfill
\begin{minipage}[t]{0.48\textwidth}
\includegraphics[width=0.75\textwidth,angle=-90]{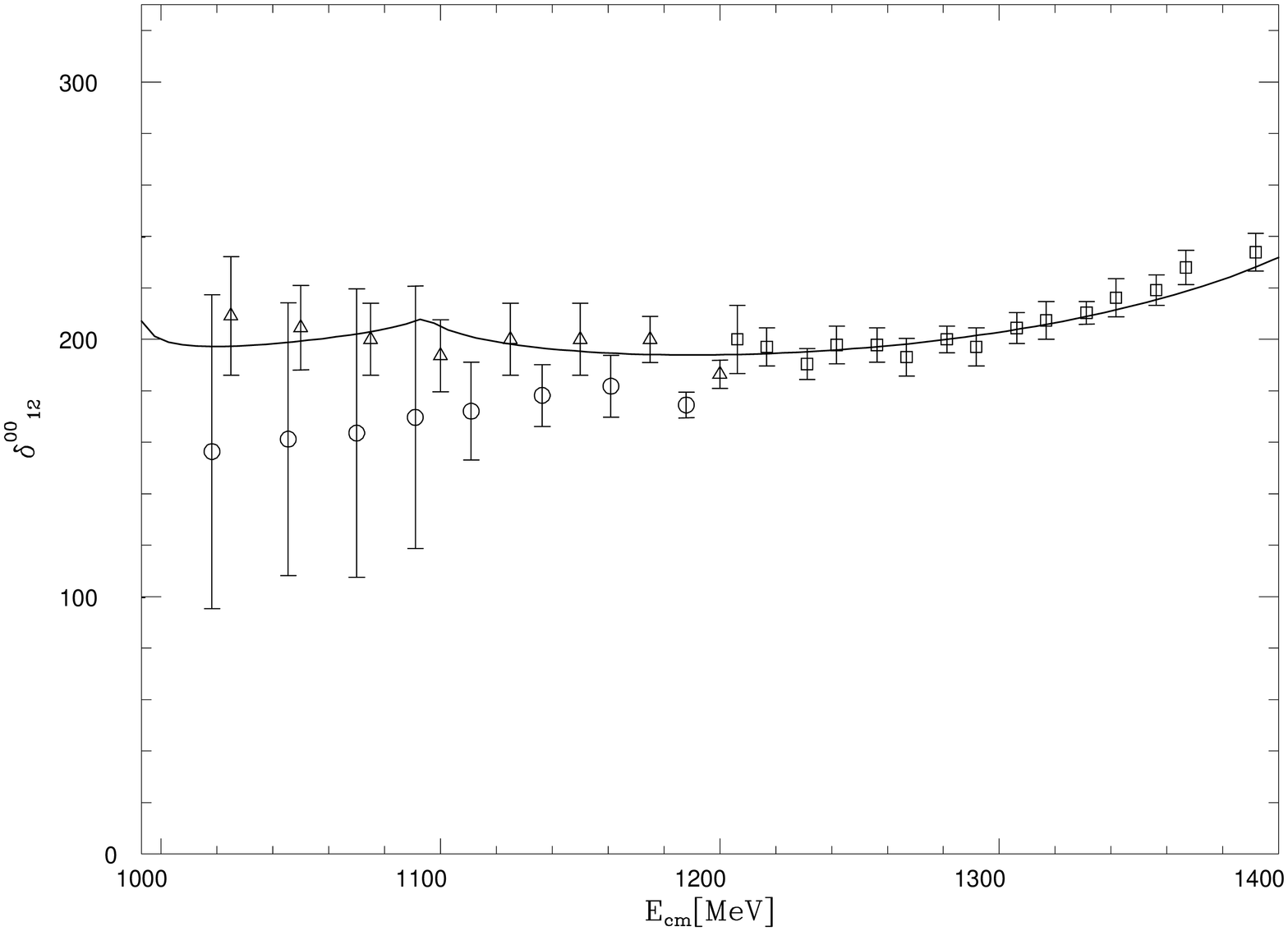}
\caption{S-wave $K\bar{K}\rightarrow \pi\pi$ isoscalar phase shifts,
$\delta^{00}_{12}$. The triangles points are from \cite{Cohen}, circles 
correspond
to the average of ref. \cite{43,44} and squares to the one of ref. \cite{Cohen,43,44}.}
\end{minipage}
\end{figure}

\begin{figure}[H]
\begin{minipage}[t]{0.48\textwidth}
\includegraphics[width=0.75\textwidth,angle=-90]{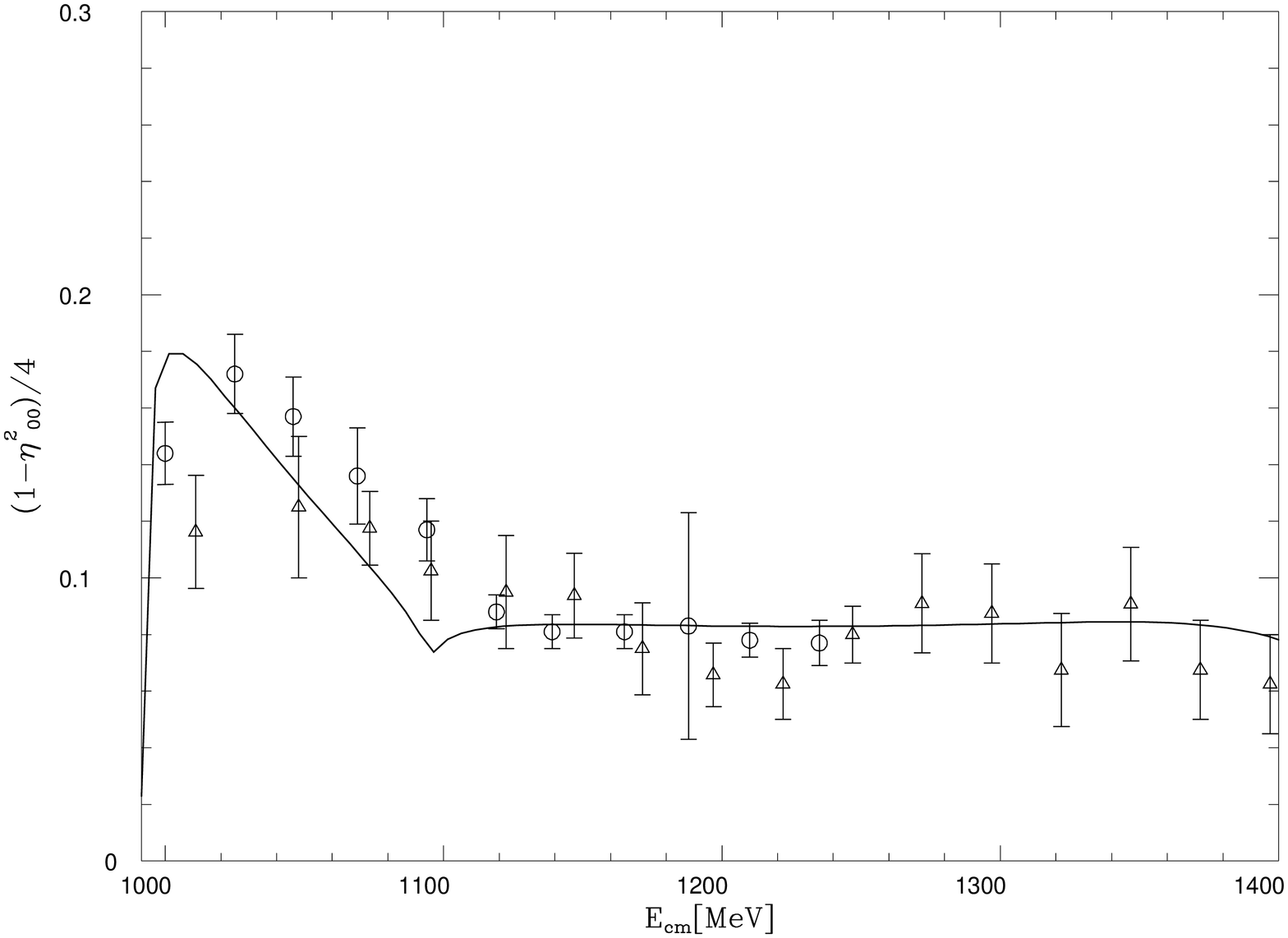}
\caption{$\frac{1-\eta^2_{00}}{4}$ with $\eta_{00}$ the $I$=$L$=0 S-wave
inelasticity. Circles \cite{Cohen}, triangles \cite{43}.}
\end{minipage}
\hfill
\begin{minipage}[t]{0.48\textwidth}
\includegraphics[width=0.75\textwidth,angle=-90]{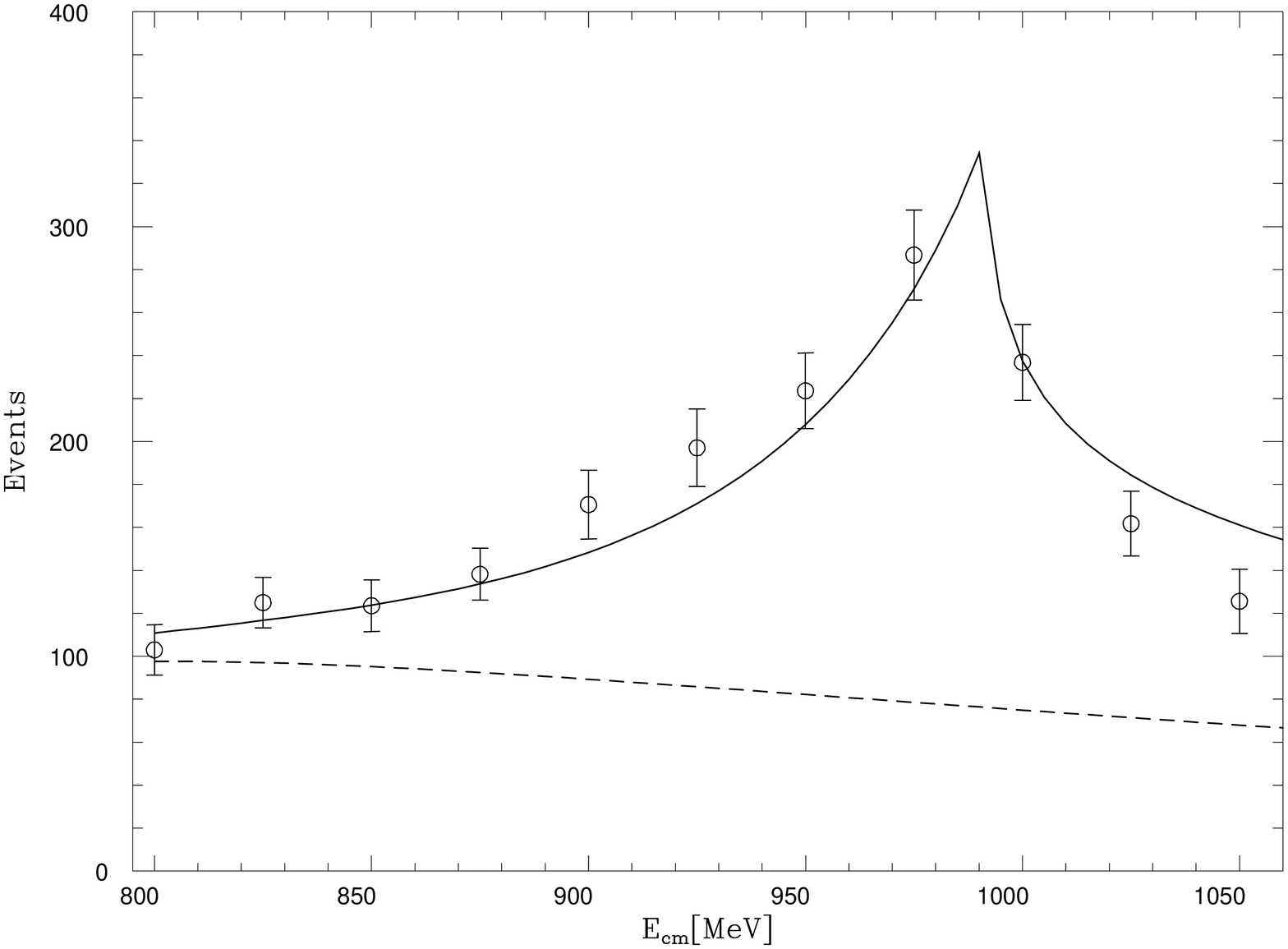}
\caption{Distribution of events around the $a_0(980)$ mass corresponding to the
central production $\pi\pi\eta$ in 300 GeV $pp$ collisions \cite{Arm}. The abscissa 
represents the $\pi \eta$ invariant mass, $E_{cm}$. The dashed
line represents the background introduced in the same reference.}
\end{minipage}
\end{figure}

\begin{figure}[ht!]
\begin{minipage}[t]{0.48\textwidth}
\includegraphics[width=0.75\textwidth,angle=-90]{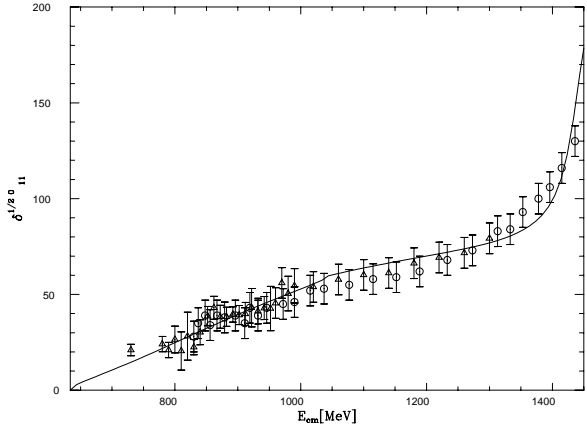}
\caption{S-wave $I$=1/2 $K\pi$ elastic phase shifts, $\delta^{1/2 \,0}_{11}$. The
triangles correspond to the average, as described in $\delta^{1/2 \,0}_{11}$
subsection, of ref. \cite{Mercer,Bingham,Estabrooks3}. Circles correspond to 
\cite{Aston}.}
\end{minipage}
\end{figure}

\begin{table}[ht!]
\begin{center}
\caption{Pole position and residues for the full amplitude.}
\begin{tabular}{|c|c|}     
\hline
&\\
\begin{math}
\begin{array}{ccc}
\sqrt{s}_{\sigma}&=&445-i\; 221 \;\, \hbox{MeV}\\
&&\\
\zeta^{\sigma}_{\pi\pi}&=&4.26 \;\, \hbox{GeV}\\
&&\\
\frac{\zeta^{\sigma}_{K\bar{K}}}{\zeta^{\sigma}_{\pi\pi}}&=&0.254\\
&&\\
\frac{\zeta^{\sigma}_{\eta\eta}}{\zeta^{\sigma}_{\pi\pi}}&=&0.036 
\end{array}
\end{math} &
\begin{math}
\begin{array}{ccc}
\sqrt{s}_{f_0}&=&987-i\; 14 \;\, \hbox{MeV}\\
&&\\
\zeta_{K\bar{K}}^{f_0}&=&3.63\;\, \hbox{GeV}\\
&&\\
\frac{\zeta_{\pi\pi}^{f_0}}{\zeta_{K\bar{K}}^{f_0}}&=&0.51\\
&&\\
\frac{\zeta_{\eta\eta}^{f_0}}{\zeta_{K\bar{K}}^{f_0}}&=&1.11
\end{array}
\end{math}\\
&\\
\hline
&\\
\begin{math}
\begin{array}{ccc}
\sqrt{s}_{{a_0}}&=&1053.13-i\; 24 \;\, \hbox{MeV}\\
&&\\
\zeta_{K\bar{K}}^{{a_0}}&=&5.48 \;\, \hbox{GeV}\\
&&\\
\frac{\zeta_{\pi\eta}^{{a_0}}}{\zeta_{K\bar{K}}^{{a_0}}}&=&0.70
\end{array}
\end{math}&
\begin{math}
\begin{array}{ccc}
\sqrt{s}_{\kappa}&=&779-i\; 330 \;\, \hbox{MeV}\\
&&\\
\zeta_{K\pi}^{\kappa}&=&4.99 \;\, \hbox{GeV}\\
&&\\
\frac{\zeta_{K\eta}^{\kappa}}{\zeta_{K\pi}^{\kappa}}&=&0.62
\end{array}
\end{math}\\
&\\
\hline
\end{tabular}
\end{center}
\end{table}

\begin{table}[ht!]
\begin{center}
\caption{Pole position and residues when the bare resonant contributions are
removed}
\begin{tabular}{|c|c|}     
\hline
&\\
\begin{math}
\begin{array}{ccc}
\sqrt{s}_{\sigma}&=&434-i\; 244 \;\, \hbox{MeV}\\
&&\\
\zeta^{\sigma}_{\pi\pi}&=&4.21 \;\, \hbox{GeV}\\
&&\\
\frac{\zeta^{\sigma}_{K\bar{K}}}{\zeta^{\sigma}_{\pi\pi}}&=&0.301\\
&&\\
\frac{\zeta^{\sigma}_{\eta\eta}}{\zeta^{\sigma}_{\pi\pi}}&=&0.033 
\end{array}
\end{math} &
\begin{math}
\begin{array}{ccc}
\sqrt{s}_{f_0}&=&\hbox{cusp effect}\\
&&\\
\zeta_{K\bar{K}}^{f_0}&=& ... \\
&&\\
\frac{\zeta_{\pi\pi}^{f_0}}{\zeta_{K\bar{K}}^{f_0}}&=&0.38\\
&&\\
\frac{\zeta_{\eta\eta}^{f_0}}{\zeta_{K\bar{K}}^{f_0}}&=&1.04
\end{array}
\end{math}\\
&\\
\hline
&\\
\begin{math}
\begin{array}{ccc}
\sqrt{s}_{{a_0}}&=&1081.95-i\; 13.3 \;\, \hbox{MeV}\\
&&\\
\zeta_{K\bar{K}}^{{a_0}}&=&5.98 \;\, \hbox{GeV}\\
&&\\
\frac{\zeta_{\pi\eta}^{{a_0}}}{\zeta_{K\bar{K}}^{{a_0}}}&=&0.74
\end{array}
\end{math}&
\begin{math}
\begin{array}{ccc}
\sqrt{s}_{\kappa}&=&770-i\; 341 \;\, \hbox{MeV}\\
&&\\
\zeta_{K\pi}^{\kappa}&=&4.87 \;\, \hbox{GeV}\\
&&\\
\frac{\zeta_{K\eta}^{\kappa}}{\zeta_{K\pi}^{\kappa}}&=&0.61
\end{array}
\end{math}\\
&\\
\hline
\end{tabular}
\end{center}
\end{table}

While for the $f_0(980)$ one has a preexisting tree level resonance with a mass
of 1020 MeV, for the other resonances present in Table 3.5 the situation is 
rather
different. 
In fact, if one removes the tree level nonet contribution from
$T^\infty$, the $a_0(980)$, $\sigma$ and
$\kappa$ poles still appear as can be seen in Table 3.6. For the $f_0(980)$, in
such a situation, one has not a pole but a very strong cusp effect in the
opening of the $K\bar{K}$ threshold. In fact, by varying a little the
value of $a^{SL}$ one can regenerate also a pole for the $f_0(980)$ from this
strong cusp effect. In Table 3.6 we have not given an absolute value for the
coupling of the $f_0(980)$ to the $K\bar{K}$ channel because one has not a pole
for the given value of $a^{SL}$. However, the ratios between the different 
amplitudes
are stable around the cusp position. As a result, the physical $f_0(980)$ will 
have two contributions: one from the bare singlet state with $M_1=1020$
MeV and the other one coming from meson-meson scattering, particularly
$K\bar{K}$ scattering, generated by the lowest order $\chi PT$ Lagrangian.

When the resonant
tree level contributions are removed from $T^\infty$, only the lowest order, 
${\mathcal{O}}(p^2)$, $\chi PT$ contributions remain. Thus, except for the
contribution to the $f_0(980)$ coming from the bare singlet at 1 GeV, the poles
present in Table 3.6 originate from a `pure potential' scattering, following the
nomenclature given in ref. \cite{Chew2}. In this way, the source of the dynamics is 
the lowest order $\chi PT$ amplitudes. The constant $a^{SL}$ can be
interpreted from the need to give a `range' to this potential so that the loop 
integrals converge. In ref. \cite{OOnd} it is also shown that these meson-meson 
states in the
 SU(3) limit, equaling the masses of the pseudoscalars, appear as a degenerate
 octet plus a singlet. 
 
 Finally, in ref. \cite{OOnd} an estimation of the unphysical cut contributions
 was done from ref. \cite{UMR} up to $\sqrt{s}\lesssim 800$ MeV for the 
 resonance $L=0$ partial waves.  As shown in Table 3.7 the influence is rather 
 small in the physical region. 

 \begin{table}[ht!]
\begin{center}
\caption{Influence of the unphysical cuts for the $I$=$L$=0 $\pi\pi$ and 
$I$=1/2, $L$=0
$K\pi$ partial waves \cite{OOnd}. The three first columns refer to $\pi\pi$ 
and the last
three to $K\pi$. $T_{Left}$ was already introduced at the end of the former
section, eq. (\ref{ournotation}).}
\vspace{0.2cm}
\begin{tabular}{c|c|c||c|c|c}
\hline
$\sqrt{s}$  & $\frac{T_{Left}}{|T_{11}|}$
& $\frac{T_{Left}}{T^\infty_{11}}$ & $\sqrt{s}$  &
 $\frac{T_{Left}}{|T_{11}|}$
& $\frac{T_{Left}}{T^\infty_{11}}$  \\
MeV & $\%$ & $\%$ & MeV & $\%$ & $\%$\\
\hline
276 & 3.7 & 4.8 &634 & 7.1 & 8.7 \\
376. & 3.5 & 5.1 &684 & 3.7 & 4.7 \\
476 & 4.1 & 5.7 &734 & 0.3 & 0.4 \\
576 & 5.7 & 6. &784 & -2.5 & -3.3 \\
676 & 8.1 & 6.1 &834 & -5.7 & -7.2 \\
776 & 11.2 & 5.6 &   & & \\
\hline
\end{tabular}
\end{center}
\end{table}

\section{Bethe-Salpeter equation for S-wave meson-meson and meson-baryon}

In this sections we report about the use of the Bethe-Salpeter equation for 
the study of the S-wave meson-meson scattering \cite{npa} and for the S-wave 
meson-baryon system with strangeness($S$)=$-$1 \cite{AO}. In both cases the 
potential is the lowest order  
$\chi PT$ amplitude. The use of the Bethe-Salpeter equation together with 
$\chi PT$ was first
considered in ref. \cite{NK} in the meson-baryon sector. 

One of the advantages of the approach in refs. \cite{npa,AO} is that the
Bethe-Salpeter equation, which is an integral equation, is reduced to an
algebraic one. This is accomplished through an analysis of the renormalization
process embodied in a Bethe-Salpeter scattering equation.

\subsection{Bethe-Salpeter equation for S-wave meson-meson scattering}

To see how the former simplification occurs let us consider eq. 
(\ref{lastndf}). In that section, 
$T^\infty$ was defined to be the sum of the lowest order $\chi PT$ amplitudes 
plus the s-channel resonance exchanges. If we consider only the contribution 
from the lowest order $\chi PT$ we will have from that equation:

\be
\label{bst}
T=\left[T_2^{-1}-g_0(s)\right]^{-1}=\left[1-T_2\cdot g_0(s)\right]^{-1}\cdot T_2
\ee
We can rewrite the previous equation in a form that will remind us of the
Bethe-Salpeter equation:

\be
\label{bse}
T=T_2+T_2\cdot g_0(s) \cdot T
\ee
with $g_0(s)$ given in eq. (\ref{5.0}). The above equation would correspond to a
Bethe-Salpeter equation with a potential given by the corresponding lowest 
order
$\chi PT$ partial wave, $T_2$. However, while a true Bethe-Salpeter equation is
an integral equation the former one is algebraic. Note that one should have
instead of $T_2\cdot g_0(s) \cdot T$ in eq. (\ref{bse}) the integral:

\begin{equation}
\label{TLalso2}
(T_2 g_0 T)_{\alpha \beta}=\sum_{j} \int \frac{d^4 q}{(2\pi)^4}
T_2(k,p;q)_{\alpha j}\frac{i}{\left(q^2-m_{1j}^2\right)
\left((P-q)^2-m_{2j}^2\right)} T(q;k',p')_{j \beta}
\end{equation}
where P is the total momentum, $k$ and $p$ represent the initial momenta of 
the ingoing mesons and $k'$
and $p'$ the final momenta of the outgoing ones. Only when $T_2$ and
$T$ are factorized on shell outside the integral in eq. (\ref{TLalso2}) one
recovers eq. (\ref{bse}). This is exactly what happens as stated by eq. 
(\ref{bst}). However, when considering this equation one has to recall also eq.
(\ref{fin/d}) in order to realize that all the parameters that appear in
$T_2$ have to be the physical or renormalized ones corresponding to the 
final positions and residues of the CDD poles or to the substraction constant 
that result from eq. (\ref{bst}).

The final result present in ref. \cite{npa} is obtained when approximating the
'physical values' of the parameters in $T_2$ by the ones dictated by
the lowest order $\chi PT$ results. That is, the constant $f$ is translated to
$f_\pi$ and for the bare masses one takes the physical ones. Both in the IAM or
in the N/D method with unphysical cuts, the parameters appearing in $T_2$ are
renormalized according to $\chi PT$ at $\Opc$.

The argumentation given in ref. \cite{npa} for factorizing on shell the potential and
the physical amplitude in eq. (\ref{TLalso2}) is discussed below and will be 
considered in further detail when discussing the Bethe-Salpeter equation 
for the meson-baryon scattering in the next subsection \cite{AO}. Briefly, in 
the former reference the potential was
splitted in two parts:

\be
\label{onshell}
T_2=V_{on-shell}+V_{off-shell}
\ee
The V$_{on-shell}(p_i)$ part, with $p_i^2=m_i^2$, factorizes out of the integral in 
eq. (\ref{TLalso2}) since it only depends of the Mandelstam variable $s$. For 
V$_{off-shell}$ it was realized that since it only involves terms proportional 
to $p_i^2-m_i^2$, they cancel one of the mesons propagators in the loop and give 
rise to tadpole-like 
contributions which can be reabsorbed in the final values of the parameters
present in $T_2$. So that, at the end, one only needs V$_{on-shell}$ and hence,
eq. (\ref{bse}) follows. This result has been recently derived from an
alternative point of view in ref. \cite{nieves}.

In ref. \cite{npa} the $g_0(s)$ function was calculated making use of a cut off
regularization. The cut off\footnote{The relation between $\Lambda$ and the
three momentum cut off $q_{max}$, introduced in {\it{section 3.1.2}}, is such that
$\Lambda=\sqrt{m_K^2+q_{max}^2}$.}, $\Lambda$, was fixed to reproduce the
experimental points, or in other words, to give the right value for the
substraction constant $a^{SL}_0$ in eq. (\ref{g(s)}). On the other hand, making
use of the IAM one also recovers eq. (\ref{bst}) when putting $T_4^P=0$ in 
eq. (\ref{t4ap}). In Fig. 3.20 we show the 
results of ref. \cite{npa} compared with 
data. The agreement 
is rather good and surprising. On the other hand, poles corresponding to
the resonances $f_0(980)$, $a_0(980)$ and $\sigma(500)$ were found and their 
masses, partial and total decay widths were also analyzed.

\begin{figure}[ht]
\centerline{\protect\hbox{
\psfig{file=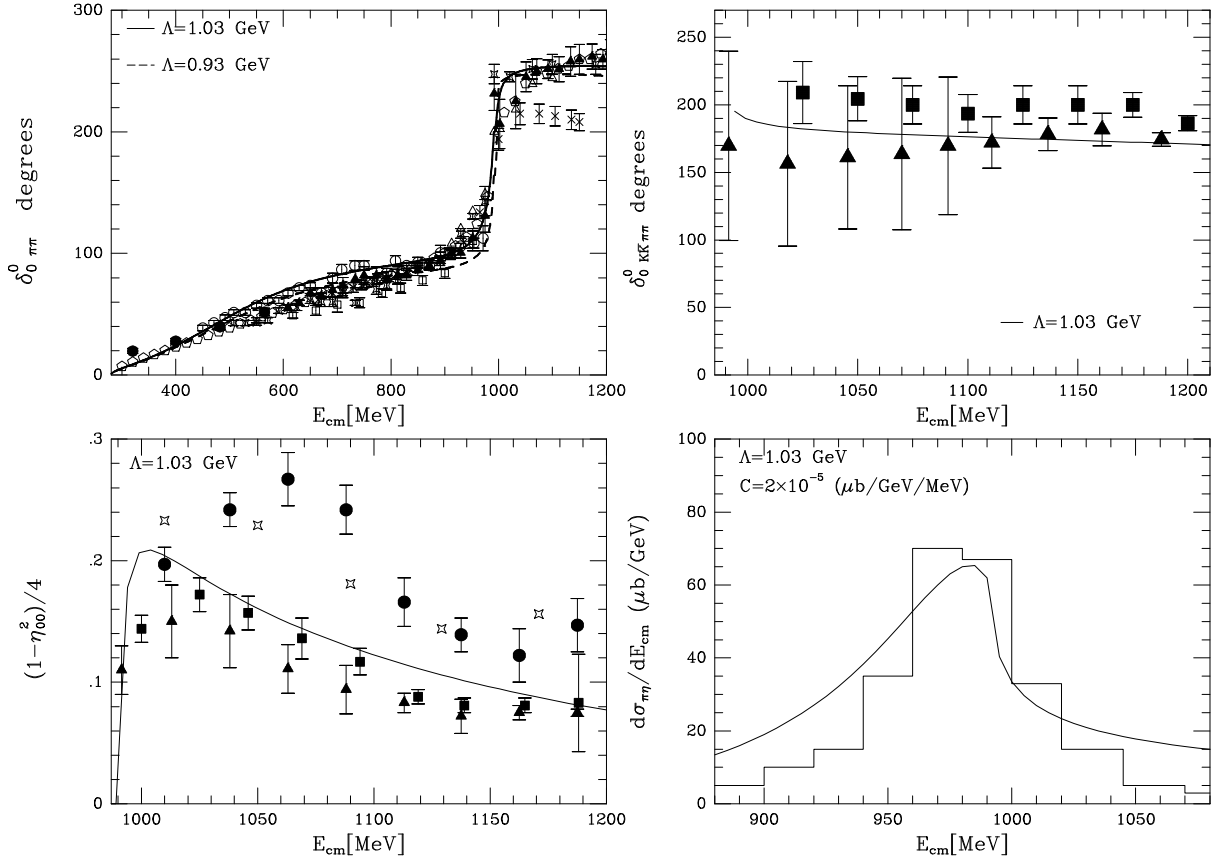,width=0.7\textwidth,angle=-90}}}
\caption{Results from ref. \cite{npa}. References to experimental data are 
also given in this paper.}
\end{figure}

Hence, we see that for the scalar sector with $I$=0,1 the
unitarization of the lowest order $\chi PT$ amplitudes plays a very
important role and also that the resonances which appear there with masses
$\leq$1 GeV will have a large meson-meson component in their 
nature. The same results are expected to hold in the $I$=1/2 S-wave amplitude
by $SU(3)$ symmetry as we have seen in {\it{section 3.2.3}}. In the P-waves, 
where the resonances that appear there are of preexisting nature, one cannot 
reproduce the substraction constant $a^L$ present in eq.
(\ref{limitd'}) by 
a reasonable cut off and the method fails. In ref. \cite{OOnd} it is argued that a
cut off around 300 GeV would be needed to reproduce the $\rho$ which is a
senseless result.

On the other hand, whereas in {\it{section 3.2.3}}, ref. \cite{OOnd}, one 
needs seven free parameters, a nonet of resonances with 6 parameters and a 
subleading constant, for describing the scalar sector, in this section, ref.
\cite{npa}, there is only one free constant, a cut off, for the $I=0$ and 1 
S-waves. This is due to: 1) in ref. \cite{OOnd} the fit is pushed up to 
$\sqrt{s}\approx 1.4$ GeV,
while in ref. \cite{npa} the fit is up to $\sqrt{s}=1.2$ GeV. In fact, the effect
of this octet around 1.4 GeV is soft enough below 1.2 GeV to be reabsorbed in
the cut off (or subleading constant). In this way, below 1.2 GeV, one would 
have needed only 4 parameters in the approach of ref. \cite{OOnd}. 2) In this 
latter reference the $\eta\eta$ channel is included and in order to
reproduce the $\displaystyle{\frac{1-\eta_{00}^2}{4}}$ data one
has had to include the singlet resonance around 1 GeV. The $\eta \eta$ channel 
was not considered, however, in ref. \cite{npa}. Should one have taken the 
available data for
$\eta_{00}$, which are measured with much worse precision than 
$\displaystyle{\frac{1-\eta_{00}^2}{4}}$, the effect of the  $\eta\eta$ channel
would have been masked by the large errors in $\eta_{00}$.

\subsection{Bethe-Salpeter equation for S-wave meson-baryon scattering}

The effective chiral Lagrangian techniques, which successfully
describe meson-meson scattering at low energies
\cite{Pich,Meissner,Eckerep,xpt}, have also proved to be an excellent tool
to study low energy properties of meson-baryon systems
when the interaction is weak, as is the case of
the S-wave $\pi N$ \cite{Meissner,Eckerep,BKM97,FMS98,Moj98,DP97,ET98} and
$K^+ N$
interactions
\cite{Lee94,Bro93}, where
the leading
term in the chiral expansion ${\cal O} (p)$
is the dominant one close to threshold. The perturbative
scheme breaks down --even at low
energies-- in the vicinity of a resonance. This is the case of
the $\bar{K} N$ system in the $S = -1$ sector with an isospin zero S-wave
resonance, the $\Lambda (1405)$.
Originally treated as a bound ${\bar K}N$ state \cite{dalitz59}, this
resonance was later interpreted as a conventional three-quark system
\cite{isgur78,arima94}. Analysis in terms of the cloudy-bag model
reinforced the idea of the $\Lambda(1405)$ being a $\bar{K}N$ bound state
\cite{veit84} and, in the framework of the bound-state soliton model,
it corresponds to a bound state of a kaon in the background potential of
the soliton \cite{schat95}.
The fact that the $\Lambda(1405)$ resonance is located 27 MeV below the
$K^- p$ threshold makes it difficult to
reproduce the scattering observables whithin the
standard chiral Lagrangian techniques, unless 
the resonance is explicitly introduced as an elementary
field \cite{CHL96a}
or one resorts to nonperturbative techniques similar to those
reviewed in
previous sections for meson-meson scattering.

A nonperturbative scheme to study the $S = -1$ meson-baryon sector,
yet using
the input of the chiral Lagrangians, was employed in
\cite{NK}. A
potential model was constructed such that, in Born approximation,
it had
the same S-wave scattering length as the chiral Lagrangian up to
order
$p^2$. This potential, which includes also finite range factors
to
regularize the integrals, was inserted in a set of
coupled-channel
Lippmann Schwinger integral equations.
The channels included were those opened around the
$K^-p$ threshold, namely $K^- p, \bar{K}^0
n, \pi^0 \Lambda, \pi^+ \Sigma^-, \pi^0 \Sigma^0$ and $\pi^-
\Sigma^+$.
By fiting five
parameters,
corresponding to, so far, unknown parameters of the second order
chiral
Lagrangian plus the range parameters of the potential, the
$\Lambda
(1405)$ resonance was generated as a quasibound meson-baryon state
and the
cross sections of the
$K^- p
\rightarrow K^- p, \bar{K}^0 n, \pi^0 \Lambda, \pi^+ \Sigma^-,
\pi^0
\Sigma^0, \pi^- \Sigma^+$ reactions at low energies, plus the
threshold
branching ratios
\begin{eqnarray}
\gamma&=&\displaystyle\frac{\Gamma(K^-p\to \pi^+ \Sigma^-)}{
\Gamma(K^-p \to \pi^-\Sigma^+)} \nonumber \\
R_c &=& \displaystyle\frac{\Gamma(K^-p\to {\rm charged\ channels})}{
\Gamma(K^-p \to {\rm all\ channels})}  \\
R_n &=& \displaystyle\frac{\Gamma(K^-p\to \pi^0\Lambda)}{
\Gamma(K^-p \to {\rm neutral\ channels})} \nonumber
\end{eqnarray}
were well reproduced.
Although the cross
sections and
position of the resonance could also be reproduced with only the
lowest
order Lagrangian and one potential range parameter, the
impossibility of
obtaining a good result for the double charge exchange ratio
$\gamma$ was the reason for the need of
including the S-wave terms of the next-to-leading order
Lagrangian.
However, a recent work \cite{AO}, which shares many
points with \cite{NK}, showed that
all the strangeness $S=-1$ meson-baryon scattering observables  near
threshold were reproduced
with the lowest
order Lagrangian and one cut off.  The main reason was
the inclusion of the $\eta \Lambda$ channel, neglected in ref. \cite{NK}.
The chiral scheme employed in
ref. \cite{AO}, and summarized below, includes all
meson-baryon
states that can be generated from the octet of pseudoscalar
mesons and the
octet of ground-state baryons, thus including in addition
the $\eta\Lambda,\eta\Sigma^0$
and the $K^+\Xi^-,K^0 \Xi^0$ channels, adding up to a total
of 10. It
should be noted that if one sets
equal baryon masses and equal meson masses in the scheme, one
should get degenerate SU(3) multiplet states. This is only
possible if all states of the $0^-$ meson and $\frac{1}{2}^+$
baryon octets are included in the scheme and one
should start from such a situation to have control on the SU(3)
breaking due to unequal masses.

At lowest order in momentum the
interaction
Lagrangian comes from the $\Gamma_{\mu}$ term in the covariant
derivative of eqs. (\ref{BaryonL}) and (\ref{CovDer})
\begin{equation}
{\cal L}_1^{(B)} = \langle \bar{B} i \gamma^{\mu} \frac{1}{4 f^2}
[(\Phi \partial_{\mu} \Phi - \partial_{\mu} \Phi \Phi) B
- B (\Phi \partial_{\mu} \Phi - \partial_{\mu} \Phi \Phi)] \rangle
\label{eq:mesbar}
\end{equation}
which leads to a common structure for the meson-baryon amplitudes of
the type
\begin{equation}
V_{ij}=-\frac{C_{ij}}{4f^2} \bar{u}(-\vec{k^\prime}\,) \gamma^\mu
(k_{\mu} +
k_{\mu}^\prime) u(-\vec{k}\,)
\label{eq:ampl}
\end{equation}
for the different
channels, where
$u, \bar{u}$ are the Dirac spinors and $k, k^\prime$ the momenta of
the
incoming and outgoing mesons in the center of mass of the meson-baryon
system.
The particular values for the $\bar{K}N$ system
of the SU(3) coefficients $C_{ij}$, which
connect the
meson-baryon channels $j$ and $i$,
can be found in ref. \cite{AO}.
At low energies the spatial components can be neglected and the
amplitudes reduce to
\begin{equation}
V_{i j} = - C_{i j} \frac{1}{4 f^2} (k^0 + k^{\prime 0}) \ .
\label{eq:ampl0}
\end{equation}
This amplitude is inserted in a coupled channel Bethe-Salpeter
equation
\begin{equation}
T_{i j} = V_{i j} + \overline{V_{i l} \; G_l \; T_{l j}}
\label{eq:bs}
\end{equation}
\noindent
with
\begin{equation}
\overline{V_{i l} \; G_l \; T_{l j}} = i \int \frac{d^4 q}{(2
\pi)^4} \,
\frac{M_l}{E_l (-\vec{q}\,)}
\, \frac{V_{i l} (k, q) \, T_{l j} (q, k')}{k^0 + p^0 - q^0 - E_l
(-\vec{q}\,)
+ i \epsilon} \, \frac{1}{q^2 - m^2_l + i \epsilon} \ ,
\label{eq:bs2}
\end{equation}
where only the positive energy component of the fermion
propagator has been kept.
The Bethe-Salpeter equation sums up automatically the series of
diagrams of Fig.
\ref{fig:bs}. The loop integral in eq. (\ref{eq:bs2}) is
logarithmically
divergent and can be
regularized by a cut off $q_{\rm max}$. The quantities
$M_l$ and $E_l$ in eq. (\ref{eq:bs2})
correspond to the
mass and energy of the intermediate baryon, while $m_l$ is the mass
of the
intermediate meson and $k^0+p^0 \equiv \sqrt{s}$ is the total energy
in
the center of mass frame.

\begin{figure}[ht]
\begin{center}
\includegraphics[width=0.8\linewidth]{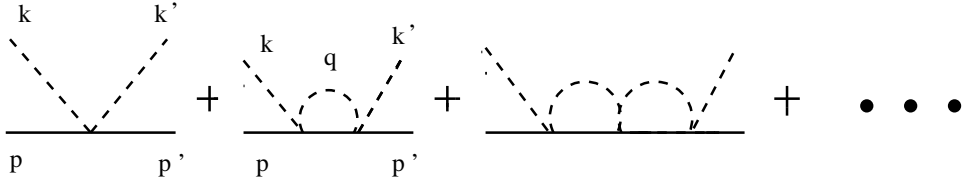}
\end{center}
\caption{
Bethe-Salpeter series for meson-baryon
scattering}
\label{fig:bs}
\end{figure}

The same arguments given in the previous section for the meson-meson sector 
allow one to retain only the on shell
part of the amplitudes
appearing in eq. (\ref{eq:bs2}), while the rest goes into
renormalization
of couplings and masses. Take, as an example, the
one loop diagram of Fig. \ref{fig:bs} with equal masses in
the external
and intermediate states for simplicity. We have
\begin{eqnarray}
V^2_{\rm off} &=& C (k^0 + q^0)^2 = C (2 k^0 + q^0 - k^0)^2
\nonumber \\
              &=& C (2 k^0)^2 + 2 C (2 k^0) (q^0 - k^0) + C (q^0
-
k^0)^2
\label{eq:off}
\end{eqnarray}
with $C$ a proportionality constant. The first term in the last
expression is the
on shell
contribution $V^2_{\rm on}$, with $V_{\rm on} \equiv C 2 k^0$.
Neglecting $p^0 - E(-\vec{q}\,)$
in eq. (\ref{eq:bs2}), a typical approximation in the heavy baryon
formalism, the one loop integral for the second term of eq.
(\ref{eq:off}) becomes
\begin{eqnarray}
&&2 i V_{\rm on}\int \frac{d^3 q}{(2 \pi)^3} \,\int \, \frac{d
q^0}{2
\pi} \,
\frac{M}{E (-\vec{q}\,)} \, \frac{q^0 - k^0}{k^0 - q^0} \,
\frac{1}{q^{02} - \omega (\vec{q}\,)^2 + i \epsilon} \nonumber \\
&\,\,\, = & - 2 V_{\rm on} \, \int \, \frac{d^3 q}{(2 \pi)^3} \,
\frac{M}{E (-\vec{q}\,)}
\,
\frac{1}{2 \omega (\vec{q}\,)} \sim V_{\rm on} \, q^2_{\rm max}
\end{eqnarray}
with $\omega (\vec{q}\,)^2 = \vec{q} \, ^2 + m^2$. This term,
proportional
to $V_{\rm on}$,
has the same structure as the tree level term in the Bethe-Salpeter
series and it can be
reabsorbed in
the lowest order Lagrangian by a suitable renormalization of the
parameter $f$.
Similarly, one of the two $(q^0
- k^0)$ factors in the last term of eq. (\ref{eq:off}) cancels the
baryon propagator in eq. (\ref{eq:bs2}) while the remaining factor
gives rise to another term proportional to $k^0$
(and hence
$V_{\rm on}$) and a term proportional to $q^0$, which vanishes for
parity reasons.

These arguments can be extended to coupled channels and higher order
loops with the
conclusion that
$V_{il}$ and $T_{lj}$ factorize with their on shell values out of
the
integral in eq. (\ref{eq:bs2}), reducing the problem to one of
solving a
set of algebraic equations, written in matrix form as
\begin{equation}
T = V + V \, G \, T
\label{eq:bs3}
\end{equation}
\noindent
with $G$ a diagonal matrix given by
\begin{eqnarray}
G_{l}(\sqrt{s}) &=& i \, \int \frac{d^4 q}{(2 \pi)^4} \, \frac{M_l}{E_l
(-\vec{q}\,)} \,
\frac{1}{\sqrt{s} - q^0 - E_l (-\vec{q}\,) + i \epsilon} \,
\frac{1}{q^2 - m^2_l + i \epsilon} \nonumber \\
&=& \int_{\mid \vec{q} \mid < q_{\rm max}} \, \frac{d^3 q}{(2
\pi)^3} \,
\frac{1}{2 \omega_l(\vec{q}\,)}
\,
\frac{M_l}{E_l (-\vec{q}\,)} \,
\frac{1}{\sqrt{s} - \omega_l(\vec{q}\,) - E_l (-\vec{q}\,) + i
\epsilon} \ .
\label{eq:loop}
\end{eqnarray}
The value of the cut off in ref. \cite{AO},
$q_{\rm max}=630$ MeV, was chosen
to reproduce the $K^- p$ threshold branching ratios \cite{To71,No78}
,
while the weak
decay constant, $f=1.15
f_\pi$, was taken in between the pion and kaon ones to optimize
the position of the $\Lambda(1405)$ resonance
\cite{Th73,Hem85,dal91}.
The branching ratios as well as the predictions of the
scattering lengths for $K^-p$ and $K^-n$ scattering are
summarized in Table \ref{tab:tab1}, where results omitting the
$\eta$ channels are also shown.
While the $\eta$ channels have a moderate effect
on the isospin $I=1$ $K^- n$ scattering length, $a_{K^- n}$, they
have a
tremendous influence
on the $K^- p$ scattering observables, especially on the ratio
$\gamma$
which changes by a factor of about 2.
As was shown in ref. \cite{AO}, it is the $I=0$ $\eta \Lambda$
channel the one that was providing most of the changes.
The $K^- p$ scattering length is essentially in
agreement
with the most recent results from Kaonic hydrogen $X$ rays
\cite{Miw97},
in qualitative agreement with the scattering lengths determined from
scattering
data in \cite{Adm81}, with an estimated error of 15\%, and in
remarkable agreement with
the result from a combined dispersion relation and
$M$-matrix analysis \cite{Adm81}.

\begin{table}[htb]          
\centering
\caption{$K^-p$ threshold ratios and $K^-N$ scattering lengths}

\begin{tabular}{|c|c|c|c|}
\hline
 & All channels & No $\eta\Lambda$,$\eta\Sigma^0$ & EXP \\
\hline
 & & & \\
$\gamma$  & 2.32 & 1.04 & 2.36$\pm$0.04
\cite{To71,No78} \\
 & & & \\
$R_c$ & 0.627 & 0.637 & 0.664$\pm$0.011
\cite{To71,No78} \\
 & & & \\
$R_n$ & 0.213 & 0.158 & 0.189$\pm$0.015
\cite{To71,No78} \\
 & & & \\
$a_{K^-p}$ (fm) & $-$1.00+i0.94 &  $-$0.68+i1.64 &
($-$0.78$\pm$0.18)+i(0.49$\pm$0.37)
\cite{Miw97} \\
 & & & $-$0.67+i0.64
\cite{Adm81} \\
 & & & $-$0.98 (from Re($a$))
\cite{Adm81} \\
 & & & \\
$a_{K^-n}$ (fm) & 0.53+i0.62 & 0.47+i0.53 & 0.37+i 0.60
\cite{Adm81} \\
 & & & 0.54 (from Re($a$))
\cite{Adm81}
\\
 & & & \\
\hline
\end{tabular}

\label{tab:tab1}
\end{table}

Finally, the $K^-p$ cross sections
for some selected channels ($K^-p
\to K^- \pi$, $\bar{K}^0n$, $\pi^+\Sigma^-$, $\pi^-\Sigma^+$) are
compared
with the low-energy scattering data
\cite{Hump,Sakitt,Kim,Kittel,Cibo,Evans}
in Fig.~\ref{fig:knscat}.
The results using the isospin basis (short-dashed line) are close to
those using the
basis of physical states (solid line) but the cusp associated to the
opening of the $\bar{K}^0 n$ channel appears in the wrong place due
to
the use of an average mass for all the members of an isospin
multiplet.
The effects of neglecting the $\eta$ channels (long-dashed line) are
much more significant.
Close to threshold, the $\pi^-\Sigma^+$ cross section is reduced
by almost a factor of 3 and
the $\pi^+ \Sigma^-$ cross section is reduced by a factor 1.3
when the
$\eta$ channels are included. This enhances the ratio $\gamma$ by
a factor 2.2 and makes the agreement with the experimental value
possible using only the lowest order chiral Lagrangian.
It is also remarkable that the $\pi^+ \Sigma^-$ cross section, which is
zero at lowest order with the chiral Lagrangians, turns out to be about
three times bigger that the analogous, allowed one, $\pi^- \Sigma^+$. The
multiple scattering with coupled channels is responsible for this.

\begin{figure}[ht!]
\begin{center}
\includegraphics[width=0.92\linewidth]{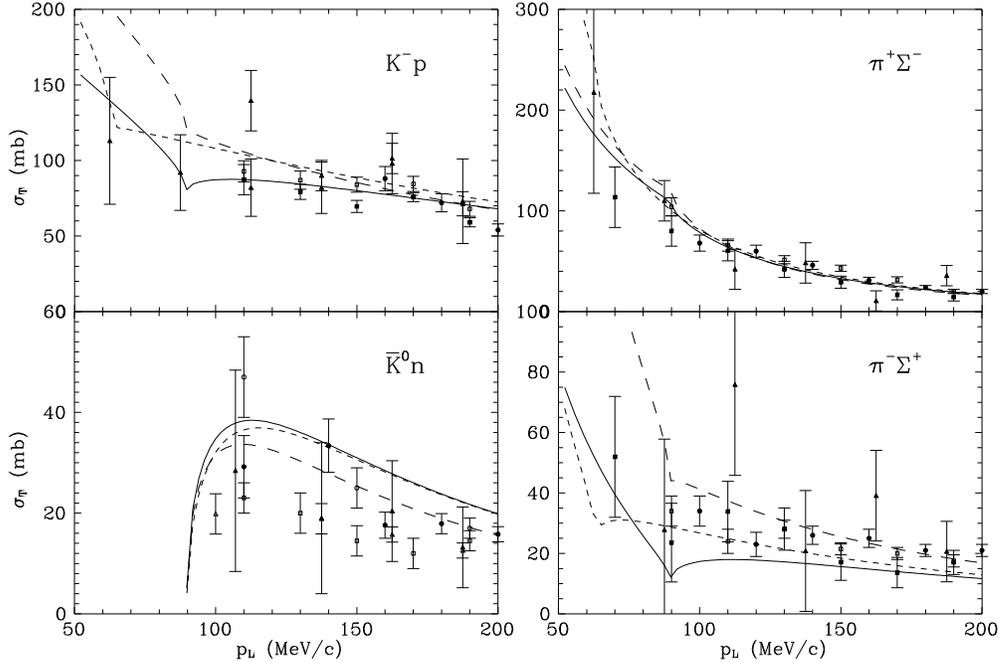}
\end{center}
\caption{
$K^-p$ scattering cross sections as functions of the $K^-$
momentum in the lab frame: with the full basis of physical states
(solid line), omitting the $\eta$ channels (long-dashed line) and
with the isospin-basis (short-dashed line). The experimental data
are
taken from refs.
\cite{Hump,Sakitt,Kim,Kittel,Cibo,Evans}
}

\label{fig:knscat}
\end{figure}

The predictions of this model for $K N$ scattering in the
strangeness $S=1$
sector at low
energies are quite satisfactory.
As shown in ref. \cite{AO}, the phase shifts in
the isospin channel $I = 1$ are about 15\% smaller
than experiment\cite{brmart}.
This
result is qualitatively similar to the one obtained in
\cite{NK}, where
it was also shown that allowing for a $K^+p$ shorter range
parameter (larger cut off)
the agreement with data improves.
The predicted $KN$ scattering lengths for isospin $I=0$ and $I=1$
are $a(S = 1, I = 0) = 2.4 \times 10^{- 7}$ fm and
$a (S = 1, I = 1) = - 0.26$ fm,
which compare favorably with present experimental data,
$0.02 \pm 0.04$ fm $(I=0)$ and
$ - 0.32 \pm 0.02$ fm $(I=1)$ \cite{Do82}.
Note that the scattering length in $I = 0$ is
zero at lowest order ($T=V$) and becomes finite, although negligibly
small,
as a consequence
of the coupling to other channels when working in the particle
basis.

We have seen that the use of only one cut off parameter and the
input of
the lowest-order Lagrangian reproduces the low energy data
in the $S=-1$ sector as satisfactorily as the model of
ref. \cite{NK}, where the $\eta$ channels were omitted and the
next-to-leading order terms of the chiral Lagrangian included. This
is
due to the fact
that at low energies
the $\eta$ meson loops only contribute to the real part of the
amplitudes and this can effectively be taken into account by
means of parameters of the second-order Lagrangian. 
Nevertheless, the values of the ${\cal O}(p^2)$ countertems of the
meson-baryon Lagrangian given in ref. \cite{NK} are affected by the
resummation of the important contributions coming from the SU(3) channels
with the $\eta$ meson and, hence, their actual values can be very
different.

The success in reproducing $\bar{K}N$ and $KN$ low energy
scattering observables with the lowest-order Lagrangian and one cut off 
\cite{AO} does not mean that this procedure can be
generalized to all meson-baryon sectors.
The richness
of information
available for meson-nucleon scattering requires the use of higher
order
Lagrangians, as it was the case in meson-meson scattering when
including
all the different channels \cite{prl,PRD}.
In fact, it has turned out to be impossible to dynamically reproduce the
S-wave
$N^*(1535)$
resonance and the low energy scattering data in the $S=0$ sector with the
lowest order Lagrangian and
only one
cut off. However, the extension of the model of ref. \cite{NK}
to pion
induced reactions in the
$S=0$ sector ($\pi^- p \to \eta\Lambda, K^0\Lambda, K^+\Sigma^-,
\pi^+p \to
K^+\Sigma^+$) produces the $N^*(1535)$ resonance as a
quasibound $K\Sigma-K\Lambda$ state \cite{Kai95b}, using the same
parameters of
their
next-to-leading order Lagrangian fitted to the low energy $\bar{K}N$
($S=-1$)
data. Simultaneously, the $\eta$ and $K$ photoproduction processes
in the $S=0$
channels were also studied ($\gamma p \to \eta p, K^+\Lambda,
K^+\Sigma^0,
K^0\Sigma^+$) and with a few more parameters a global
reproduction of the strong and electromagnetic cross section was
obtained\cite{KWW97}.
The method is being extended to higher partial waves, to gain access
to higher
energies, other resonances and polarization observables \cite{Kai}.

Recently, the unitarization of the Heavy Baryon $\chi PT$ amplitudes at
${\mathcal{O}}(p^3)$ \cite{FMS98,Moj98} has regained interest. In ref.
\cite{pelaez99},
this is done making use of the IAM method, giving rise to a reasonable 
account of
the scattering data up to around 1.2 GeV, including the region of the
$\Delta(1232)$ resonance in the $P_{33}$ partial wave. In ref.
\cite{nieves00}
it is argued that a new rearrangement of the Heavy Baryon $\chi PT$
series,
in terms of which the IAM is once again applied, leads to much better
results
than making use of the more straightforward version of the IAM used in
\cite{pelaez99}. This seems to be the case for the $P_{33}$
partial wave, although a convincing argumentation for this rearragement is
lacking in ref. \cite{nieves00}, particularly when considering other partial
waves
\footnote{With respect to this point, J.A.O. acknowleges 
very fruitful and enlightening discussions
with Jos\'e Ram\'on Pel\'aez.}.  On
the other hand, in ref. \cite{ulfoller} the unitarization of the elastic
$\pi N$ scattering is accomplished using an adapted version of the method
described in {\it section 3.2.1} for the meson-baryon sector, in a fully 
relativistic way.
Explicit resonance fields are included in ref. \cite{ulfoller} and a
matching with the ${\mathcal{O}}(p^3)$
Heavy Baryon $\chi PT$ $\pi N$ amplitudes \cite{FMS98,Moj98} is given. The data
are reproduced up to 1.3 GeV, where new channels would have to be
introduced.

The field is at a stage where rapid progress is being done and a clearer 
and broader picture of the role of chiral dynamics in
meson-baryon scattering can be expected in the near future.

%% file: fsi.tex
\chapter{Final state interactions in meson pairs}

In this chapter we present how to use the previous meson-meson strong 
amplitudes to
calculate processes with mesons in the final state. In many of them, the final
state interactions between these mesons are crucial and give rise to 
corrections of even orders of magnitude in describing the physics involved.

\section{The {\boldmath $\gamma \gamma \rightarrow$} meson-meson reaction}

The $\gamma \gamma \rightarrow$ meson-meson reaction 
provides interesting information concerning the structure of hadrons,
their spectroscopy and the meson-meson interactions, given the sensitivity 
of the reaction to the hadronic final state interactions (FSI)
\cite{10,Feindt}. In this sense, the study of these processes constitutes a very
interesting test of consistency of the approaches \cite{npa,PRD,OOnd} for the 
scalar sector.  

In ref. \cite{gamanpa}, a unified theoretical
description of the reactions $\gamma \gamma \rightarrow \pi^+ \pi^-$, 
$\pi^0 \pi^0$, $K^+ K^-$, $K^0 \bar{K}^0$, $\pi^0 \eta$ up to about 
$\sqrt{s} = 1.4$ GeV was presented for the first time. The agreement with the
experimental data was very good as can be seen in Fig. 4.4.
 
 For calculating the above processes one needs to correct for FSI the tree 
 level amplitudes coming from Born terms, Fig. 4.1, in the 
 case of the charged 
 channels, and also from the exchange of vector and axial resonances in the
 crossed channels \cite{Po}, Fig. 4.2. 

\begin{figure}[H]
\begin{minipage}[t]{0.48\textwidth}
\includegraphics[width=0.2\textwidth,angle=-90]{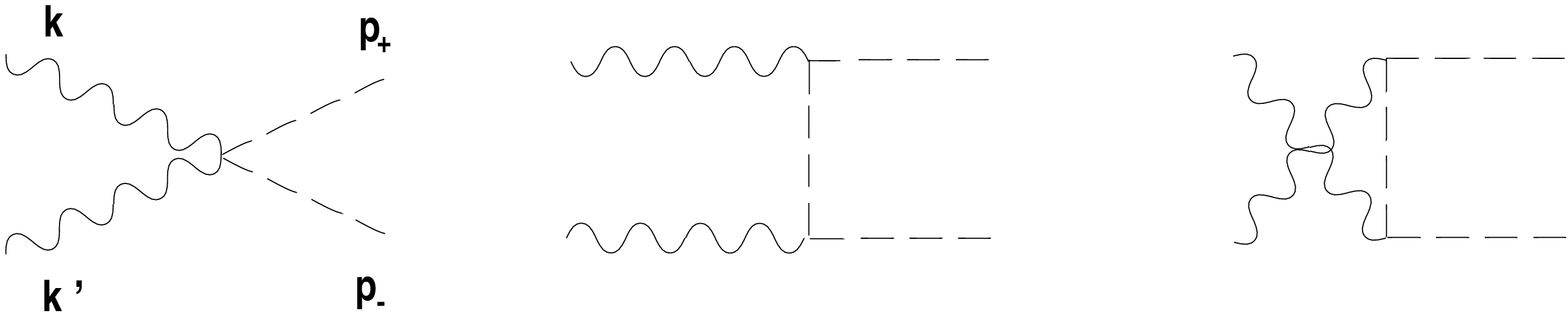}
\caption{Born term amplitude for $\gamma\gamma \rightarrow M^+M^-$. $k$ and 
$k'$ are the momenta of the incoming photons and $p_+$($p_-$) the momentum 
of the positively(negatively) charged meson.}
\end{minipage}
\hfill
\begin{minipage}[t]{0.48\textwidth}
\includegraphics[width=0.3\textwidth,angle=-90]{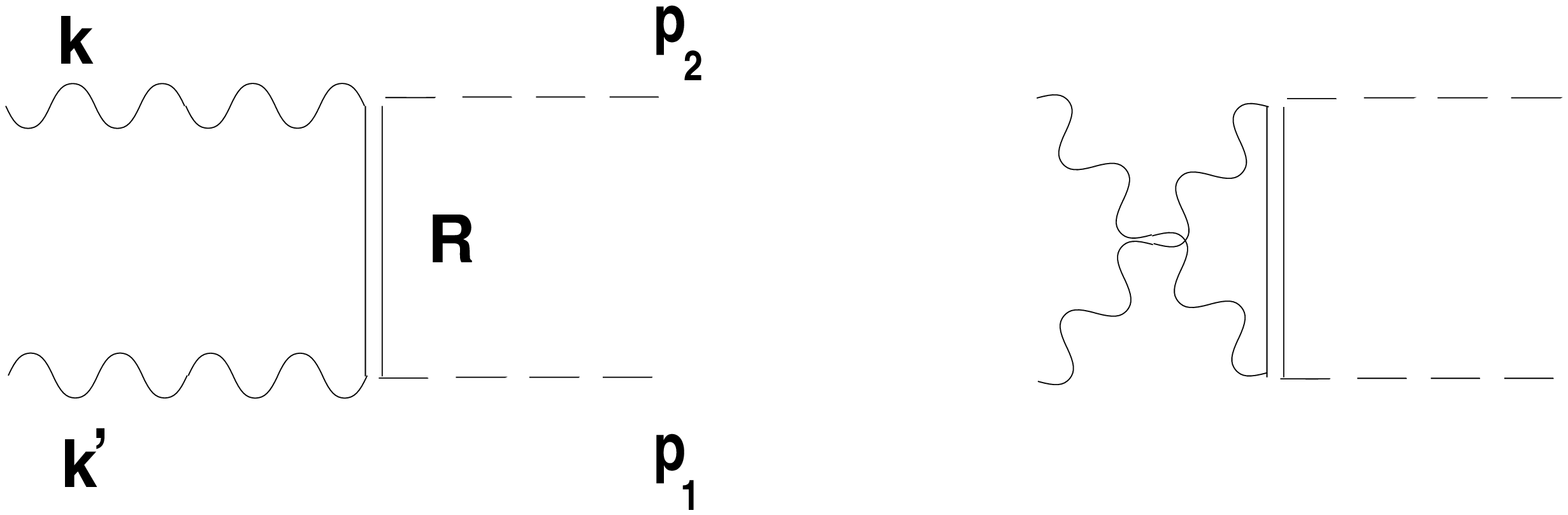}
\caption{Tree level amplitude for  $\gamma\gamma\rightarrow M_1M_2 $ through 
the exchange of a resonance $R$(axial or vectorial) in the t,u channels. 
}
\end{minipage}
\end{figure}

\subsection{FSI: S-wave}

In ref. \cite{gamanpa} the one loop corrections of the tree level amplitudes is first
considered and 
then this result is extended to the string of loops represented in Fig. 4.3.
 
The one loop contribution generated from the Born terms with 
intermediate charged mesons can be directly taken from the $\chi P T$ 
calculations \cite{xpt001,xpt002} of the $\gamma \gamma 
\rightarrow \pi^0 \pi^0$ amplitude 
at $\Opc$. The important point is that the $\Opd$ $\chi PT$ amplitude 
connecting the charge particles with the $\pi^0\pi^0$ factorizes on shell
outside the loop. One can schematically represent this situation by:

\begin{equation}
\label{gaxpt}
\sum_{a}L(s)_{a}\, T^{(2)}_{ab}(s)
\end{equation}
where the subindex $a$ represents the pair of intermediate charged mesons, 
$b$ the final ones and $T^{(2)}$ the on shell $\Opd$ amplitude.

The contribution of ref. \cite{gamanpa} beyond this first loop is to include 
all meson loops (Fig. 4.3) generated by
the coupled channel Bethe-Salpeter equations of {\it{section 3.3.1}}, ref.
\cite{npa}. We also saw there that the on shell $\Opd$ $\chi PT$ amplitudes 
factorize
outside the loop integrals. Thus, the immediate consequence of
introducing these loops is to substitute the on shell $\Opd$ $\pi \pi$ 
amplitude in eq. (\ref{gaxpt}), by the on shell 
meson-meson amplitude, $T_{ab}(s)$, evaluated in ref. \cite{npa}.

\begin{figure}[H]
\includegraphics[width=0.1\textwidth,angle=-90]
{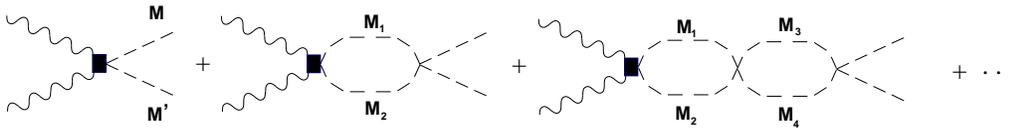}
\caption{Diagrammatic series which gives rise to the FSI from a general $\gamma
\gamma \rightarrow$M M' vertex, represented by the full square.}
\end{figure}

A similar procedure can be done to account for the FSI in the case of the tree
level diagrams with the exchange of a resonance (vector or axial). As explained
in ref. \cite{gamanpa} one can justify the accuracy of factorizing the strong
amplitude for the loops with crossed exchange of resonances, since this result
is correct for $M_R^2 \rightarrow \infty$. Because we are dealing with real
photons the intermediate axial or vector mesons are always off shell and the
large mass limit is a sensible approximation. The errors were estimated to be
below the level of $5\%$ for $M_R$ about 800 MeV.

\subsection{D-wave contribution}

For the $(2, 2)$ component we take the results of ref. \cite{26}, obtained
using dispersion relations

\begin{equation}
t_{BC}^{(2,2)} = \biggl[ \frac{2}{3} \, \chi_{22}^{T=0} \, e^{i \delta_{20}} +
 \frac{1}{3} \, \chi_{22}^{T=2} \, e^{i \delta_{22}} \biggr]\, t_B^{(2, 2)}
\end{equation}
where the functions $\chi_{ij}(s)$ are just first order polynomials in the $s$ 
variable.
 
For the $\gamma \gamma \rightarrow K^+ K^-$ reaction the non resonant 
D-wave contribution is not needed because one is close to the $K\bar{K}$ 
threshold and furthermore the functions $\chi_{ij}$ 
are nearly zero close to the mass of the $f_2$ and $a_2$ 
resonances, which are also in the energy region we are considering.

The {\it resonance contribution} in the $D-$wave coming from the 
$f_2 (1270)$ and $a_2 (1320)$ resonances is parametrized in the standard 
way of a Breit-Wigner as done in ref. \cite{5}. The parameters of these 
resonances are completely compatible with the ones coming from the Particle 
Data Group \cite{PDG}.

Once the FSI for the S- and D-waves have been taken into account, which 
completely dominate the
$\gamma \gamma \rightarrow$meson-meson reactions up to the energies considered
\cite{gamanpa,26}, one can compare with several experimental data.

\subsection{Total and differential cross sections}

The experimental data correspond to total and differential cross sections. 
As can
be seen in Fig. 4.4, the agreement is very good in all the
channels considered. It
is worth mentioning that the results presented are not a fit,
since the parameters of the axial, vector and tensor resonances were taken from
the literature.

It is also worth remarking that in the figure corresponding to the $\gamma \gamma \rightarrow 
K^+K^-$ reaction, the Born term, indicated by the long-dashed line, reduces 
to the short-dashed line when taking into account the FSI. This implies a large
reduction of this Born term thanks to which a good reproduction of the data is
obtained, hence solving a long standing problem \cite{Feindt}.

\subsection{Partial decay widths to two photons of the {\boldmath $f_0(980)$} 
and {\boldmath $a_0(980)$}}

The same procedure as in {\it{section 3.1.2}} is followed in ref. \cite{gamanpa} in
 order to calculate 
the partial decay widths of the $f_0(980)$ and $a_0(980)$ in terms of the strong
\cite{npa} and
photo-production \cite{gamanpa} amplitudes. From the amplitudes with isospin $I = 1$ and 0
\cite{npa}, 
one considers the terms which involve the strong
$M \bar{M} \rightarrow M \bar{M}$ amplitude. Then, one isolates the part of 
the
$\gamma \gamma \rightarrow M \bar{M}$ process which proceeds via the 
resonances $a_0$
and $f_0$ respectively. In the vicinity of the resonance the amplitude 
proceeds
as $M \bar{M} \rightarrow R \rightarrow M \bar{M}$. Hence, eliminating 
the $R \rightarrow M \bar{M}$ part of the amplitude plus the $R$ 
propagator and removing the proper
isospin Clebsch Gordan coefficients for the final states (1 for $\pi^0 \eta$
and $- 1/ \sqrt{2}$ for $K^+ K^-$), one obtains the coupling of the previous 
resonances to the $\gamma \gamma$ channel.

The results are:

\begin{equation}
\label{widthsgama}
\begin{array}{lll}
\Gamma_{a_0}^{\gamma \gamma} = 0.78 \; {\rm KeV} \, ;&
\Gamma_{a_0}^{\gamma \gamma} 
\displaystyle\frac{\Gamma_{a_0}^{\eta \pi}}{\Gamma_{a_0}^{tot}}
= 0.49 \; \, {\rm KeV} \, ;&
\Gamma^{\gamma \gamma}_{f_0} = 0.20 \;  \, {\rm KeV}
\end{array}
\end{equation}

The calculated width for the $f_0(980)$ is smaller than the average
value of ($0.56\pm 0.11$) KeV reported in
the PDG \cite{PDG}. In doing this average the
PDG refers to the work by Morgan and Pennington \cite{26} where they quote
a width of $(0.63\pm 0.14)$ KeV. However, in a recent
work by Boglione and Pennington they quote the much smaller width ($0.28
^{+0.09}_{-0.13}$) KeV \cite{hep-ph-BM}. When taking into account the errors, 
the former result and the one from ref. \cite{gamanpa} are compatible.

The value given above for the second magnitude 
in eq. (\ref{widthsgama}) is larger
than the value of ($0.28\pm 0.04 \pm 0.1)$ KeV given in PDG. 
However, this value
comes from references where a background is introduced in order to fit the data.
In the analysis we have discussed in this section \cite{gamanpa}, no background is
included and hence, in a natural way, the strength
of the $a_0(980)$ to two photons is increased.

\subsubsection{Conclusions} 
As important features of the approach of ref. \cite{gamanpa},
we can remark: 

1) The resonance $f_0 (980)$ shows up weakly in 
$\gamma \gamma \rightarrow \pi^0 \pi^0$ and barely in 
$\gamma \gamma \rightarrow \pi^+ \pi^-$.

2) In order to explain the angular distributions of the
$\gamma \gamma \rightarrow \pi^+ \pi^-$  reaction there is not need of the
hypothetical $f_0 (1100)$ broad resonance suggested in other works \cite{26}.
 This also solves the puzzle of why it did not show up in the 
$\gamma \gamma \rightarrow \pi^0 \pi^0$ channel. Furthermore, such resonance
does not appear in the theoretical work of ref. \cite{npa}, while the 
$f_0 (980)$ showed up clearly as a pole of the $T$ matrix in $I = 0$.

3) The resonance $a_0$ shows up clearly in the 
$\gamma \gamma \rightarrow \pi^0 \eta$ channel and the experimental results 
are well reproduced without the need of an extra background from a 
hypothetical $a_0(1100-1300)$ resonance suggested in ref. \cite{Feindt}.

4) One can explain the {\it{drastic reduction}} of the 
Born term in the $\gamma \gamma \rightarrow  K^+ K^-$  reaction in terms of
final state interaction of the $K^+ K^-$ system.

\begin{figure}[H]
\includegraphics[width=0.6\textwidth,angle=-90]
{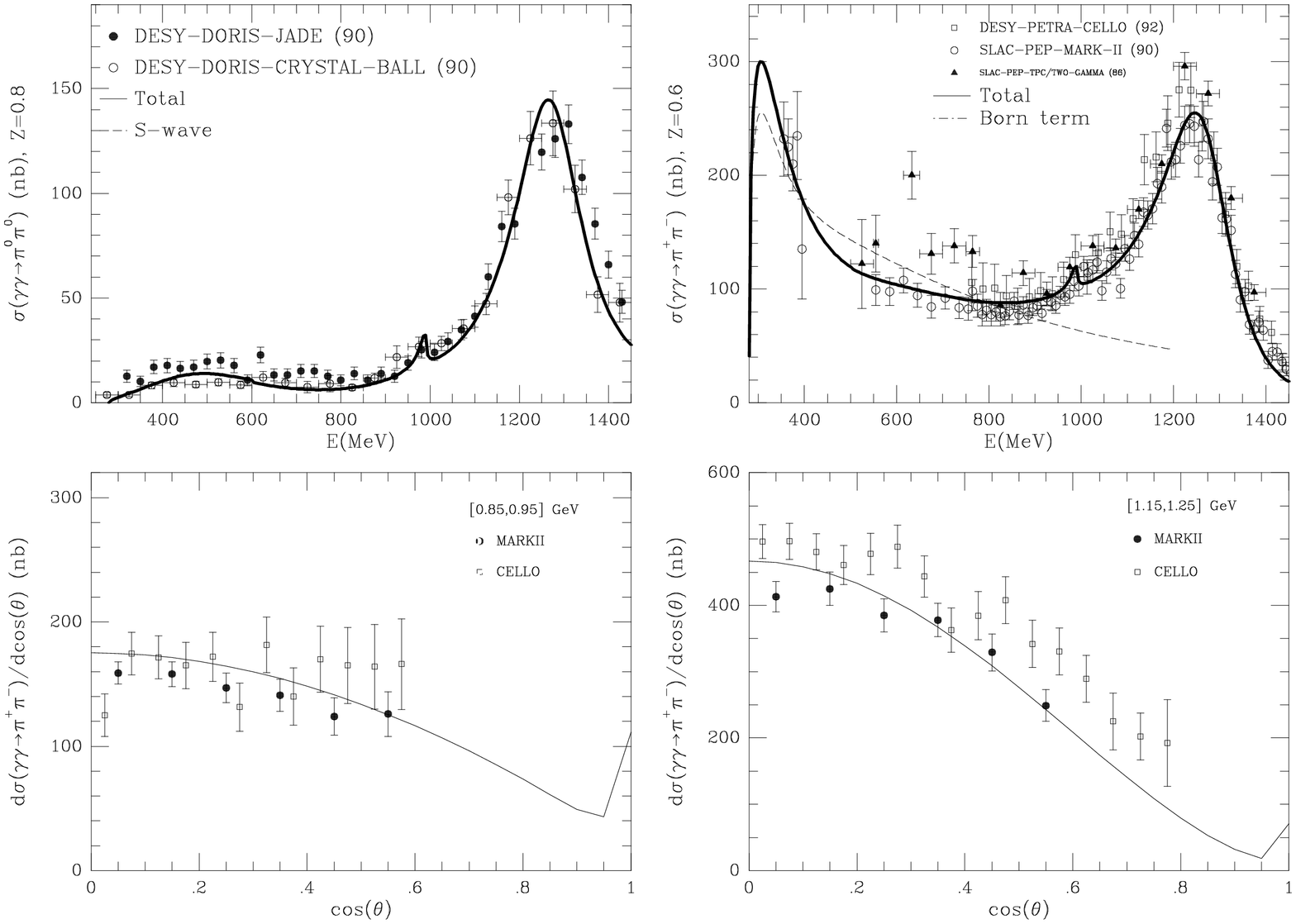}
\end{figure}
\begin{figure}[H]
\includegraphics[width=0.6\textwidth,angle=-90]{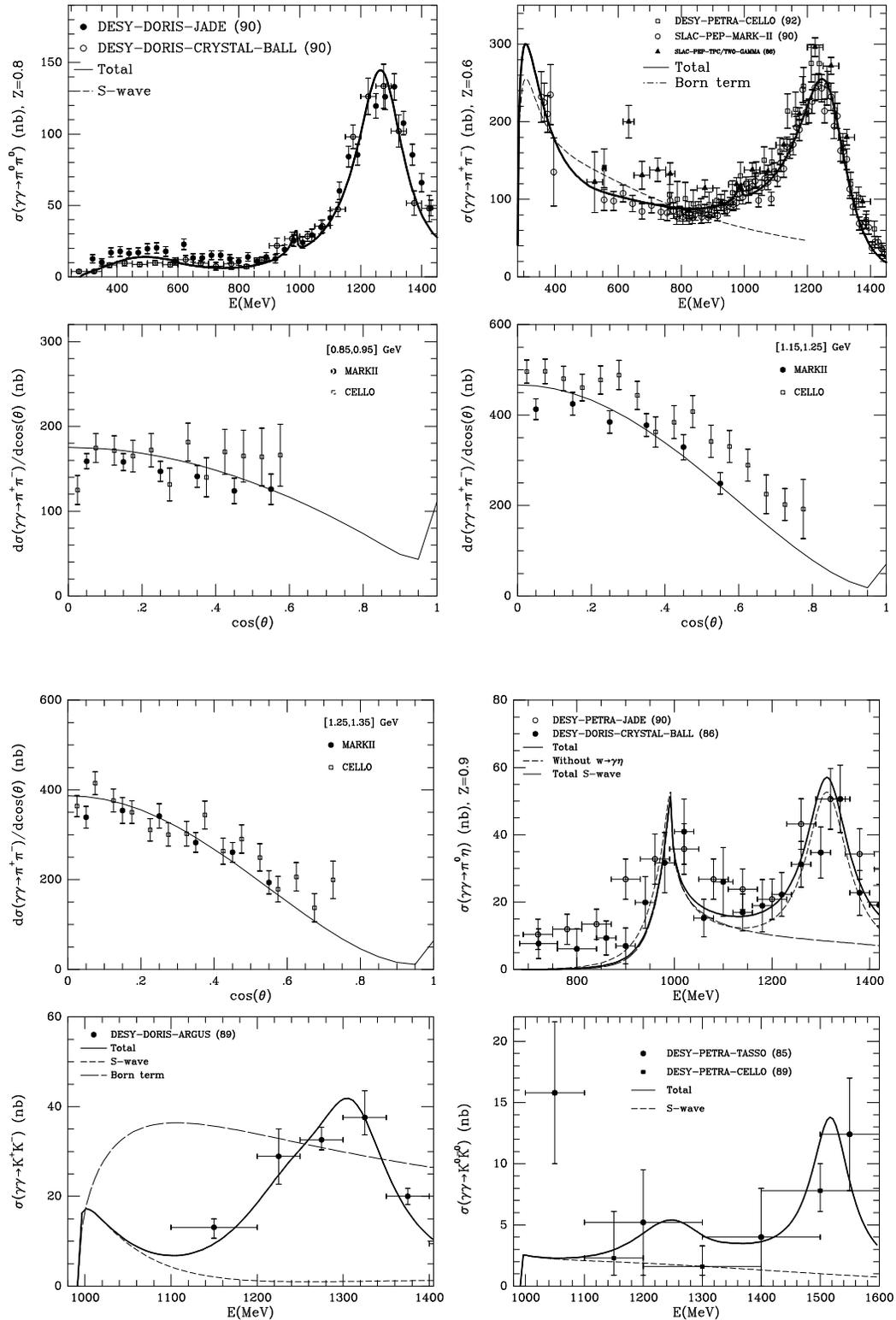}
\caption{Total and differential cross sections for several photoproduction
processes. The references to the experimental data are given in
ref. \cite{gamanpa}.}
\end{figure}

\section{The {\boldmath $\phi\rightarrow \gamma K^0\bar{K}^0$}, 
{\boldmath $\gamma \pi^0 \pi^0$} and {\boldmath $\gamma
\pi^0 \eta$} decays}

We first discuss the decay of the $\phi$ meson to $\gamma K^0\bar{K}^0$
following ref. \cite{PLB}. With the previous formalism fixed, we will consider 
the
decay of the $\phi$ to $\gamma \pi^0\pi^0$ and $\gamma \pi^0 \eta$ 
\cite{Emarco}. 

\subsection{The {\boldmath $\phi\rightarrow \gamma K^0\bar{K}^0$} decay}

The study of the process $\phi \rightarrow \gamma K^0 \bar{K^0}$ 
is an interesting subject since it provides a background to the reaction 
$\phi \rightarrow 
K^0 \bar{K^0}$. This latter process has been proposed as a way to study 
CP violating decays to measure the small ratio $\epsilon ' / \epsilon $ 
\cite{10C}, but, since this implies seeking for very small effects, a 
BR($ \phi \rightarrow \gamma K^0 \bar{K^0} $ )$ \geq 10^{-6} $ will limit the 
scope of these perspectives. There are several calculations of this 
quantity \cite{1C,2C,3C,5C,muli}. In ref. \cite{4C} it is estimated for a {\it non 
resonant} decay process without including the $f_0$ and $a_0$ resonances. The 
issue was revisited in ref. \cite{C}. 

The approach introduced in ref. \cite{npa} to treat the $I=0,1$ scalar 
meson-meson sector was the one used in ref. \cite{PLB}. The formalism,
reviewed in {\it{section 3.3.1}},
will allow us to consider simultaneously the influence of the 
$f_0(980)$ and the $a_0(980)$ resonances, as well as their mutual
interference, 
in a way that takes into account the energy 
dependence of their widths and coupling constants to the $K \bar{K}$ system. 
Furthermore, other possible contributions, non resonant, are also considered. 
The final state interactions will be taken into account following the
way of ref. \cite{gamanpa} and discussed 
in the former section.

\begin{figure}[ht]

\centerline{
\protect
\hbox{
\psfig{file=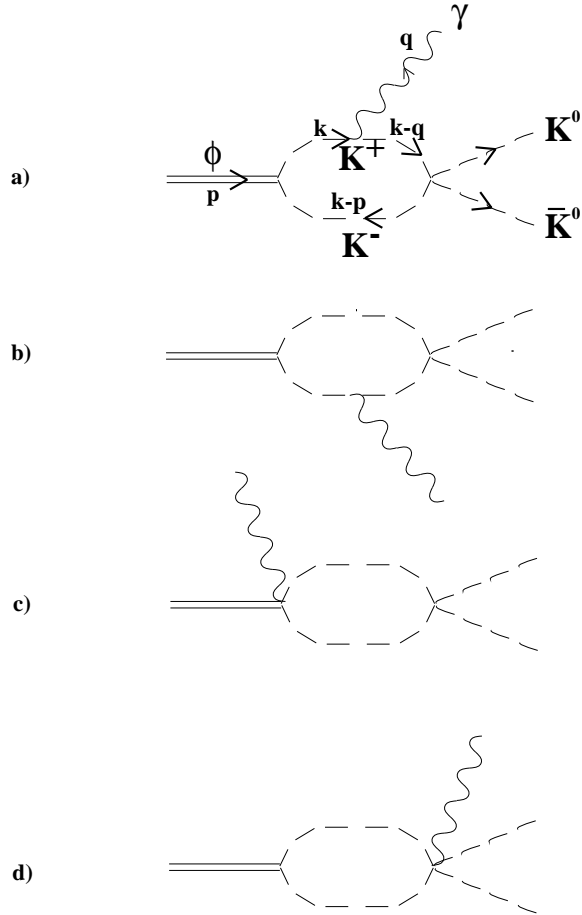,width=0.75\textwidth,angle=-90}}}
\caption{The loop radiation (a,b) and contact (c,d) contributions.}

\label{3.5}
\end{figure}

As in previous works \cite{1C,2C,3C,5C}, in ref. \cite{PLB} the process 
$\phi \rightarrow \gamma K^0 \bar{K^0}$ is calculated through an intermediate 
$K^+ K^-$ 
loop which couples strongly to the $\phi$ and the scalar resonances, 
see Fig. \ref{3.5}.

For calculating the contribution of these loop diagrams one uses the minimal 
coupling to 
make the interaction between the $\phi$ and the $K^+ K^-$ 
mesons gauge invariant, then we have 
\begin{equation}
\label{Hint}
H_{int}=(e A_\mu + g_\phi \phi_\mu)i(\partial^{\mu}
K^+ K^- - K^+ \partial^\mu K^- ) - 2 e g_{\phi} A^{\mu} \phi_{\mu} K^+ K^- \ ,
\end{equation}      
where $g_{\phi}$ is the coupling constant between the $\phi$ and the $K^+K^-$ 
system \footnote{There is an extra contribution that does not come from minimal
coupling and is given by the term proportional to $F_V$ in eq. (\ref{V}).
However, this term only originates a correction to the coupling constant
proportional to the momentum of the photon (see {\it{section 4.2.2}}) which, 
because we are very close to the threshold of the $\phi$, is very tiny.}.

\begin{figure}[ht]

\centerline{
\protect
\hbox{
\psfig{file=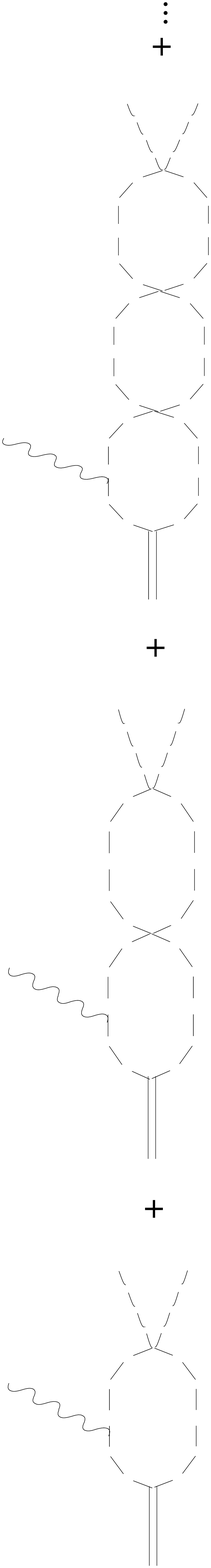,width=.13\textwidth,angle=-90}}}
\caption{Diagrammatic series representing the FSI from a general 
loop of Fig. \ref{3.5}.}
\label{3.6}
\end{figure}

An essential ingredient to evaluate the loop in Fig. \ref{3.5} is the strong amplitude 
connecting $K^+ K^- $ with $K^0 \bar{K^0}$. As we said before, the amplitude 
calculated in ref. \cite{npa} is the one used in ref. \cite{PLB} . This implies the sum 
of an infinite series of diagrams which is represented in Fig. \ref{3.6} for 
the diagram of 
Fig. \ref{3.5}a, and the analogue sums corresponding to Figs. \ref{3.5}b,c,d. This way of taking
into account the S-wave final state interactions is the same as shown above in 
Fig. 4.3 for the $\gamma \gamma \rightarrow$meson-meson reactions.

This series gives rise to the needed corrections due to final state 
interactions and in fact, from the vertex connecting the $K^+ K^-$ with 
the $K^0 \bar{K^0}$, this series is the same one as that in ref. \cite{npa}
which gives 
rise to the S-wave strong amplitude $K^+ K^- \rightarrow K^0 \bar{K^0}$, see eq.
(\ref{bse}). In this approach the vertex between 
the loops correspond to the on shell lowest order $\chi PT$ amplitude 
\cite{xpt}. Note that an analogous series before the loop 
with the emission of the photon is absorbed in the infinite series of 
diagrams contained in the $\phi$ resonance propagator and its effective
coupling. 

First of all, let us see that the strong amplitude connecting $K^+ K^-$ with 
$K^0 \bar{K^0}$ calculated in the way shown in Fig. \ref{3.6}, ref. \cite{npa}, must 
factorize out of the integral.

In order to see this, following ref. \cite{PLB}, consider the diagrams in 
Fig. \ref{3.5} but with the ${\cal O}(p^2)$ $\chi$PT 
amplitude connecting the kaons. This amplitude is given by
\begin{equation}
\label{potencial2a}
<K^0 \bar{K^0} |t| K^+ K^->= \frac{1}{2} \left[ t_{I=0} - t_{I=1} \right]
=-\frac{1}{4f^2}
\left[s+ \frac{4 m_K^2 - \sum_i p_i^2}{3} \right] \ ,
\end{equation}
where $f$ is the pion decay constant, $f\simeq 93$ MeV, $I$ refers to the 
isospin channel of the amplitude and the subindex $i$ runs from $1$ to 
$4$ and refers to any of the four kaons involved in the strong interaction. If 
the particle is on shell then $p_i^2=m_K^2$. In the present case 
$p_{K^0}^2=p_{\bar{K^0}}^2=m_K^2$ so one has
\begin{equation}
\label{potencial2b}
-\frac{1}{4f^2}\left[s+\frac{(m_{K}^2-p_{K^+}^2)+(m_{K}^2-p_{K^-}^2)}{3}\right] \ .
\end{equation}
The important point for the sequel is that the off shell part, which 
should be kept inside the loop integration, will not contribute. 

First of all let us note that, due to gauge invariance, 
the physical amplitude for $\phi \rightarrow \gamma K^0 \bar{K^0}$ has 
the form
\begin{equation}
\label{gaugea}
M(\phi (p) \rightarrow \gamma (q) K^0 \bar{K^0})=[g^{\mu \nu} (p \cdot q)-
p^\mu q^\nu]\epsilon_\mu ^\gamma \epsilon_\nu^\phi H(p\cdot q,Q^2,q\cdot Q) \ ,
\end{equation}
where $\epsilon_\mu ^\gamma$ and $\epsilon_\nu^\phi$ are the polarization 
vectors of the photon and the $\phi$ meson, $Q=p_{K^0}+p_{\bar{K^0}}$ 
and $H$ is 
an arbitrary scalar function. In 
the calculation of this loop contribution the problem is the presence of divergences 
in the loops represented in Fig. \ref{3.5}. Following refs. \cite{1C,2C,3C} we will
take into account the 
contribution of $p^\mu q^\nu$ of Figs. \ref{3.5}a,b, since 
Figs. \ref{3.5}c,d do not give such type of terms. Then, by gauge invariance, 
see formula (\ref{gaugea}), the coefficient for $(p\cdot q) g^{\mu \nu}$ is 
also fixed. In fact, as shown in ref \cite{1C,2C,3C,C}, the $p^\mu q^\nu$ 
contribution will be finite since the off shell part of the strong amplitudes 
does not contribute, as we argue below, and then one is in the same situation 
than in the latter references. On the other hand, depending on the
renormalization scheme chosen, additional tadpole like terms can appear
\cite{Bramon}. However, they do not contribute to the $p^\mu\,q^\nu$ structure
and hence can be ignored. 

If we take the diagrams of Figs. \ref{3.5}a,b, which give the same contribution, we find
the following amplitude for the sum of both diagrams:
\begin{eqnarray}
\label{factora}
M'&=& \epsilon_\mu ^\gamma \epsilon_\nu^\phi \frac{2 e g_\phi }{i} 
\int \frac{d^4k}{(2 \pi)^4} \frac{(2k_\nu-p_\nu)(2k_\mu-q_\mu)}{
(k^2-m_K^2+i \epsilon )( (k-q)^2-m_K^2 + i\epsilon)( (k-p)^2-m_K^2+i \epsilon)}
  \nonumber \\
 &\times& \frac{(-1)}{4f^2}\left[Q^2+ \frac{(m_K^2-p_{K^+}^2)
 +(m_K^2-p_{K^-}^2)}{3}\right] \ .
\end{eqnarray}
The momentum for each particle in the loop is indicated in Fig. \ref{3.5}a and so 
one has $p_{K^+}=k-q$, $p_{K^-}=k-p$. Concentrating in the 
off shell part of the strong amplitude, one has the following integral
\begin{eqnarray}
\label{factorb}
&& \hspace*{-0.8cm}\int \frac{d^4k}{(2\pi)^4} \frac{(2k_\nu-p_\nu)(2k_\mu-q_\mu)}{
(k^2-m_K^2+i\epsilon)((k-q)^2-m_K^2+i\epsilon)((k-p)^2-m_K^2+i\epsilon)}
\nonumber \\
&& \hspace*{-0.8cm}~~~~\times~
[(k-q)^2-m_K^2 + (k-p)^2-m_K^2 ]  =  \\
&& \hspace*{-0.8cm}\int  \frac{d^4k}{(2\pi)^4} \frac{(2k_\nu-p_\nu)(2k_\mu-q_\mu)}{
(k^2-m_K^2+i\epsilon)((k-p)^2-m_K^2+i\epsilon)} + \int \frac{d^4k}
{(2\pi)^4} \frac{(2k_\nu-p_\nu)(2k_\mu-q_\mu)}{(k^2-m_K^2+i\epsilon)
((k-q)^2-m_K^2+i\epsilon)} \nonumber
\end{eqnarray}
Taking into account that 
\begin{equation}
\label{vector}
\epsilon_\mu^\phi \cdot p^\mu=0 \; ; \; \epsilon_\nu^\gamma \cdot q^\nu=0 \; 
\; (\hbox{Feynman gauge})
\end{equation}
then one only has
\begin{equation}
\label{factorc}
\int \frac{d^4k}{(2\pi)^4}\frac{4 k_\mu k_\nu}{(k^2-m_K^2+i\epsilon)
((k-p)^2-m_K^2+i\epsilon)}+\int\frac{d^4k}{(2\pi)^4}\frac{4 k_\mu k_\nu }
{(k^2-m_K^2+i\epsilon)((k-q)^2-m_K^2+i\epsilon)} 
\end{equation}

The above integrals do not give contribution to $q^\mu p^\nu$ since 
in each integral there is only one of the two vectors $q$ or $p$. In this 
way we see that the strong amplitude ${\cal O}(p^2)$ factorizes out on shell in 
(\ref{gaugea}). Note that the important point in the former argumentation 
is the form of the off shell part of the S-wave strong amplitude at 
${\cal O}(p^2)$ and this is common to any other S-wave meson-meson amplitude 
at this order, as one can see in ref. \cite{npa}.  

Next we consider the sum of all the infinite series represented in Fig. 
\ref{3.6}. In
ref. \cite{PLB} this was accomplished by noting that at the one loop level the
strong $\Opd$ $\chi PT$ amplitude factorizes on shell, as we have seen. Then, 
one can apply the
same technique as for the $\gamma \gamma \rightarrow$meson-meson 
and substitute the $\Opd$ amplitude by the full one calculated in ref. \cite{npa}. 
Then to all orders in the approach of ref. \cite{npa} one has the amplitude
\begin{equation}
\label{strong}
t_S=\frac{1}{2}\left[t_{I=0}-t_{I=1}\right] \ .
\end{equation}
Note that the amplitude obtained in ref. \cite{npa} contains also the 
resonances $f_0(980)$ and $a_0(980)$ which are generated dynamically.
The final expression for the amplitude 
$ \phi(p) \rightarrow \gamma(q) K^0 \bar{K^0}$, as given in ref. \cite{PLB}, is 
then
\begin{equation}
\label{amplitude}
M = \epsilon_\mu^\gamma \epsilon_\nu^\phi \frac{2 e g_\phi }{i} 
t_{S}  \int \frac{d^4 k}{(2 \pi )^4} 
\frac{(2k_\nu -p_\nu )(2k_\mu -q_\mu )}
{(k^2-m_K^2+i\epsilon )((k-q)^2-m_K^2+i\epsilon )((k-p)^2-m_K^2+i\epsilon )}
\end{equation} 
This integral has been evaluated in ref. \cite{1C} using dimensional regularization 
and confirmed in ref. \cite{C}, with the result
\begin{equation}
\label{finalamplitude}
M=\frac{e g_\phi}{2 \pi^2 i m_K^2}\, I(a,b)\, [(p \cdot q)(\epsilon_\gamma 
\cdot \epsilon_\phi) - (p \cdot \epsilon_\gamma)(q\cdot \epsilon_\phi)]t_S \ ,
\end{equation}
with $a=M_\phi^2/m_K^2$, $b=Q^2/m_K^2$ and
\begin{equation}
\label{Iab}
I(a,b)=\frac{1}{2(a-b)}-\frac{2}{(a-b)^2}\left[f\left(\frac{1}{b}\right)-
f\left(\frac{1}{a}\right)\right]+
\frac{a}{(a-b)^2}\left[g\left(\frac{1}{b}\right)-g\left(\frac{1}{a}\right)
\right] \ ,
\end{equation} 
where 
\begin{eqnarray}
\label{fx}
f(x)&=&\left\{  
\begin{array}{cl}
-\arcsin^2\left(\frac{1}{2 \sqrt{x}}\right)  & x>\frac{1}{4}
\\[1.5ex]
\frac{1}{4}\left[\ln\left(\frac{\eta_+}{\eta_-}\right)-i\pi\right]^2 & 
x<\frac{1}{4}
\end{array}
\right. \nonumber 
\\
 g(x)&=&\left\{  
\begin{array}{c l}
(4x-1)^\frac{1}{2} \arcsin\left(\frac{1}{2 \sqrt{x}}\right) & x>\frac{1}{4}
\\[1.5ex]
\frac{1}{2}(1-4x)^\frac{1}{2}\left[\ln\left(\frac{\eta_+}{\eta_-}\right)
-i\pi\right] & x<\frac{1}{4}
\end{array}
\right. 
\\
\eta_{\pm}&=&\frac{1}{2x}\left(1 \pm (1-4x)^{\frac{1}{2}}\right)
\nonumber
\end{eqnarray}

After summing over the final polarizations of the photon, averaging  
over the ones of the $\phi$ and taking into account the phase space for 
three particles \cite{PDG} one obtains
\begin{equation}
\label{widthform}
\Gamma (\phi \rightarrow \gamma K^0 \bar{K^0})=\int \frac{dm_{12}^2 dQ^2}
{(2 \pi)^3 192 M_{\phi}^3} \left| e g_{\phi} \frac{I(a,b)}{2 \pi^2 m_K^2} 
\right|^2
(M_\phi^2 - Q^2 )^2  |t_S|^2
\end{equation}
where $m_{12}^2=(q+p_{K^0})^2$. 
Taking $\displaystyle\frac{g_\phi^2}{4\pi}=1.66$ from its width to $K^+K^-$, $M_\phi=1019.41$
MeV, $\Gamma (\phi)=4.43$ MeV, BR( $\phi \rightarrow K^0 \bar{K^0})=
0.34$ and using the mass of the $K^0$ for the phase 
space considerations, ref. \cite{PDG}, one gets 
\begin{equation}
\label{fullresult}
\begin{array}{rl}
\Gamma( \phi \rightarrow \gamma K^0 \bar{K^0})= & 2.22 \times 10^{-7} \, {\rm MeV}
\\[2ex]
BR(\phi \rightarrow \gamma K^0 \bar{K^0})= & 0.50 \times 10^{-7}
\\[2ex]
\displaystyle\frac{ \Gamma( \phi \rightarrow \gamma K^0 \bar{K^0})}{\Gamma ( \phi 
\rightarrow K^0 \bar{K^0})}= & 1.47 \times 10^{-7}
\end{array}
\end{equation}
The uncertainties coming from the range of the possible values for the 
cut off give a relative error around $20 \%$. 
Taking only into account the $I=0$ contribution
\begin{equation}
\label{noa0}
\begin{array}{rl}
\Gamma( \phi \rightarrow \gamma K^0 \bar{K^0})= & 8.43 \times 10^{-7} \, {\rm MeV}
\\[2ex]
BR(\phi \rightarrow \gamma K^0 \bar{K^0})= & 1.90 \times 10^{-7}
\\[2ex]
\displaystyle\frac{\Gamma( \phi \rightarrow \gamma K^0 \bar{K^0})}
{\Gamma (\phi \rightarrow K^0 \bar{K^0})}= & 5.58 \times 10^{-7}
\end{array}
\end{equation}
and with only the $I=1$ 
\begin{equation}
\label{nof0}
\begin{array}{rl}
\Gamma( \phi \rightarrow \gamma K^0 \bar{K^0})= & 2.03 \times 10^{-7} \, {\rm MeV}
\\[2ex]
BR(\phi \rightarrow \gamma K^0 \bar{K^0})= & 4.58 \times 10^{-8}
\\[2ex]
\displaystyle\frac{\Gamma( \phi \rightarrow \gamma K^0 \bar{K^0})}{\Gamma (\phi \rightarrow
 K^0 \bar{K^0})}= & 1.35 \times 10^{-7}
\end{array}
\end{equation}
We see that the process is dominated by the $I=0$ contribution and 
that the interference between both isospin channels is destructive. 
From the former results one concludes that the 
$\phi \rightarrow \gamma K^0 \bar{K^0}$ background will {\it not} be too 
significant for the purpose of 
testing CP violating decays from the $\phi \rightarrow K^0 \bar{K^0}$ process 
at DA$\Phi$NE in the lines of what was expected in ref. \cite{C}. All these 
calculations have been done in a way that both the resonant and non-resonant 
contributions are considered at the same time and taking into account also 
the different isospin channels.


\subsection{The {\boldmath $\phi\rightarrow 
 \pi^0 \pi^0 \gamma$} and {\boldmath $\pi^0 \eta \gamma$} decays}

 Radiative $\phi$ decay into neutral mesons has been a subject of interest often
 advocated as a source of information on the nature of the scalar meson
 resonances. Calculations in the line of the former section have been done
 \cite{Bramon} using the amplitude for the $K \bar{K} \rightarrow
 \pi^0 \pi^0$ amplitude from $\chi PT$. Other calculations have concentrated
 on the possibility of using the reactions to decide the nature of the $f_0$
 resonance between several models like a $K \bar{K}$ molecule, a  $q \bar{q}$
 state or a $q \bar{q} q \bar{q}$ structure \cite{5C}. The laboratories of
 Frascati and Novosibirsk have been actively pursuing research on this topic and
 very recently some novel results have been reported by two Novosibirsk groups
 \cite{Novo,CMDpi0pi0,pi0eta,CMDpi+pi-}.
 In ref. \cite{Emarco} calculations along the lines reported in the previous section
 have been conducted following however the tensor approach to the vector mesons
 of ref. \cite{EPR}
 reported in {\it section 2.4}. This approach had been previously used for the
 $\rho \rightarrow \pi^+ \pi^- \gamma$ decay in the absence of final state
 interaction in ref. \cite{Huber}.
  The novelties with respect to the approach of the former section
  can be summarized in the
  basic couplings involved in Fig. \ref{3.5} given by 
\begin{eqnarray}
t_{\phi K^+ K^-} &=& \frac{G_V M_{\phi}}{\sqrt{2} f^2} (p_{\mu}-p'_{\mu})
\epsilon^{\mu} (\phi) \nonumber \\
t_{\phi \gamma K^+ K^-} &=&
-\sqrt{2} e \frac{G_V M_{\phi}}{f^2} \epsilon_{\nu} (\phi) \epsilon^{\nu}
(\gamma) \nonumber \\
&&- \frac{\sqrt{2} e}{M_{\phi} f^2} \left(\frac{F_V}{2} -G_V\right)
P_{\mu} \epsilon_{\nu} (\phi)
[k^{\mu} \epsilon^{\nu} (\gamma) -k^{\nu} \epsilon^{\mu} (\gamma)]
\label{eq:couplings}\\  \nonumber
\end{eqnarray}
with $p_{\mu}$, $p'_{\mu}$ the $K^+, K^-$ momenta, $P_{\mu}$,
$k_{\mu}$ the $\phi$ and photon momenta and $f$ the pion decay
constant.

The couplings of eq. (\ref{eq:couplings}) are easily induced from the 
Lagrangian of eq.
(\ref{V}). The $\phi$ meson is introduced in the scheme by means of a singlet,
$\omega_1$, going from SU(3) to U(3) through the substitution
$V_{\mu \nu} \rightarrow V_{\mu \nu} +I_3 \frac{\omega_{1,\mu \nu}}{\sqrt{3}}$,
with $I_3$ the $3 \times 3$ diagonal matrix. Then, assuming ideal mixing
for the $\phi$ and $\omega$ mesons

\begin{eqnarray}
  \sqrt{\frac{2}{3}} \omega_1 + \frac{1}{\sqrt{3}} \omega_8 &\equiv& \omega
    \nonumber \\
  \frac{1}{\sqrt{3}} \omega_1 - \frac{2}{\sqrt{6}} \omega_8 &\equiv& \phi
\end{eqnarray}
one obtains the required vertices after substituting in eq. (\ref{V}) 
$V_{\mu \nu}$
by $\tilde V_{\mu \nu}$, given by

\begin{equation}
\tilde V_{\mu \nu} \equiv \left(\begin{array}{ccc}
\frac{1}{\sqrt{2}} \rho^0_{\mu \nu} + \frac{1}{\sqrt{2}} \omega_{\mu \nu}
 & \rho^+_{\mu \nu} & K^{* +}_{\mu \nu} \\
\rho^-_{\mu \nu}& -\frac{1}{\sqrt{2}} \rho^0_{\mu \nu}
+ \frac{1}{\sqrt{2}} \omega_{\mu \nu} & K^{* 0}_{\mu \nu} \\
K^{* -}_{\mu \nu} & \bar{K}^{*0}_{\mu \nu} & \phi_{\mu \nu}
\end{array}
\right)
\end{equation}

 As we can see, with respect to the approach of the former section
 the contact term contains an
 extra part, the term proportional to $\displaystyle\frac{F_V}{2}-G_V$, 
which is gauge
 invariant by itself. The first part of the contact term proportional to $G_V$
 is not gauge invariant and this requires the addition of the diagrams (a) and
 (b) of Fig. \ref{3.5} to have a gauge invariant set, as discussed in the former
 section. This
 means that in addition to the terms discussed there one gets now an additional
 term proportional to $(\displaystyle\frac{F_V}{2}-G_V) k$, where $k$ is
the photon momentum in the $\phi$ rest frame. This term
appears only with the structure of diagram (c). This has as a consequence
 that in the treatment of the final state interaction the first loop contains
 only the two meson propagator and hence is the same function $g(M_I)$ defined
  in
 eq. (\ref{5.0}). 
This technical detail plus the use of the $K^+ K^- \rightarrow \pi^0
 \pi^0$ and $\pi^0 \eta$ amplitudes instead of the  $K^+ K^- \rightarrow \ K^0
 \bar{K^0}$ in the former section are the basic modifications needed in this
 work. The values $G_V=55$ MeV and $F_V=165$ MeV
which are suited to the $\phi \rightarrow K^+ K^-$ and
$\phi \rightarrow e^+ e^-$ decay widths respectively have been used in
ref. \cite{Emarco}.

\begin{figure}[H]
\centerline{
\includegraphics[width=0.55\textwidth,angle=-90]{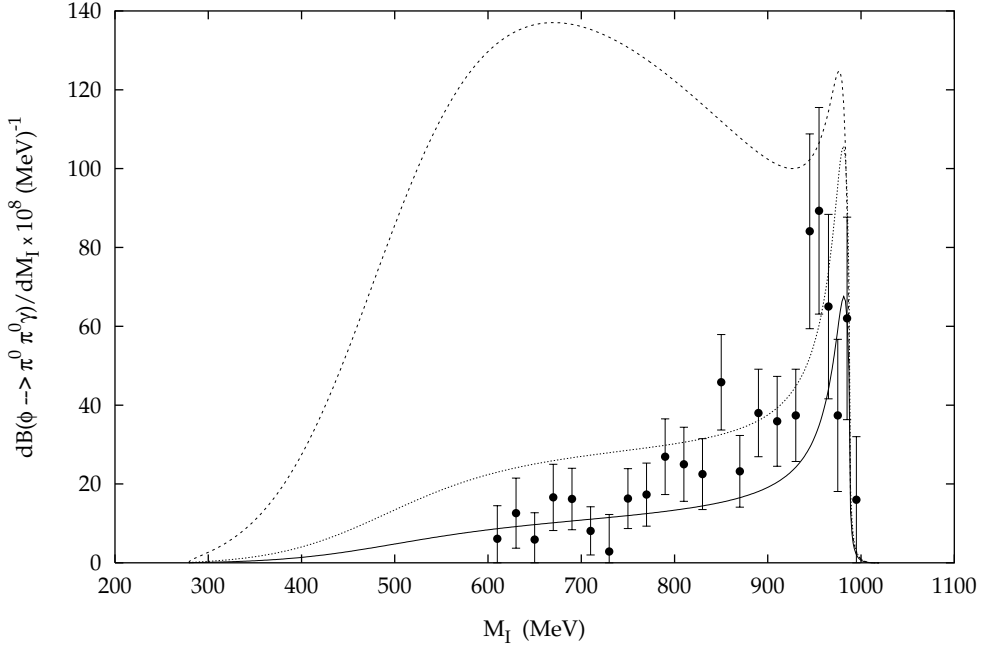}
}
\caption{Distribution $dB /dM_I$ for the decay
$\phi \rightarrow \pi^0 \pi^0 \gamma$, with $M_I$ the invariant mass of
the $\pi^0 \pi^0$ system. Solid line: our prediction, with $F_V G_V >0$. 
Dashed line: result taking $F_V G_V <0$. The intermediate dotted line
corresponds to $G_V=67$ MeV and $F_V=154$ MeV \cite{jaojr}, the values 
of the parameters
for $\rho$ decay.
The data points are from ref. \cite{Novo} and only
statistical errors are shown. The systematic errors are similar to the
statistical ones \cite{Novo}.
}
\label{fig3}
\end{figure}

In Fig.~\ref{fig3} we show
the distribution $dB/dM_I$ for $\phi \rightarrow \pi^0 \pi^0 \gamma$
which allows one to see the
$\phi \rightarrow f_0 \gamma$ contribution since the $f_0$ is
the important scalar resonance appearing in the
$K^+ K^- \rightarrow \pi^+ \pi^-$ amplitude \cite{npa}.
The solid curve shows the results with $F_V G_V >0$, the sign
predicted by vector meson dominance \cite{EPR}.
The intermediate dotted line corresponds to taking $G_V=67$ MeV
and $F_V=154$ MeV \cite{jaojr}, which are the values of the parameters for
$\rho$ decay, usually assumed as standard values. The two curves
give us an idea of the theoretical uncertainties.
The upper dashed curve is obtained considering $F_V G_V <0$.
These results are compared in the figure with the recent ones of the
Novosibirsk experiment \cite{Novo}.
We can see that the shape of the spectrum is relatively well reproduced
considering statistical and systematic errors (the latter ones not shown in
the figure). The results considering $F_V G_V <0$ are
in complete disagreement with the data.

The finite total branching ratio  for the $\phi \rightarrow \pi^0 \pi^0 \gamma$
 is $0.8 \times 10^{-4}$ . This latter
number is slightly smaller than
the result given in  ref. \cite{Novo},
$(1.14\pm 0.10\pm 0.12)\times 10^{-4}$, where the first error is statistical
and the second one systematic. The result given in
ref. \cite{CMDpi0pi0} is
$(1.08\pm 0.17\pm 0.09)\times 10^{-4}$, compatible with the prediction.
If the values for $F_V$, $G_V$ of the $\rho$ decay are used one obtains 
$1.7 \times 10^{-4}$ \cite{jaojr}.
The branching ratio obtained for the case
$\phi \rightarrow \pi^0 \eta \gamma$ is $0.87 \times 10^{-4}$. The
results obtained at Novosibirsk are \cite{pi0eta} $(0.83 \pm 0.23)
\times 10^{-4}$
and \cite{CMDpi0pi0}
$(0.90\pm 0.24\pm 0.10)\times 10^{-4}$. 
If the values of the $\rho$ decay are used one
obtains $1.6\times 10^{-4}$ \cite{jaojr}. The spectrum, not shown,
is dominated by the $a_0$ contribution.

 The results of this section for these radiative decays are a striking success of
 the chiral unitary approach reported here . The evidence given by 
Fig. \ref{fig3} in
 favour of the sign $F_V G_V >0$ is much stronger than the one given in
 ref. \cite{Huber} from the tail of the $\rho\rightarrow \pi^+ \pi^- \gamma$
 distribution.

\section{Vector and scalar pion form factors}

The scalar and vector form factors of the pion are defined respectively as

\be
\left<\pi^a (p') \pi^b (p) \, \hbox{out} \left| {\hat{m}} (\bar u u+ \bar d d) 
\right| 0  \right>=\delta^{ab} m_\pi^2 \Gamma(s)
\ee

\noindent
and
\be
\left< \pi^i (p') \pi^l (p) \, \hbox{out} \left| \bar q \gamma_{\mu} 
\left(\frac{\tau^k}{2}\right) q \right| 0 \right>=i \, \epsilon^{ikl} 
(p'-p)_{\mu} \, F_V (s)
\ee

\noindent
with $\hat{m} =(m_u+m_d)/2$ and $\epsilon^{ijk}$ the total antisymmetric tensor 
with three indices.

Assuming elastic unitarity (valid up to the $K {\bar K}$ threshold and 
neglecting multipion states) and making use of the Watson final state theorem 
\cite{Wa1} the phase of $\Gamma(s)$ and $F_V (s)$ is fixed to be the one of the 
corresponding partial wave strong amplitude:
\begin{eqnarray}
\hbox{Im} \, \Gamma (s+i\epsilon)&=& \tan \delta^0_0 \, \hbox{Re} \Gamma (s)  
\nonumber  \\
\hbox{Im} \, F_V (s+i\epsilon)   &=& \tan \delta^1_1 \, \hbox{Re} F_V (s)
\label{elasunit}
\end{eqnarray}
The solution of (\ref{elasunit}) is well known and corresponds to the Omn\`es 
type \cite{Mu1,Om1}:
\begin{eqnarray}
\Gamma (s)&=& P_0 (s) \, \Omega_0 (s)   \nonumber \\
F_V (s)   &=& P_1 (s) \, \Omega_1 (s)
\label{ffactors}
\end{eqnarray}
with
\be
\Omega_i (s)= \exp \left\{ \frac{s^n}{\pi} \int^{\infty}_{4m^2_{\pi}} \,
\frac{ds'}{{s'}^n} \, \frac{\delta^i_i (s')}{s'-s-i\epsilon} \right\}
\label{disp}
\ee

\noindent
In (\ref{ffactors}) $P_0 (s)$ and $P_1 (s)$ are polynomials of degree fixed by 
the number of 
subtractions done in $\ln \{ \Omega_0 (s) \}$ and $\ln \{ \Omega_1 (s)
\}$ minus one, and the zeros of $F_V$ and $\Gamma$.
For $n=1$, $P_i (s)=1$. This follows from the normalization requirement that 
$\Gamma (0)=F_V (0)=1$ and the absence of zeros for those
quantities.\footnote{In principle $\Gamma(0)\neq 1$. However, since this is
the leading $\chi PT$ result and we are at low energies, the difference with 
respect one is very small \cite{ulfff}.}

In ref. \cite{NPB} the previous dispersion integrals, eq. (\ref{disp}), are evaluated 
making use of the phase shifts calculated in the same reference, 
see {\it{section 3.1.1}}. The resulting vector and scalar form factors are shown in 
Figs. 4.8 and 4.9 respectively. The 
Omn\`es solution assumes the phase of the form factor to be that of the 
scattering amplitude, and that is true exactly only until the first inelastic 
threshold. 
The first inelastic threshold is the $4\pi$ one. However, as it was already 
said, its influence, in a first approach, is negligible. The first important 
inelastic threshold is the $K {\bar K}$ one around 1 GeV. This is essential in 
$I=L=0$ but negligible in $I=L=1$. This inelastic threshold, as 
discussed above, has been included in the approach and it is mostly responsible 
for the appearance of the $f_0 (980)$ resonance.

\begin{figure}[ht]
\begin{minipage}[t]{0.48\textwidth}
\includegraphics[width=0.6\textwidth,angle=-90]{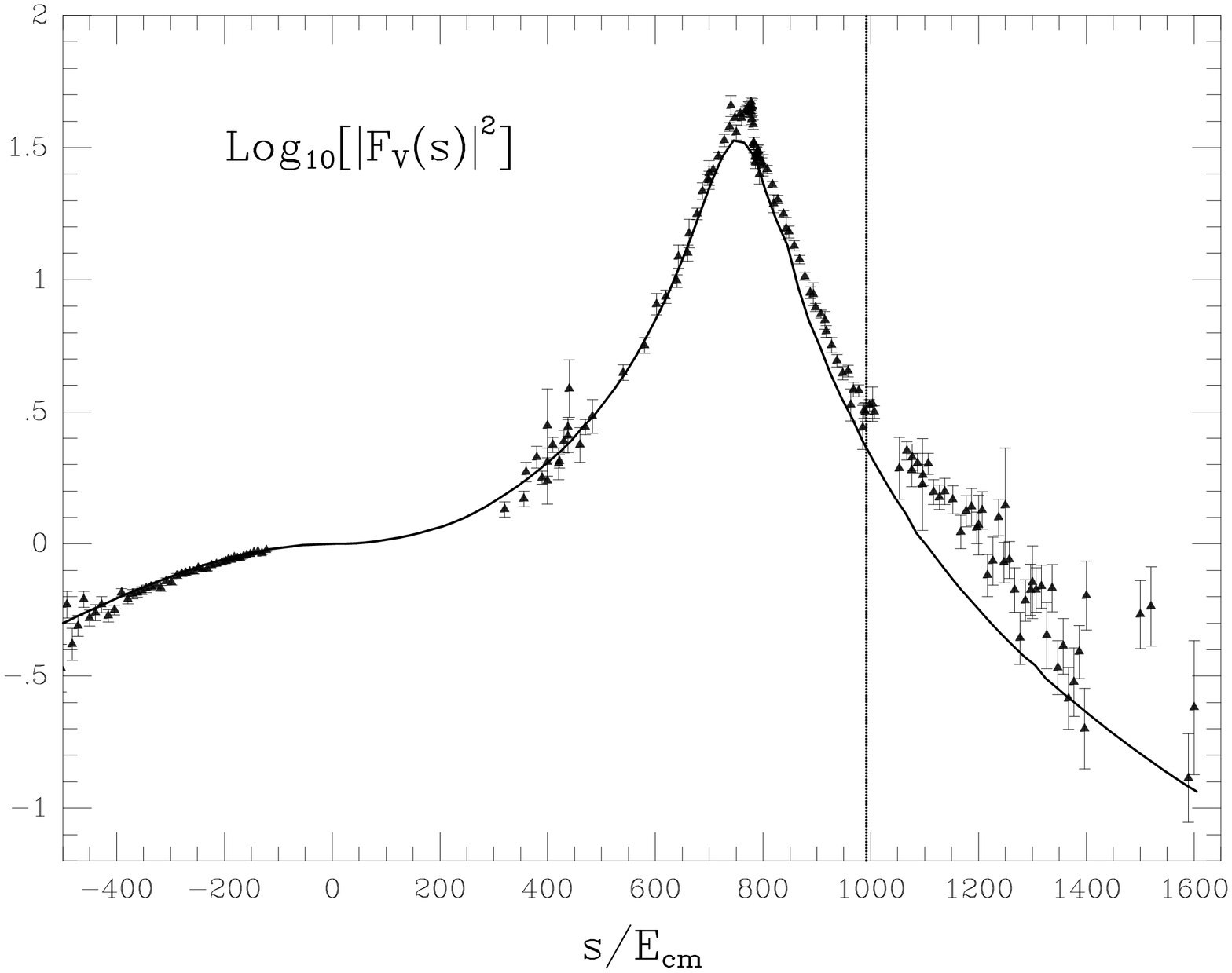}
\caption{Vector pion form factor. The vertical line shows
the opening of the $K {\bar K}$ threshold. Data from ref. \cite{Bar1}.}
\end{minipage}
\hfill
\begin{minipage}[t]{0.48\textwidth}
\includegraphics[width=0.6\textwidth,angle=-90]{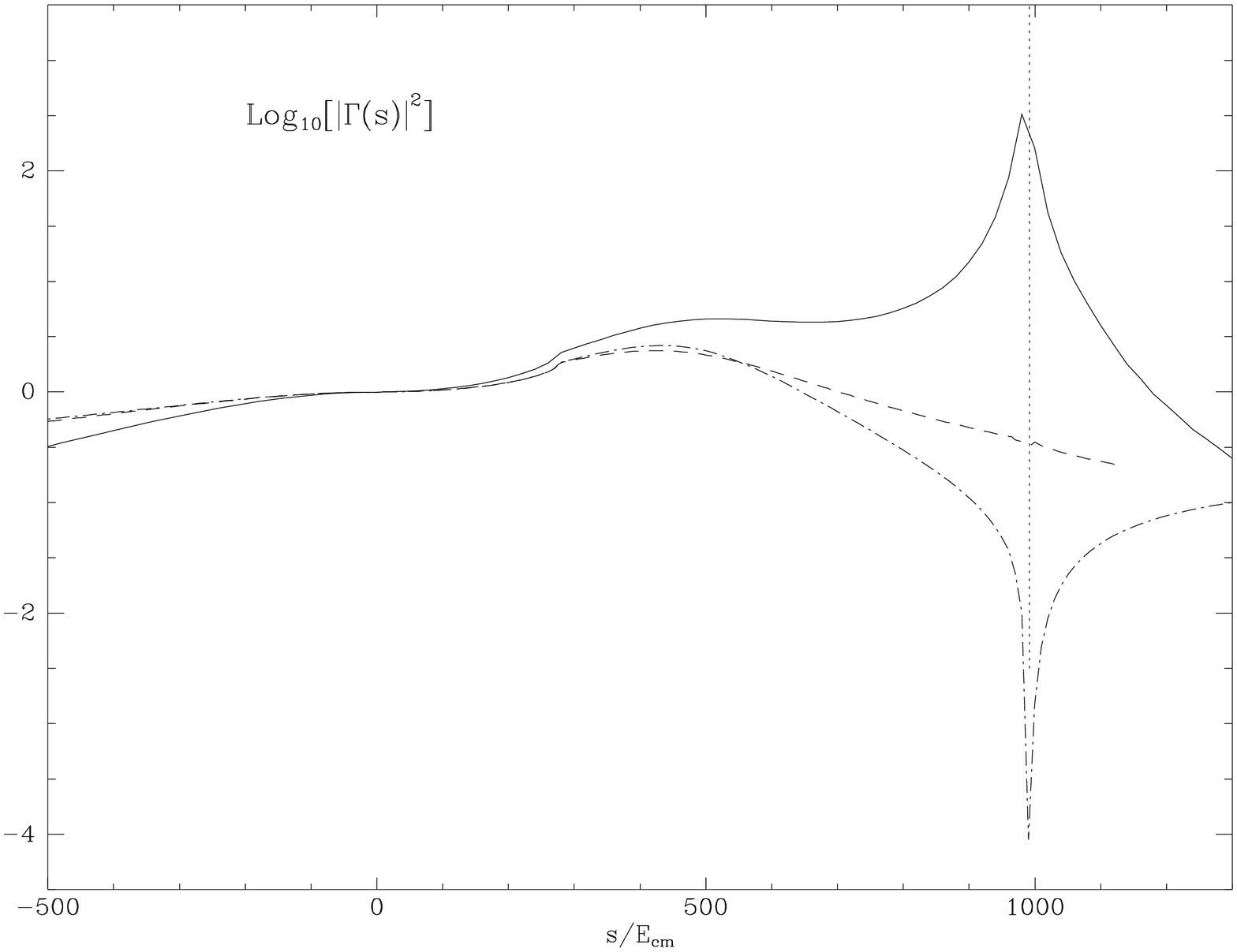}
\caption{Scalar form factor. The dashed curve is the result unitarizing
only with pions. The solid line is the full result with both pions and kaons
in the intermediate state integrating up to infinity in eq. (\ref{disp}). The
dotted-dashed line is the same as the solid one but integrating only up to the
openning of the $K{\bar K}$ threshold. The vertical line shows the opening
of this threshold}
\end{minipage}
\end{figure}

 In the case of the $F_V (s)$  the agreement with 
existing data is 
quite satisfactory, with the dominant role played by the $\rho(770)$ resonance. 
On the other hand, taking into account possible 
uncertainties coming from orders higher than $p^4$ in $\chi PT$, the result obtained 
for the vector form factor is similar to the one obtained in ref.
\cite{F2} using another phase shift expression. For the 
$I=L=0$ channel the most dramatic effect is the openning of the $K\bar{K}$ channel. In 
Fig. 4.9 the continuous line corresponds to the use of eq. (\ref{ffactors}) 
integrating up to infinity in the Omn\`es formula (\ref{disp}). On the other hand, 
the dashed-dotted line
corresponds to integrating
 only up to the openning of the $K\bar{K}$ threshold. The
differences between both options are tremendous making evident that a coupled channel
approach is necessary in order to describe properly the scalar form factor for energies
above
400--500 MeV. Finally, the dashed line represents the 
scalar form factor unitarizing only with pions to obtain the $\delta_{00,
\pi\pi}$ phase shift, in the line of the works \cite{T1,T2,H1,nieves}.

\section{
{\boldmath $\gamma p \rightarrow p$}  meson-meson}

\begin{figure}[ht]
\centerline{
\includegraphics[width=0.6\textwidth,angle=-90]{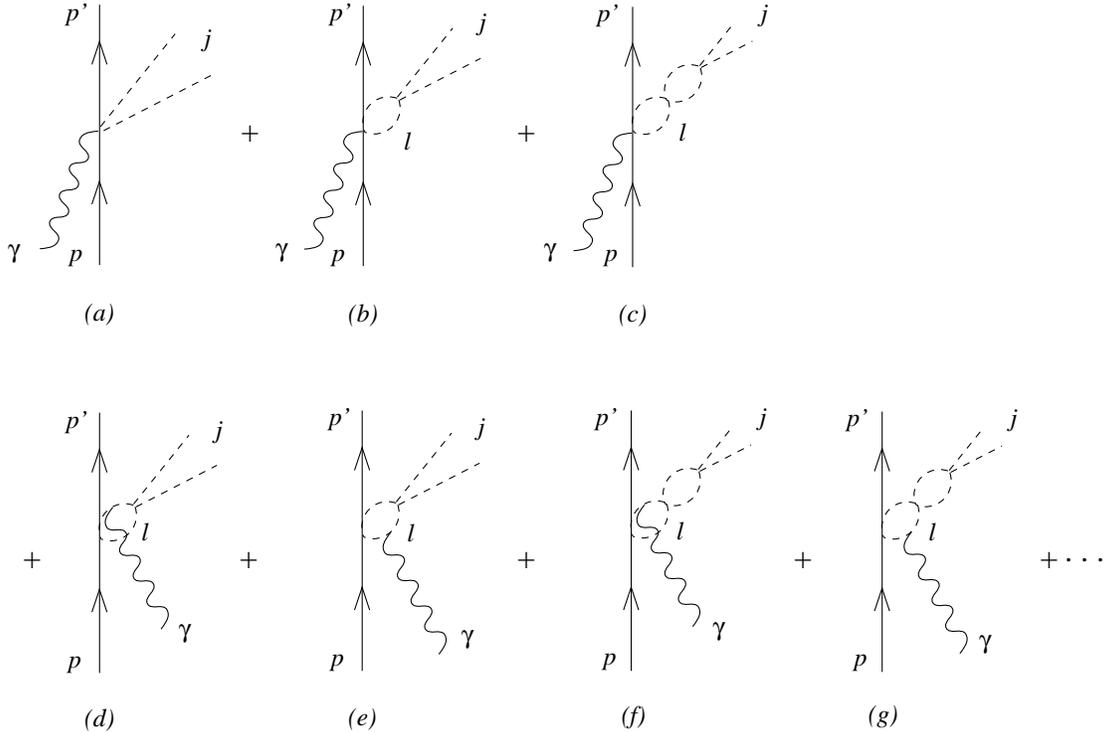}
}

\caption{ Tree level and iterated terms from the contact term and from
mesonic Bremsstrahlung.}
\label{fig:gpmmp}
\end{figure}

  An interesting example of final state interaction of two mesons appears in the
 photoproduction of pairs of mesons in the $\gamma p$  reaction when the
 pair is produced with an invariant mass close to the mass of one resonance. An
 example of it is given in ref. \cite{ugemeson} where the photoproduction of scalar
 meson is studied. The process is depicted in Fig. \ref{fig:gpmmp} where the leading tree
 level process appears in diagram (a) and is a contact term induced from minimal
 coupling from the meson-baryon Lagrangian of eq. (\ref{eq:mesbar}).
 This term is given by

  \begin{equation}        \label{eq:elec_vertex}
V^{\gamma}_{\pi (K)} = -C_{\pi (K)} \frac{e}{2f^2} \bar u(p') \gamma^\mu
u(p) \epsilon_{\mu} \ ,
\end{equation}
where $C_\pi$ =1 and $C_K$=2 stand for the case of the production of a $\pi^+
\pi^-$ or a $K^+ K^-$ pair respectively.

The final state interaction of the mesons
is accounted for by means of the rest of the diagrams in the figure and,
similarly to the case of the $\phi$ radiative decay, the photon lines must also be
coupled to the lines in the loops to guarantee gauge invariance. Missing in the
figure are the tree level diagrams of Bremmsstrahlung where the photon is
coupled to any of the external meson lines in diagram (a). This is justified in
the case of meson production close to threshold (for this purpose a photon beam
of energy 1.7 GeV is suggested in ref. \cite{ugemeson}) or in the case of production of
neutral meson pairs like $\pi^0 \pi^0$ or $\pi^0 \eta$.  At higher energies
there can be more involved production mechanisms \cite{Adam,Fries,Titov} 
and also at lower energies there are background terms which are
important particularly for the case of $\pi^+ \pi^-$ production 
\cite{Tejedor,Ochi}. 
Yet the resonant production should show up as a bump on top of the
background and allow one to study the resonances in a different setup, and in
addition
study the production of the resonances in nuclei which we will address below.
The technique to include final state interaction here follows closely the steps
described in the study of the $\gamma \gamma \rightarrow$ meson-meson reaction
in ref. \cite{gamanpa} and of the $\phi$ radiative decay in ref. \cite{Emarco},
which have been reported above. There is only one difference since the
square of the momentum transferred by the nucleon (which plays here the role
of the $\phi$ mass squared in the $\phi$ decay) can be here negative. In
this case one has to
extrapolate analytically the $I(a,b)$ function of eq. (\ref{Iab}) and detailed
expressions are given in ref. \cite{ugemeson}.

\begin{figure}[ht]
\centerline{
\includegraphics[width=0.55\textwidth,angle=-90]{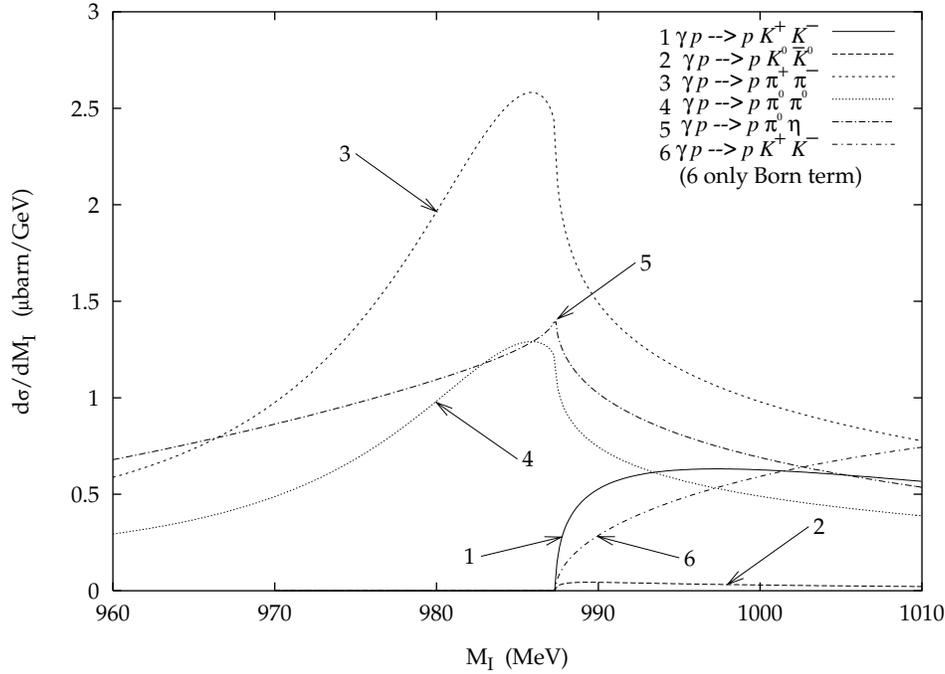}
}

\caption{Results for the cross section on protons
as a function of the invariant
mass of the meson-meson system.}
\label{fig:variosprod}
\end{figure}

  The results obtained for the invariant mass distribution of the two mesons are
given in Fig. \ref{fig:variosprod} for different pairs of mesons in the final state. The
figure shows clear peaks  for the production of the $f_0$ and $a_0$ resonances
in the $\pi^+ \pi^-$, $\pi^0 \pi^0$  and $\pi^0 \eta$ production respectively.
In these cases the ratio of the resonance signal to background is found to be
optimal for the $\pi^0 \pi^0$ and $\pi^0 \eta$ cases. The figure also shows the
cross section for $K^+ K^-$ production close to threshold which is appreciably
renormalized by final state interaction with respect to the Born contribution.
The $K^0 \bar{K^0}$ production is found to be very small.

  It is interesting to note that the signal for the $\pi \pi$ production around
  the $f_0$ resonance shows up with a peak. This is a novelty with respect to
  the case in $\pi \pi$ scattering where the $f_0$ shows up with a minimum in
  the cross section. The different structure of the loop function in the first
  loop in the figures with respect to the $g(s)$ function of the plain two
  meson propagator is responsible for this different interference pattern with
  the background from the $\sigma$ pole contribution. This was also the case
  in the radiative $\phi$ decay where the $f_0$ resonance also appeared as
  a peak in the invariant mass distribution.

\subsection{
 Scalar meson production in nuclei}

  It is also relevant to see that the previous reaction can be suited for the
  observation of the resonance modification in nuclei. The $f_0$ and $a_0$
  couple strongly to $K \bar{K}$ but they are narrow because there is little
  phase space for the decay into that state. However, the kaons are apprecibly
  renormalized in nuclei and hence one expects that these resonances will also
  be appreciably modified inside a nuclear medium.

  The interaction of $K$, $\bar{K}$ with nuclei is a subject that has attracted
much attention 
\cite{WT76,CT77,T7778,D80,RT80,CE92,JEC95,JK92,MHT94,FGB94,BGF97,alberg76,star87,Koch94}. 
Interesting developments
have been done recently in 
$K^- N$ scattering using the chiral Lagrangians \cite{AO,NK}, which have
allowed to tackle the problem of the $K$, $\bar{K}$ nucleus interaction
with some novel results \cite{WKW96,Waas97,Lutz98,RO99}.
The issue is not yet settled since there
are still important discrepancies between the different results.

The first thing which we
observe is that if one looks for a proton in the final state,
one can also have the
$\gamma n \rightarrow p \pi^- \eta (K^- K^0)$ reactions and approximately
one would expect a cross section

\begin{equation}
\left.\frac{d\sigma}{dM_I}\right|_A \simeq Z\frac{d\sigma}{dM_I} (p)
+ N\frac{d\sigma}{dM_I} (n) \, .
\end{equation}
The latter cross section can proceed through the
meson channels $K^- K^0$ and $\pi^- \eta$, both in $I=1$.
These cross sections are found in ref. \cite{ugemeson} to be one order of
magnitude smaller than those on the proton target.
Hence, in nuclei we should expect a cross section roughly
$Z$ times the one of the proton, unless the properties of
the resonances $a_0$ and $f_0$ are drastically modified
in the medium, which is, however, what one expects. As noted above,
 the relatively small width of the $f_0$ resonance is 
due to the small coupling to the
$\pi \pi$ channel. The coupling of $a_0$ to $\pi^0 \eta$ is comparatively
much larger. These resonances,
however, couple very strongly to the $K \bar{K}$ system but the decay
is largely inhibited because the $K \bar{K}$ threshold
is above the resonance mass. Only the fact that the $f_0$ and $a_0$ 
resonances have
already a width for $\pi \pi$ and $\pi \eta$ decay, respectively,
allows the $K \bar{K}$ decay through the tail of the
resonance distribution. If the $K^-$ develops a large width on
its own this enlarges considerably the phase space for $K \bar{K}$
decay and the $a_0$, $f_0$ width should become considerably larger.

Given the interest that the
modifications of meson resonances in nuclei, like the $\sigma$
\cite{Schuck,Aouissat}, $\rho$ \cite{Soyeur,Hermann}, etc., is
raising, the study of the
modifications of the $f_0$ and $a_0$ is bound to offer us
some insight into the nature of these resonances, that has
been so much debated, and into the chiral unitary
approach to these resonances which we are discussing in this work.
Preliminary results using the $K,\bar{K}$ self-energies in the medium
discussed in {\it section 6.2} are already available \cite{manolonew} and
indicate a large increase of the $f_0$ width in the medium.

%% file: isifsi.tex
\chapter{Initial and final state interaction in meson-baryon
reactions}

In {\it section 3.3.2} we studied the meson-baryon interaction around
the region of the $\Lambda(1405)$ and $N^*(1535)$ resonances.  
We saw
there how the unitarization in coupled channels was essential to
reproduce the scattering data. 
In this chapter we show some
examples of physical reactions where the meson-baryon interaction
appears in the initial or final states. We shall show how the proper
consideration of the initial state interactions along the lines discussed
in {\it section 3.3.2} brings a natural solution to one problematic reaction,
the  $K^-  p$  radiative capture, where the ratio of $\Lambda$ to
$\Sigma^0$ production is abnormally low.  
The
coupled channel unitary techniques will be also applied to study
reactions in which the
resonances appearing in the meson-baryon interactions are now generated
in the final state.
In addition we shall also devise how the techniques
can be used to evaluate static properties of those resonances.

\section{{\boldmath $K^-$} proton radiative capture: {\boldmath 
$K^- p \to \gamma \Lambda, \gamma \Sigma^0$}}

The near threshold $K^- p \rightarrow \gamma Y$ reaction with $Y=
\Lambda, \Sigma^0$ has long attracted a lot of interest, mainly
because
of the possibility of using this reaction to resolve the
debates\cite{jones,darewych,veit,zhong,kaxiras,he,arima} over
the structure of the $\Lambda(1405)$ resonance. Most of the
earlier theoretical
investigations\cite{workman} neglected the initial strong $K^- p$
interactions.
It was first demonstrated by Siegel and Saghai\cite{saghai}
that the initial $K^- p$ interactions can drastically change the
predicted
capture rates and thus can significantly alter the interpretation
of the data.
With the phenomenological separable
potentials, they, however, needed an about $30 - 50 \%$ deviation
of the coupling
constants from the SU(3) values to obtain an accurate description
of the data.

This unsatisfactory situation was revised in ref. \cite{LOOR98},
where it
was investigated whether the data could be well described by
treating
the initial
$K^-p$ interactions within the unitary coupled-channel chiral
approach of refs. \cite{AO,NK}, which has been discussed in detail in
{\it section 3.3.2}. In the present section we report on the results
obtained 
in ref. \cite{LOOR98} and we will show that, indeed, a satisfactory
agreement with the experiment can be obtained without the need of 
SU(3) breaking.

\begin{figure}[ht!]
\begin{center}
\includegraphics[width=0.45\linewidth,angle=-90]{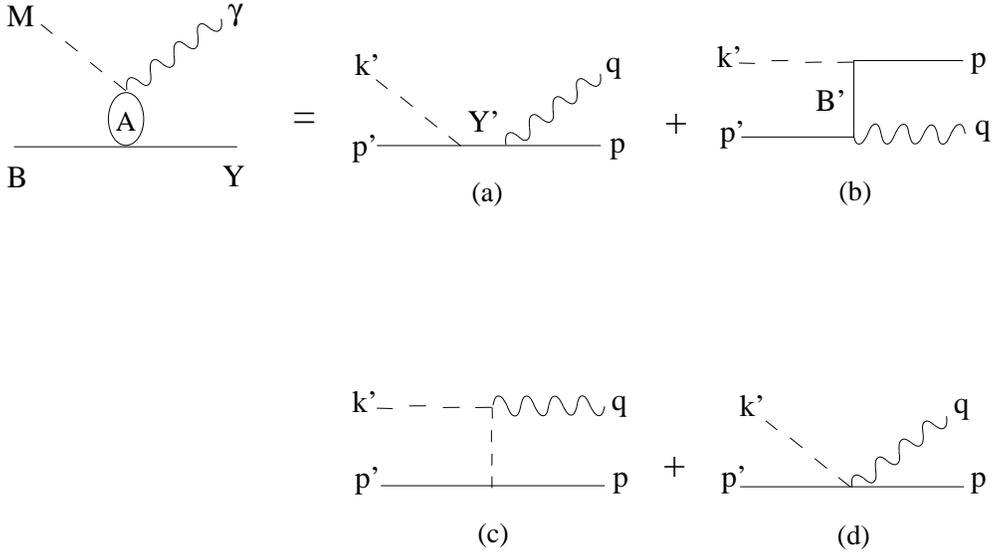}
\end{center}
\caption{Elementary amplitudes for the reaction $K^- p \to \gamma
Y$}
        \label{fig:kpgamma}
\end{figure}

A detailed
derivation of the $K^- p \to \gamma Y$ reaction can be found in 
ref. \cite{LOOR98}, where standard electromagnetig
vertices are used and the initial state
strong $K^- p$ interaction is described through the
coupled-channel Bethe-Salpeter
equation. The resulting amplitude
in the center of mass frame $P=(\sqrt{s},\vec{0})$ reads
\begin{eqnarray}
T_{\gamma Y, K^- p}(q,k') &=& Q_{\gamma Y,K^- p}(q,k')
+[QGT]_{\gamma Y,K^- p}(q,k')+ \Delta_{\gamma Y,K^- p}(q,k') \ ,
\label{eq:gammat} 
\end{eqnarray}
where the first term
\begin{eqnarray}
Q_{\gamma Y,MB}(q,k^\prime)= i e_M
[\vec{\sigma}\ \vec{\epsilon}\, ]
\frac{C_{Y,MB}}{2f}
\left( 1 -
\frac{\omega_M(k^\prime)}{2q}+\frac{\mu^2_M}{4qk^\prime} 
\ln\frac{\omega_M(k^\prime)+k^\prime}
{\omega_M(k^\prime)-k^\prime} \, \right) \ , 
\label{eq:gammaborn}
\end{eqnarray}
with $MB\equiv K^- p$,
collects the Born terms for the elementary $K^- p \to \gamma Y$
reaction 
displayed in Fig. \ref{fig:kpgamma}. 
Note that the above expression contains only 
the contact (Fig. \ref{fig:kpgamma}d) and the meson exchange
(Fig. \ref{fig:kpgamma}c) terms.
In the heavy-baryon approximation
one can show that
the baryon pole term (Fig. \ref{fig:kpgamma}a) contributes only
to the
meson-baryon P-wave states, while
some S-wave contributions from the baryon-exchange term (Fig.
\ref{fig:kpgamma}b) also
vanish at threshold ($K^-$ capture at rest). The charge and mass
of the
meson $M$ are denoted, respectively, by $e_M$ and $\mu_M$.
The SU(3) coupling constants $C_{Y,MB}$ for
the $MB \leftrightarrow Y$ transition are given by
\begin{eqnarray}
C_{Y,MB}= X_{Y,MB}(D+F) + Z_{Y, MB} (D-F) \ ,
\end{eqnarray}
where the values of the $X$
and $Z$ coefficients are easily evaluated from the chiral
Lagrangians and can be found in ref.
\cite{LOOR98}.
The initial state strong interactions are present in the second
term
of Eq. (\ref{eq:gammat})
\begin{eqnarray}
[QGT]_{\gamma Y,K^-p}(q,k') &=& \sum_{MB}
\int
\frac{d\vec{k}}{(2\pi)^3}\frac{M_B}{E_B(\vec{k})}\frac{1}{2\omega
_M(\vec{k})}
\frac{Q_{\gamma
Y,MB}(q,k)}{\sqrt{s}-E_B(\vec{k})-\omega_M(\vec{k})
+i\epsilon} \nonumber \\
&\times & T_{MB, K^- p}(\vec{k}_{MB}, \vec{k}',\sqrt{s}) \ ,
\end{eqnarray}
where $T_{MB,K^- p}$ is the amplitude for the strong $K^- p \to
MB$ transition, as well as in the third term
\begin{eqnarray}
\Delta_{\gamma Y,K^- p}(q,k') &= &\sum_{MB}
\int \frac{d\vec{k}}{(2\pi)^3} \frac{M_B}{E_B(\vec{k})}
\frac{1}{2\omega_M(\vec{k}-\vec{q}\,)} 
\frac{1}{\sqrt{s}-E_B(\vec{k}\,)-q^0
-\omega_M(\vec{k}-\vec{q}\,)+i\epsilon}
\nonumber \\
& \times &
ie_M \frac{C_{Y,MB}}{2f} 
\frac{2 [\vec{k}\vec{\epsilon}\,]
[\vec{\sigma}(\vec{k}-\vec{q})\,]} 
{(q^0+\omega_M(\vec{q}-\vec{k}))^2-\omega_M^2(\vec{k})} 
T_{MB,K^-p}(\vec{k}_{MB},\vec{k}',\sqrt{s}) \ , 
\end{eqnarray}  
which is related to the exact treatment of the meson
propagator. It corresponds to the contribution of the pole of the
meson propagating between the emitted photon and the final
hyperon $Y$
in  Fig. \ref{fig:kpgamma}c and in the loop diagrams generated by
the
initial state interactions.
Note that, in the above expressions, the allowed
intermediate states are the charged particle
channels ($MB = K^-p, \pi^+ \Sigma^-, \pi^- \Sigma^+$ and $ K^+
\Xi$).
Moreover, the strong amplitude, $T_{MB, K^- p}$,
appears with the on shell momentum $k_{MB}$ and factors out of
the integral
because, as shown in
the appendix of ref. \cite{LOOR98}, the
off shell piece can be absorbed in 
the renormalization of the charge.

The results for the branching ratios, defined by
\begin{eqnarray}
B_{K^- p \rightarrow \gamma Y} = \frac{\sigma_{K^- p \rightarrow
\gamma Y}
(\sqrt{s}_{\rm th})}
{\sigma_{K^- p \rightarrow {\rm all}}(\sqrt{s}_{\rm th})} \, ,
\label{eq:ratios}
\end{eqnarray}
where $Y= \Lambda, \Sigma^0$ and $\sqrt{s}_{\rm th}\rightarrow
\mu_{K^-} +
M_{p}$, are compared to the experimental data \cite{gammadat} 
in Table \ref{tab:ratios}.

\begin{table}[ht!]
\centering
\caption{
$K^- p\rightarrow \gamma\Lambda,
\gamma\Sigma^0$ branching ratios defined in
eq. (\ref{eq:ratios}) (in unit of $10^{-3}$)} 

\label{tab:ratios}
\begin{tabular}{|l|c|c|c|}
\hline
\multicolumn{1}{|c|}{Amplitude}&$B_{K^-p\rightarrow
\gamma\Lambda}$&
$B_{K^-p\rightarrow \gamma\Sigma^0}$&$R=B_{K^-p\rightarrow\gamma
\Lambda}/B_{K^-p\rightarrow \gamma \Sigma^0}$\\
\hline
   $Q $ & 1.12 & 0.073 & 16.4 \\
   $Q + QGT  $ &1.31 & 0.95&1.38 \\
   $Q + QGT + \Delta $ &1.58 &1.33 &1.19    \\
   $\, [Q + QGT+\Delta \,]_{{\rm no}\ \eta}  $ &2.47 &1.27&1.94
\\
   $\, [ Q + QGT+\Delta \, ]_{{\rm with}\ \Lambda_\pi}$ &1.10 &
1.05 & 1.04 \\
EXP \protect\cite{gammadat} &$0.86\pm0.16$ & $1.44\pm0.31$&$0.4 -
0.9$
\\
\hline
\end{tabular}
\end{table}

Neglecting initial meson-baryon
interactions (first row in Table
\ref{tab:ratios}) gives a very weak
branching ratio for $\gamma \Sigma^0$ production and
the predicted ratio between the two production rates is
an order of magnitude larger than the data. This is in
agreement with the findings of Siegel and Saghai\cite{saghai}.
When
the strong coupled-channel effects are included
(third row in Table \ref{tab:ratios}) the ratio is close to the
experimental value.
The predicted branching ratio for the $\gamma \Lambda$ production
is about 50\% larger than the experimental value, but it is
within
the experimental uncertainty for the $\gamma \Sigma^0$
production.
As shown in ref.  \cite{LOOR98}, the enhancement of the $\gamma
\Sigma^0$
production is essentially coming from the coupling of the photon
to intermediate
$\pi^+ \Sigma^-$ and $\pi^- \Sigma^+$ states.
The exact treatment of the meson propagator in Fig.
\ref{fig:kpgamma}c
leads to a contribution from a second meson pole, $\Delta_{\gamma
Y,
K^-p}$, which
can change the $\gamma \Sigma^0$
branching ratio by about 40\% and brings 
the predicted ratio closer to the experimental
value. The influence of the $\eta$ channels on the strong
$T_{MB,K^-p}$ amplitudes has, as for the case of the low energy
$K^- p$
scattering data \cite{AO}, a significant effect here. Comparing
the
third and fourth rows
in Table \ref{tab:ratios} one sees that the predicted
branching ratio for $\gamma \Lambda$ production is increased by
about
60\% if the $\eta$ channels are omitted in the calculation of the
strong amplitudes. It is thus clear that
including the
$\eta$ channels is also crucial in using this reaction to test
the chiral SU(3) symmetry. We note that the $\eta$ channels were
omitted in the model of ref. \cite{saghai} and, at the same time,
the
couplings had to be substantially changed with
respect to their SU(3) values in order to obtain a good fit to
the data. In retrospective one can say that the deviation of the
coupling constants from their SU(3) value is in fact trying to
restore
the breaking that was induced by the omission of the
$\eta$ channels.

Finally, we note that the strong meson-baryon-baryon vertex in
each of the
photoproduction amplitudes should in principle have a form factor
because hadrons are composite particles. 
The results obtained with a monopole form factor with a
cut off of $\Lambda_\pi=1$ GeV, a value which is commonly
accepted, are shown in the fifth row of Table \ref{tab:ratios}
and agree
roughly with
the data within experimental errors,
which are of the order of 20\%. If one compares with the central
values of
the
experimental branching ratios, the results are on the upper edge
of the $B_{K^- p \to \gamma \Lambda}$ ratio while those for
$B_{K^- p \to \gamma \Sigma^0}$ are on the lower edge. Looked at
it
in the context that the coupled channels and unitarization have
reduced the ratio $R$ by a factor 14, differences of the order of
10--20\% are not so significative. 
Note that all coupling constants are consistent with the chiral
SU(3)
symmetry and the model depends on only the cut off parameter,
which was
fixed in the
study of $S=-1$ meson-baryon reactions \cite{AO}.
In this approach, neither the meson-baryon nor the
photoproduction
mechanisms involve the explicit consideration of excited hyperon
states
since the $\Lambda(1405)$ resonance, which plays a key role in
these
reactions,  is generated dynamically, hence 
strengthening the interpretation of the
$\Lambda(1405)$ as a quasi-bound meson-baryon system with $S=-1$,
as
already
supported by the study of the strong interactions in
\cite{AO,NK}.

\section{Photoproduction of the {\boldmath $\Lambda(1405)$} on protons and
nuclei}
  
  As we saw in {\it section 3.3.2}, the $\Lambda(1405)$ resonance
is
produced 
  dynamically by using the Bethe-Salpeter equation and the lowest
order
  Lagrangian for meson-baryon interaction in S-wave suggesting
that
  this resonance is like a quasibound meson-baryon state rather
than a
genuine
  $3q$ state. Further tests on the nature of the resonance can be
done by
  studying different production processes. One of them was
studied in
  ref. \cite{nacherphot} by means of photoproduction on the
proton, i.e. 
\begin{equation}
\gamma p\rightarrow  K^+ \Lambda(1405) \ .
\end{equation}     

  The study of the reaction in nuclei is also of much interest
since
different
  studies predict sizeable changes of the resonance in nuclei
  \cite{Koch94,WKW96,Waas97,Lutz98,RO99} which would have also
repercussions
on the
  properties of $\bar{K}$ inside a nuclear medium. This is a hot
topic since it
  relates to the possibility of having kaon condensation in stars
and to the 
  puzzle of the strong $\bar{K}$ attraction needed to explain the
$K^-$ atoms 
  which seems to violate the low density theorem. 
  
\begin{figure}[ht!]
\centerline{
\includegraphics[width=0.3\textwidth,angle=-90]{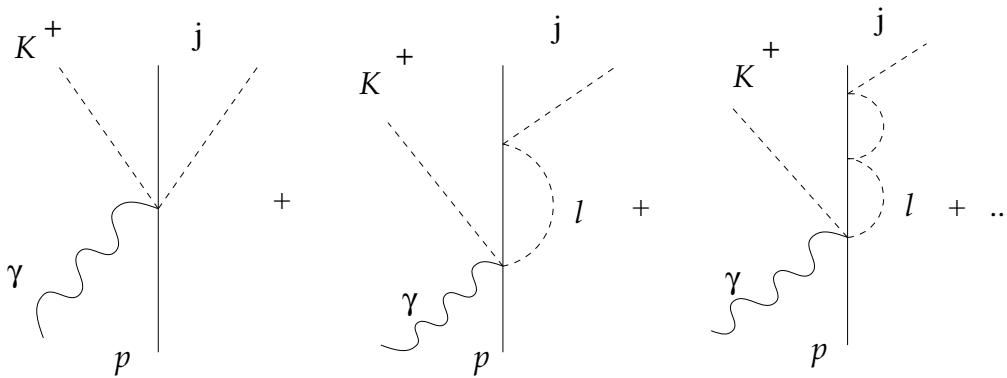}
}
\caption{Diagrammatic representation of the meson-baryon final
state interaction in the
 $\gamma p\rightarrow
K^+\Lambda$(1405) process. }
\label{fig:fdiagram}
\end{figure}

    In ref. \cite{nacherphot} a study was done along lines
similar to
those exposed
in {\it section 4.3.1}, however, in this case a meson and a
baryon combine
through
final state interaction to give the $\Lambda(1405)$.
Diagrammatically the
mechanism for the production is depicted in Fig.
\ref{fig:fdiagram}.

 As we can see there, a $K^+$ is produced together with another
meson and a
 baryon which combine through final state interaction with the
coupled channels
 in the $S=-1$ sector and give rise to the $\Lambda(1405)$
resonance.
  Once again one needs the vertex with two mesons a photon and a
baryon line
  which is obtained via minimal coupling from eq. (\ref{eq:ampl})
and
is given by 
  
   \begin{equation}
V_{ij}^{(\gamma)} = C_{ij}\frac{e}{4f^2}(Q_i + Q_j)\, 
\overline{u}(p')\gamma^\mu u(p)\epsilon_\mu ,
\end{equation} 
where $Q_i$, $Q_j$ are the initial and final meson
charges and $\epsilon_\mu$ the photon polarization vector and
$C_{ij}$ the
coefficients of eq. (\ref{eq:ampl}).

 Once again Bremsstrahlung terms on the meson lines of the tree
diagram are
 negligible if the reaction is done close to threshold. In
ref. \cite{nacherphot} the
 photon energy was chosen 1.7 GeV in the lab frame.
 
\begin{figure}[ht!]
\centerline{
\includegraphics[width=0.55\textwidth,angle=-90]{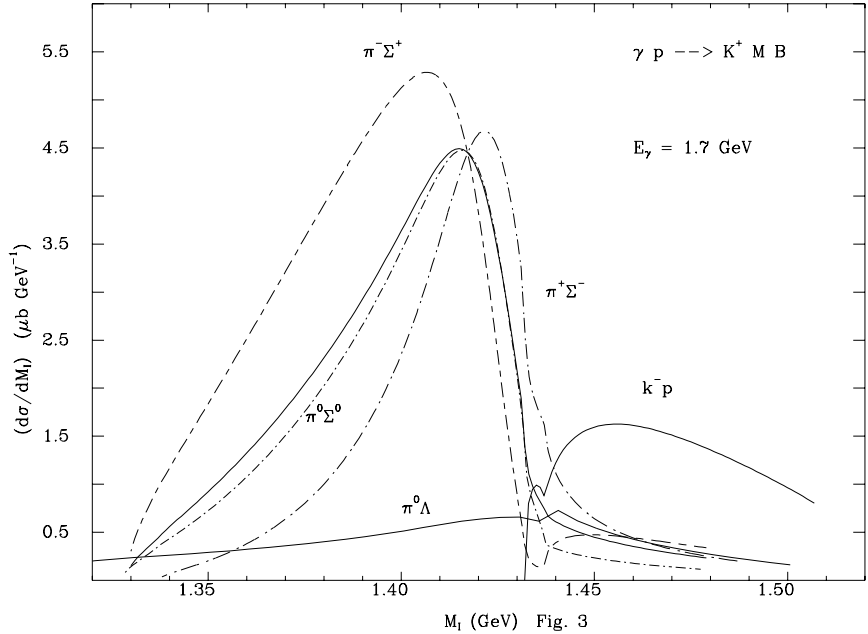}
}
\caption{ Mass distribution in the $\gamma p \to K^+ MB$ reaction
for the
different channels. The dashed lines show the $\Sigma^+\pi^-$,
$\Sigma^-\pi^+$ and $\Sigma^0\pi^0$ distributions. The solid line
with the resonance shape is the sum of the
three $\Sigma\pi$ channels divided by three. The distributions
for
$\pi^0\Lambda$ and $K^-p$ production are also given, while that
for
$\bar{K^0}n$
production is small and not shown in the figure.
}
\label{fig:photlam}
\end{figure}

    In Fig. \ref{fig:photlam} we show ${d\sigma}/{dM_I}$
for the
different channels. 
While all coupled channels collaborate to the building up of the 
$\Lambda$(1405) resonance, most of them open up at higher
energies and the resonance
shape is only visible in the
 $\pi^+\Sigma^-$,
     $\pi^-\Sigma^+$,
   $\pi^0\Sigma^0$ channels. The  $\bar{K}N$ production occurs at
   energies slightly above the resonace and the $\pi^0 \Lambda$,
with isospin one, only
   provides a small background below the resonance.
     

     It is interesting to see the different shapes of the three
$\pi\Sigma$ 
     channels.
     This can be understood in terms of the isospin decomposition
of the states
\begin{equation}
    |\pi^+\Sigma^-\rangle = - \frac{1}{\sqrt{6}}|2, 0\rangle -
\frac{1}{\sqrt{2}}|1, 0\rangle -
     \frac{1}{\sqrt{3}}|0, 0\rangle
\end{equation}
\begin{equation}
    |\pi^-\Sigma^+\rangle = - \frac{1}{\sqrt{6}}|2, 0\rangle +
\frac{1}{\sqrt{2}}|1, 0\rangle -
     \frac{1}{\sqrt{3}}|0, 0\rangle
\end{equation}
\begin{equation}
|\pi^0\Sigma^0\rangle =  \sqrt{\frac{2}{3}}|2, 0\rangle -
\frac{1}{\sqrt{3}}|0, 0\rangle 
\end{equation}

   Disregarding the $I = 2$ contribution which is negligible, the
cross sections
   for the three channels are proportional to the modulus squared
of the
   amplitude and hence they go as:
\begin{equation}
\frac{1}{2}|T^{(1)}|^2 + \frac{1}{3}|T^{(0)}|^2 
+ \frac{2}{\sqrt{6}}{\rm Re\,}(T^{(0)} T^{(1)*})\,
;\hspace{0.5cm} \pi^+\Sigma^-
\label{eq:pipsm}
\end{equation}
\begin{equation}
\frac{1}{2}|T^{(1)}|^2 + \frac{1}{3}|T^{(0)}|^2 
- \frac{2}{\sqrt{6}}{\rm Re\,}(T^{(0)} T^{(1)*})\,
;\hspace{0.5cm} \pi^-\Sigma^+
\label{eq:pimsp}
\end{equation}
\begin{equation}
\frac{1}{3}|T^{(0)}|^2\, ;\hspace{0.5cm} \pi^0\Sigma^0
\label{eq:pi0s0}
\end{equation}

    The crossed term $T^{(0)}T^{(1)*}$ is what makes these cross
sections different. We can
    also see that
\begin{equation}
 3\frac{d\sigma}{dM_I}(\pi^0\Sigma^0)\simeq
\frac{d\sigma}{dM_I}(I=0)
\end{equation}
\begin{equation}
\frac{d\sigma}{dM_I}(\pi^0\Sigma^0) +
\frac{d\sigma}{dM_I}(\pi^+\Sigma^-) +
\frac{d\sigma}{dM_I}(\pi^-\Sigma^+)\simeq
\frac{d\sigma}{dM_I}(I=0) 
+  \frac{d\sigma}{dM_I}(I=1)
\end{equation}

   This means that the real shape of the resonance must be seen
in either the
   $\pi^0 \Sigma^0$ channel or in the sum of the three
$\pi\Sigma$ channels,  provided the I
   = 1
   cross section (not the crossed terms which are relatively
large) is small
   as it is the case. Incidentally, eqs.
(\ref{eq:pipsm}),(\ref{eq:pimsp})
also show that the difference between the
   $\pi^+ \Sigma^-$ and $\pi^- \Sigma^+$ cross sections gives the
crossed term and hence
   provides some information on the $I = 1$ amplitude.

\begin{figure}[ht!]
\centerline{
\includegraphics[width=0.55\textwidth,angle=-90]{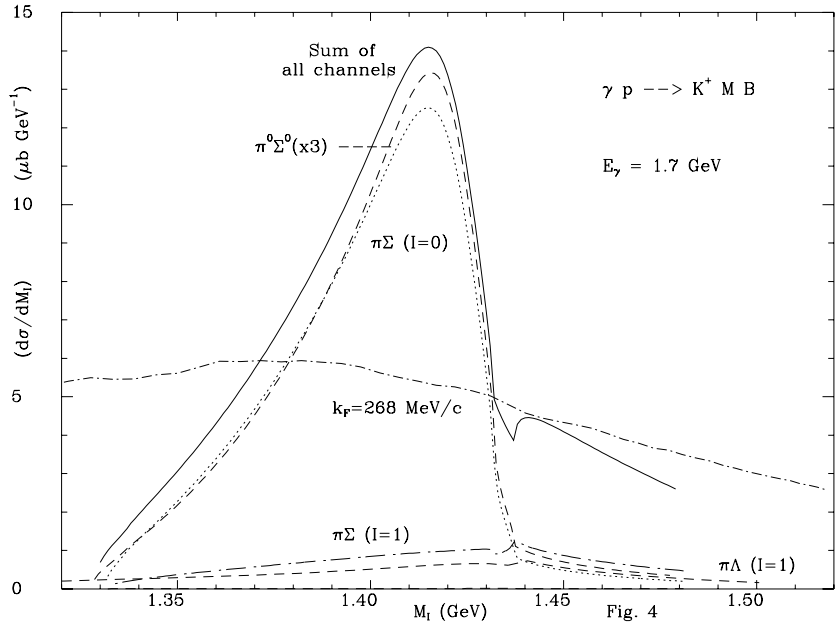}
}

\caption{ Mass distributions for the $\gamma p \to K^+ MB$
reaction.
Dashed line (resonant
shape): $\Sigma^0\pi^0$
 distribution multiplied by three. Dotted line (resonant shape):
pure $I = 0$ contribution from
 the $\Sigma\pi$ channels. Solid line (resonant shape): Sum of
the cross sections for all the
 channels. The $I = 1$ background contribution from the
$\Sigma\pi$
 and $\pi^0\Lambda$ channels is also shown. Short-dash-dotted
line: Effects of
 the Fermi motion with $k_F = 268$ MeV/c $(\rho = \rho_0)$ where
the free
space
 $\Lambda(1405)$ distribution is assumed in the calculation.}
\label{fig:combi}
 \end{figure}

     In Fig. \ref{fig:combi} the results are recombined in a
practical
way from the 
     experimental point of
     view. They 
      show the $I = 0$ contribution, the $\Sigma^0 \pi^0$
contribution and the sum of all channels including
     the $\pi^0 \Lambda$, and we see that they are all very
similar and the total contribution is just the
     $\Lambda$(1405) contribution plus a small background. In
practical 
     terms this result means
     that the detection of
     the $K^+$ alone (which sums the contribution of all
channels) is sufficient to determine the shape and the
     strength of the $\Lambda$(1405) resonance in this reaction.

     The study of the reaction in nuclei requires special care.
Indeed, assume one uses the same
     set up as before with a nuclear target and measures the
outgoing $K^+$. There the invariant
     mass will be given by
\begin{eqnarray}    
\label{eq:invmass}
 M_I\, ^2(p) = (q + p - k)^2 & = & M^2 + m_K^2 - 2 q^0 k^0 +
  2\vec{q}\cdot\vec{k} + \\ \nonumber
& & 2 p^0 (q^0 - k^0) -
2 \vec{p}\cdot(\vec{q} - \vec{k})
\end{eqnarray}  
with $q, k, p$ the momenta of the photon, $K^+$ and initial
proton respectively.
     Since $\vec{q}-\vec{k}$ has a large size, there will be a
large spreading of invariant
     masses due to Fermi motion for a given set up of photon and
$K^+$ momenta, unlike
     in the free proton case where $M_I\, ^2$ is well
determined\footnote{
     We are endebted to T. Nakano and J. K. Ahn for calling us
the attention on this point}. The nuclear cross section
     normalized to the number of protons (the neutrons through
$K^-n$ and coupled channels
     only contribute to $I = 1$ with a small background) would be
given by the convolution 
     formula

\begin{equation}
\frac{1}{Z}\frac{d\sigma}{dM_I}\bigm|_A\simeq\frac{2}{\rho_p}\int
\frac{d^3p}{(2\pi)^3} \,
\frac{d\sigma}{dM_I(\vec{p}\,)}\hspace{.6cm} ; \hspace{1.cm}
\rho_p =\frac{k_F\, ^3}{3\pi^2}
\end{equation} 
where the integral over $\vec{p}$ ranges up to the Fermi momentum
$k_F$.
     
     In order to show the effects of the Fermi motion, values of
$\vec{k}$ 
     corresponding to forward $K^+$ in the CM (and hence largest
value of $\vec{k}$ 
     in the lab frame) are chosen in ref. \cite{nacherphot}.
These
components 
     would minimize the spreading of the $M_I\, ^2(\vec{p}\,)$ in
eq. 
(\ref{eq:invmass}).
     Even then, 
the spreading of the invariant masses is so large that one looses
     any trace of the original resonance, as
one can see in Fig. \ref{fig:combi}. The $M_I$ in the
     x axis of the figure in this case is taken for reference
from eq.
(\ref{eq:invmass}) for a nucleon at
     rest. This result simply means that in order to see genuine
     dynamical effects one would have to look at the invariant
mass of the resonance
     from its decay product, $\pi\Sigma$, tracing back this
original invariant mass with appropiate
     final state interaction corrections.

          One interesting thing here is that the $\Lambda(1405)$
resonance is produced
          with a large momentum in the nuclear lab frame. Because
of that, Pauli
          blocking effects in the resonance decay, which are so
important for the resonance
          at rest in the nucleus, become now irrelevant. Hence
medium modifications
          of the resonance in the present situation should be
attributed to other
          dynamical effects \cite{Lutz98,RO99}.

     Experiments on this reaction are now done at TJNAF and are
being
     analysed. They are also scheduled to run with priority at
LEPS of 
     SPring8/RCNP. 
      These experiments  will allow to test
     current ideas on chiral symmetry for the elementary
reaction. When 
     used with nuclear targets they should provide us with 
      much needed information on the in medium properties of the
     $\Lambda$(1405) resonance and the $K^-$ meson. This should
help
     resolve questions like $K^-$ condensation and the origin of
the attraction seen in 
        $K^-$ atoms.

\section{Radiative production of the {\boldmath $\Lambda(1405)$} resonance in
{\boldmath $K^-$} collisions on
protons and nuclei}

   One of the problems which we encountered in the former section
regarding the
   study of the properties of the $\Lambda(1405)$ resonance in
nuclei is that
   the detection of the $K^+$ alone did not allow one to observe
the shape of
   the resonance because the effect of Fermi motion of the
nucleons in the
   nucleus produced a large spread of the invariant mass of the
meson-baryon
   system building up the resonance. One had to reconstruct the
invariant mass
   from the $\pi\Sigma$ decay products. In this section we report
on an
   alternative reaction to produce the $\Lambda(1405)$ resonance
which is
   dynamically quite different from the photoproduction process,
hence offering
   extra tests of the chiral symmetry ideas in the baryon sector.
Furthermore, it
   has an attractive feature since in this case an easy
experimental set up is
   still sufficient to investigate the resonance properties in
nuclei.
    
     The reaction reported here is the
$K^-p\rightarrow\Lambda(1405)\gamma$ 
     at low $K^-$ energies, which was studied in ref.
\cite{nacherrad}.
     
\begin{figure}[ht!]
\centerline{
\includegraphics[width=0.75\textwidth,angle=0]{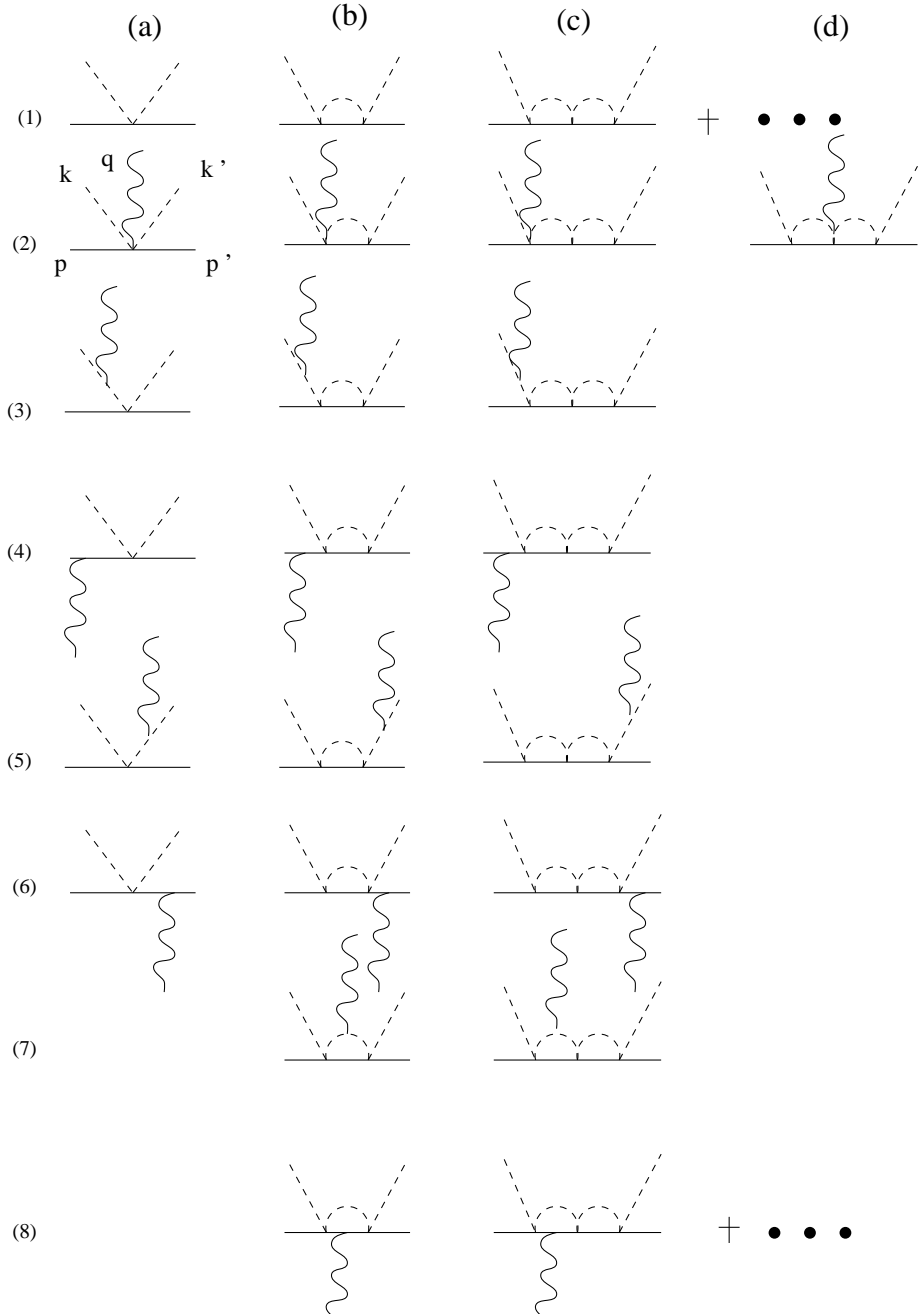}
}
\caption{Feynman diagrams used in the model for the $K^-
p\rightarrow \Lambda(1405)\gamma$ reaction. }
\label{fig:otherdiag}
\end{figure}

  Although the present reaction corresponds to a crossed channel
of the 
   $\gamma p\rightarrow K^+\Lambda(1405)$ reaction studied in
ref. \cite{nacherphot}
   and reported above, the two
   processes
   are rather different dynamically in their respective physical
channels,
with the dominant mechanisms in the photoproduction reaction
being
negligible in the present
   one, and others which could be proved negligible in the
photoproduction 
one
   becoming now dominant. The set of diagrams considered
   in ref. \cite{nacherrad} is depicted in Fig.
\ref{fig:otherdiag}. The
first line simply
shows the
   diagrams of the Bethe-Salpeter equation which are utilized to
generate the
   meson-baryon scattering $T$-matrix with coupled channels. The
channels
   considered here are the same 10 channels considered in {\it
section 3.3.2}.
 The rest of the diagrams in Fig. \ref{fig:otherdiag} stand for
the
radiative production of the 
$\Lambda(1405)$. 

Apart from the strong $MB\rightarrow M'B'$ vertices of eq.
(\ref{eq:ampl}) 
we also need
the coupling of the photon to the baryons, the mesons, plus the
contact
   term of diagram (2.a) of Fig. \ref{fig:otherdiag} required by
gauge
invariance. These vertices
   are standard and after the nonrelativistic reduction of the
$\gamma$
   matrices, are given in the Coulomb gauge, $\epsilon^0 = 0$, 
   $\vec{\epsilon}\cdot\vec{q}=0$, with $\vec{q}$ the photon
momentum, by

\begin{equation}
a) -it_{M'M\gamma} = 2ieQ_M\vec{k}'\cdot\vec{\epsilon}
\end{equation}
for the coupling of the photon to the mesons, with $e$ electron
charge, 
$Q_M$ the charge of the meson, $k'$ the momentum of the outgoing
meson
and $\epsilon_{\mu}$ the photon polarization vector, 

\begin{equation}
b) -it_{B'B\gamma} = ie(Q_B\frac{\vec{p}+\vec{p}\, '}{2M_B} -i
\frac{
\vec{\sigma}\times\vec{q}}{2M_B}\mu_B)\vec{\epsilon} 
\end{equation}
for the coupling of the photon to the baryons, with $Q_B$ the
charge 
of the baryon,
$\vec{p}$, $\vec{p}\, '$ the incoming, outgoing baryon momenta
and $M_B$,
$\mu_B$ the mass and magnetic moment of the baryon, and

\begin{equation}
c) -it_{B'M'B M\gamma} = iC_{ij}(Q_i +
Q_j)\{\frac{\vec{p}+\vec{p}\, '}{2\bar{M}} 
-i \frac{
\vec{\sigma}\times(\vec{p}-\vec{p}\,
')}{2\bar{M}}\}\vec{\epsilon} 
\label{eq:contact}
\end{equation}
for the contact term of diagram (2.a) of Fig.
\ref{fig:otherdiag}, with
$C_{ij}$ the coefficients
of eq. (\ref{eq:ampl}), $i,j$ standing for a $MB$ state, $Q_i$,
$Q_j$ the
charges of the
mesons, $\bar{M}$ an average mass of the baryons and $\vec{p}$,
$\vec{p}\, '$ the
momenta of the incoming, outgoing baryons.

In ref. \cite{nacherrad} one is concerned with $K^-$ with momenta
below
500 MeV/c in the
lab frame. In this energy domain it is easy to see that the
Bremsstrahlung
 diagrams from mesons and baryons (diagrams (3.a), (4.a), (5.a),
(6.a))
are of the same
 order of magnitude and that the contact term (diagram (2.a)) is
of order
 $q/2M$ of the corresponding meson Bremsstrahlung diagrams ((3.a)
and
(5.a)). With CM
 photon momenta $q$ of the order of 150 MeV or below, the terms
of row 2 
represent
 corrections below the 8$\%$ level and are neglected. In
addition, terms
 like in diagram (2.d), where the photon couples to internal
vertices of
the
 loops, vanish for parity reasons.
 
 The diagrams in row 7 of Fig. \ref{fig:otherdiag} where the
photon
couples to mesons in the
 loops vanish due to the gauge condition
$\vec{\epsilon}\cdot\vec{q} = 0$
 and the same happens to the diagrams in row 8, where the photon
couples
with 
 the dielectric part to
 the baryons inside the loops ($(\vec{p}+\vec{p}\, ')$ term of
eq. (\ref{eq:contact})). The magnetic coupling of the photons in
row 8 survives.

 Hence, the process is given, within the approximations
mentioned, by the
 diagrams in rows 3, 4, 5, 6 plus the magnetic part in row 8.
This
situation is opposite to the
 one found in ref. \cite{nacherphot} for $\Lambda(1405)$
photoproduction
close to threshold, where
 the dominant terms came from the contact term and the
Bremsstrahlung diagrams were
 negligible.
 
 If one inspects the series of terms in rows 3 and 4 of
Fig.~\ref{fig:otherdiag} one can see that
 the strong part of the interaction to the right of the
 electromagnetic vertex involves the series of terms of the 
Bethe-Salpeter equation  and generates the 
  $T$-matrix from the initial $MB$ state to the
 final $M'B'$ state after losing the energy of the photon, this
is, with an
 argument $M_I$, where $M_I$ is the invariant mass of the $M'B'$
state.
 Similarly, in the rows (5) (6) the strong $T$-matrix factorizes
before the
 electromagnetic vertex with an argument $\sqrt{s}$, with $s$ the
Mandelstam
 variable for the initial $K^- p$ system. In the diagrams of row
(8) we have
 a loop with one meson and two baryons. The strong interaction to
the left
 originates $T(\sqrt{s})$ and the one to the right $T(M_I)$. The
loop function
 of row (8) contains two baryon propagators and for the small
energies involved
 here can be obtained by 
 differentiating the $G(\sqrt{s})$ function of a meson-baryon
loop, eq.
(\ref{eq:loop}), with respect to $\sqrt{s}$, which duplicates the
baryon
propagator.

By
 choosing an appropiate pair ($\epsilon_1, \epsilon_2$) of
orthogonal photon
 polarization vectors, also orthogonal to $\vec{q}$, and summing
over
 final photon and baryon polarizations plus averaging over the
initial proton
 polarizations, one 
  obtains
  the cross section for the process given by
 ($\sigma$ the
 cross section for each $i, j$ transition)
 
\begin{equation}
\frac{d\sigma}{dM_Id\varphi} = \frac{1}{2\pi}\frac{d\sigma}{dM_I}
+
\frac{d\sigma_I}{dM_Id\varphi}\cos\varphi\, ,
\label{eq:diffc}
\end{equation}
with $\varphi$ the azimutal angle formed by the plane containing
the $\vec{k}'$
 and $\vec{q}$ vectors and the one containing the $\vec{k}$ and
$\vec{q}$
 vectors. The only dependence on the azimutal angle $\varphi$
comes in the
 $\cos\varphi$ dependence which accompanies the interference
cross
section,
 $\sigma_I$, in eq. (\ref{eq:diffc}), which means that both
$d\sigma/dM_I$
and $d\sigma_I/
 dM_Id\varphi$ do not depend on the angle $\varphi$.
 Explicit expressions for $d\sigma/dM_I$ and
$d\sigma_I/dM_Id\varphi$
 are given in ref. \cite{nacherrad}.
 
\begin{figure}[ht!]
\centerline{
\includegraphics[width=0.6\textwidth,angle=-90]{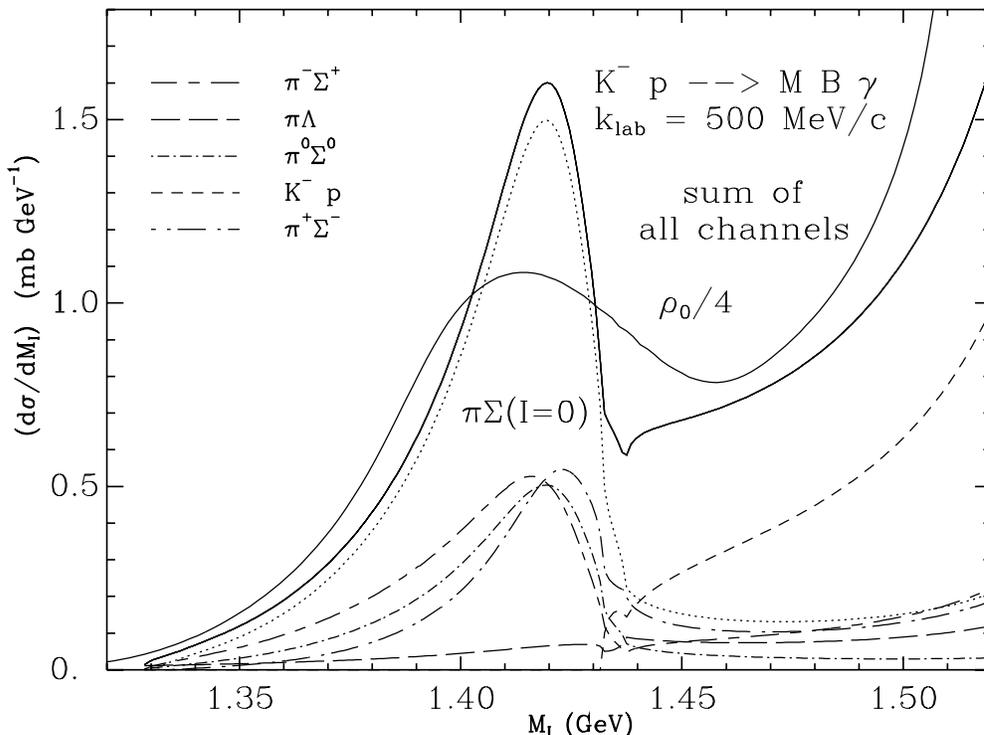}
}
\caption{Mass distribution for the different channels of the $K^-
p \to
MB \gamma$ reaction. The solid line
with the resonance  shape is the sum of cross sections for all
channels.
Dotted line: pure $I=0$ contribution from the $\Sigma\pi$
channels.
The effects of the Fermi motion with ($\rho=\rho_0/4$) is shown
with solid
line. The labels for the other lines are shown in the figure. }
\label{fig:frad}
\end{figure}

 The results for $d\sigma/dM_I$ are shown in Fig. \ref{fig:frad}.
There one can see the results for the
  cross sections in the $K^- p\rightarrow\pi^-\Sigma^+\gamma, 
 \pi^+\Sigma^-\gamma, \pi^0\Sigma^0\gamma, \pi^0\Lambda\gamma$,
$K^- p \gamma$ 
 channels. The cross section for $K^-p\rightarrow\bar{K}^0
n\gamma$ is very
  small, around 0.1 mb\, GeV$^{-1}$ in the range $1.44 - 1.52$
GeV, and is not
  plotted in the figure. The $\Lambda(1405)$ peak appears clearly
in the $\pi\Sigma$
 spectrum. It is interesting to notice the difference between the
cross
 sections
 for the different $\pi\Sigma$ channels. The origin of this is
the same
 one discussed in ref. \cite{nacherphot} and reported in the
former
section 
 due to the different isospin combinations of the
 three charged states and the crossed products of the $I=1$,
$I=0$ amplitudes
 which appear in the cross section. The $\pi^0\Sigma^0$ has no
$I=1$ component
 and since the $I=2$ component is negligible, the $\pi^0\Sigma^0$
distribution is very
 similar to the $I=0$ $\Lambda(1405)$ distribution.  The $I=0$
contribution alone,
 coming from the excitation of the $\Lambda(1405)$, can be
obtained using a
 combination of the three $\pi\Sigma$ amplitudes
\begin{equation}
(t_{K^- p\rightarrow\pi^-\Sigma^+} + t_{K^-
p\rightarrow\pi^+\Sigma^-} + 
 t_{K^- p\rightarrow\pi^0\Sigma^0})/\sqrt{3}\ .
\end{equation} 

 The results for the pure $I=0$ excitation (dotted line) shown
 in Fig.~\ref{fig:frad} look very similar to the total strength
around the
 $\Lambda(1405)$ peak.
 Below the $K^- p$ threshold there is some strength for $I=1$,
 $\pi^0\Lambda$ excitation, which is also very small. As
 a consequence of that, the sum of all channels in the
$\Lambda(1405)$
 region, which requires exclusively the detection of the photon,
has
 approximately the $\Lambda(1405)$ shape and strength.
 
 It is interesting to observe the fast rise of the cross section
in the
 $K^-p\rightarrow K^-p\gamma$ channel, showing the Bremsstrahlung
infrarred
 divergence at large $M_I$ (small photon momentum). The other
channels also 
 would show the infrarred divergence at higher
 energies,
  when the photon momentum goes to zero. The relative larger 
  weight of the $K^- p\rightarrow K^- p\gamma$ reaction at these
  energies, with
  respect to the other ones, is a reflection of the fact that the
  $K^-p\rightarrow K^-p$ cross section at values of $M_I$ or
$\sqrt{s}$ of the
  order of 1500 MeV is much bigger than the other
$K^-p\rightarrow M'B'$ cross
  sections. 
 
 There is a lower limit, which happens around 200 MeV/c for the
$K^-$ lab momentum, where the
 tails  of the distribution, reflecting the Bremsstrahlung
properties of the
 reaction, overlap with the peak of the resonance and hence the
information on
 the $\Lambda(1405)$ is lost.

   Let us now turn the attention to nuclei where we would
consider the reaction 
   $K^- A\rightarrow\Lambda(1405) \gamma (A-1)$. In this case if
one
  detects only the photon one has a distribution of invariant
masses due to
  Fermi motion since now $M_I^2 = (k + p_N - q)^2$ and $p_N$ runs
over all nucleon
  momenta of the
  occupied states.  One can fold the results of $d\sigma/dM_I$ 
with the
  distribution of $M_I$ coming from a Fermi sea of nucleons. The
  results in this case are shown in Fig. \ref{fig:frad} for $\rho
=
\rho_0/4$, a likely
  effective density for this reaction, taking into account the
distortion
  of the initial $K^-$ through the nucleus.
  We can see a widening of the $\Lambda(1405)$ distribution, with
the shape
  only moderately changed, such that other effects from genuine
changes
  of the $\Lambda(1405)$ properties in the medium, predicted to
be quite
  drastic \cite{Koch94,WKW96,Waas97,Lutz98,RO99}, could in
principle be
visible. Certainly, the detailed
  measurement of the final meson-baryon in coincidence with the
photon
would
  allow a 
  much better determination of the $\Lambda(1405)$ properties
than just the
  photon detection, and ultimately these exclusive measurements
should also be
  performed. But the fact that the simple detection of the photon
can provide
  interesting information is a welcome feature from the
experimental point
  of view.

 The  reactions discussed can be easily implemented at present
facilities like
 KEK or Brookhaven. In Brookhaven some data from recent $K^- p$
 experiment with detection of photons in the final state are in
the process of
  analysis \cite{peterson}. The present results should encourage
the detailed
 analysis of the particular channels discussed here.

%% file: nuclear.tex
\chapter{Further nuclear applications}

In {\it sections 5.2} and {\it 5.3} we already discussed how the $\Lambda(1405)$
resonance could be generated in a nuclear environment and the additional
information that this would bring on the nature of the resonance, plus
the repercussions of these findings on the interaction of $\bar{K}$ with
nuclei.  In this chapter we will show some examples where the
combination of the chiral unitary approach with the many body techniques
proves to be a rather powerful tool to clarify issues which have
remained so far controversial, particularly the problems of 
the $\pi \pi$ 
interaction in the nuclear medium and the interaction of $K^-$ with
nuclei.
   
\section{The isoscalar {\boldmath $\pi \pi$} interaction in a nuclear medium}

The $\pi \pi$ interaction in a nuclear medium in the $J=I=0$
channel ($\sigma$ channel) has stimulated much theoretical work
lately.
It was realized that the attractive P-wave interaction of the
pions
with the nucleus led to a shift of strength of the $\pi \pi$
system
to low energies and eventually produced a bound state of the two
pions around $2 m_\pi  -  10$  MeV \cite{Schuck}. This state
would
behave like a $\pi \pi$
Cooper pair in the medium, with repercussions in several
observable
magnitudes in nuclear reactions \cite{Schuck}. The possibility
that such
effects could have already been observed in some unexpected
enhancement
in the ($\pi, 2 \pi$) reaction in nuclei \cite{2} was also
noticed there.
More recent experiments where the enhancement is seen in the
$\pi^+ \pi^-$
channel but not in the $\pi^+ \pi^+$ channel \cite{3} have added
more
attraction to that conjecture.

Yet, it was early realized that constraints  of chiral symmetry
in the 
amplitude at low energies might
affect those conclusions \cite{Schuck}. In order to investigate
the
influence of chiral constrain\-ts in $\pi \pi$ scattering in the
nuclear
medium two different models \cite{JL86,6} 
for the $\pi \pi$ interaction were used
in ref. \cite{4}.
One of them \cite{JL86} did not satisfy the chiral constraints,
while
another one \cite{6} produced an amplitude behaving
like $m_\pi$ in the limit of small pion masses. The conclusion of
ref.
\cite{4} was that, although in the chirally constrained model
the building up of $\pi \pi$ strength at low energies was
attenuated, it
was still important within  the approximations done in their
calculations.
Among these approximations there is the use of only
$\Delta h$ excitation with zero $\Delta$ width to build up the
$\pi$
nuclear interaction. Warnings
were also given that results might depend on the off shell
extrapolation of the $\pi \pi$ scattering matrix.

Further refinements were done in ref. \cite{7}, where the width
of the
$\Delta$ and coupling to $1p \, 1h$ and $2p \, 2h$ components
were
considered. The coupling of pions to the $ph$ continuum led to a
dramatic
re-shaping of the $\pi  \pi$ strength distribution, but the
qualitative conclusions about the accumulated strength at 
low energies remained.

In ref. \cite{Aouissat} the importance of the coupling to the
$ph$ components
was reconfirmed and the use of more accurate models for the $\pi
\pi$ 
interaction, as the J\"{u}lich model based on meson exchange
\cite{9},
did not change the conclusions on the enhanced $\pi \pi$ strength
at low
energies. However, the use of a linear and nonlinear models for
the 
$\pi \pi$ interaction, satisfying the chiral constraints at small
energies,
led to quite different conclusions and showed practically no
enhancement of the $\pi \pi$ strength at low energies. The same
conclusions were reached using the J\"{u}lich model with a
subtracted dispersion relation so as to satisfy the chiral
constraints.
The latter model employed the Blakenbecler-Sugar equation in
which the
$2 \pi$ intermediate states were placed on shell. The conclusion
of this paper was that the imposition of chiral constraints in
the $\pi \pi$ amplitude prevented the pairing 
instabilities shown by the other models not satisfying those
constraints.

In a further paper \cite{Rapp} the authors showed, however, that
the
imposition of the chiral  constraints by themselves did not
prevent the pairing instabilities and uncertainties remained
related to
the off shell extrapolation of the $\pi \pi$ amplitude and the
possible ways to implement the minimal chiral constraints. The
situation,
as noted in ref. \cite{Rapp}, is rather ambiguous, but the
studies done
have certainly put the finger in the questions that should
be properly addressed: chiral symmetry, off shell extrapolations,
unitarity, etc.

 The chiral unitary methods discussed in {\it section 3} address
automatically
 all these questions and seem most appropriate to tackle the
problem discussed
 above. That task was undertaken in ref. \cite{chiang} and we
report here on their
 results. Since one is concerned about the S-wave $\pi \pi$
interaction, the
 Bethe-Salpeter approach is the most economical one to follow and
this is what is
 done in ref. \cite{chiang}. In the nuclear medium the pions will
get renormalized
 and hence their propagators will differ from the free ones. In
addition the
 point four meson vertices also get renormalized as we see below.

\begin{figure}[ht!]
\centerline{
\includegraphics[width=0.32\textwidth,angle=-90]{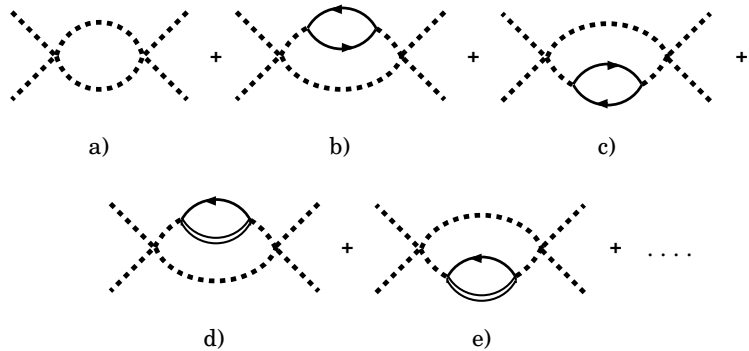}
}
\caption{ Terms appearing in the scattering matrix allowing the
pions to excite
$ph$ and $\Delta h$ components }
\label{fig:pipi1}
\end{figure}

\begin{figure}[ht!]
\centerline{
\includegraphics[width=0.42\textwidth,angle=-90]{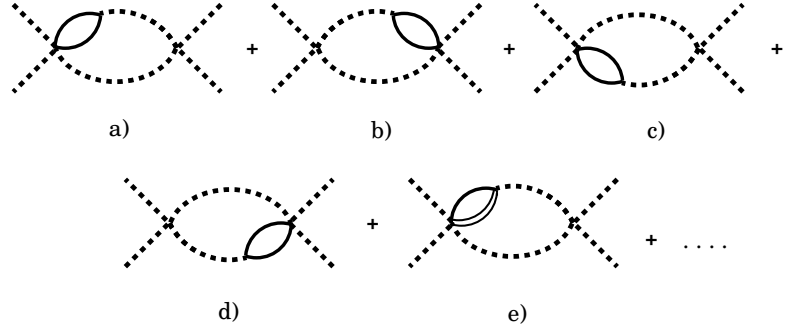}
}
\caption{ Terms of the $\pi \pi$ scattering series in the nuclear
medium related
to three meson baryon contact terms from the Lagrangian of eq.
(\ref{L2})}
\label{fig:pipi2}
\end{figure}

\begin{figure}[ht!]
\centerline{
\includegraphics[width=0.25\textwidth,angle=-90]{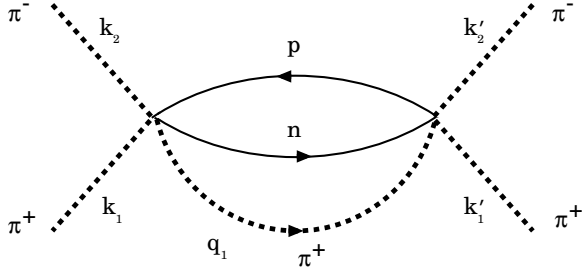}
}
\caption{ Diagram involving the three meson baryon contact terms
of fig. 
\ref{fig:pipi2} 
in each of the vertices }
\label{fig:pipi3}
\end{figure}

   In order to illustrate the medium modifications in the $\pi
\pi$ amplitude we
  show in Figs. \ref{fig:pipi1}, \ref{fig:pipi2} and
\ref{fig:pipi3}
 the diagrams which are now involved at the level of just
  one loop. The pion is dressed by allowing it to excite $ph$ and
$\Delta h$
  components, as is usually done in pion nuclear physics
\cite{wolferic}.
   
  The diagrams in Fig. \ref{fig:pipi1} involve the usual $ph$ and
$\Delta h$ excitation of
  the pion, which acquires a self-energy leading to a
modification of the pion
  propagator, and the two pion loop function. In ref.
\cite{chiang} only the pions
  were renormalized but the kaons were also taken into account in
the coupled
  channel Bethe-Salpeter equation. In Fig. \ref{fig:pipi2}
another sort of diagrams
  appears. 
These
  diagrams,
which qualify as vertex corrections of the four meson
vertex, 
come from the Lagrangian of Eq. (\ref{BaryonL}), and
involve a baryon line and
three mesons. The need to consider this contact three meson
term in connection
  with diagrams involving a pion pole, like one has in Fig.
\ref{fig:pipi1},  was already 
  known prior to the developments of $\chi PT$ \cite{15} and has
been
  systematically used in studies of the $\pi N \rightarrow \pi
\pi N$ reaction
  in refs. \cite{16,19,19a,20,21}.  The advent of $\chi PT$ has
made it easier to extend
  these ideas to the strangeness sector \cite{toniulf} where the
effective
   Lagrangians were not available. The presence of the three
meson contact term
   leads also to the term in Fig. \ref{fig:pipi3} in a natural
way.  The interesting
   feature of these diagrams is that when one separates the four
pion vertex 
appearing in Figs. \ref{fig:pipi1}, \ref{fig:pipi2}
   into an on shell and an off shell part, as done in {\it
section 3.3.1}, eq. (\ref{onshell}),
   there is an exact cancellation between the off shell parts and
the three
   meson contact terms of Figs. \ref{fig:pipi2}, \ref{fig:pipi3},
such that at the end only diagrams of the
   type of Fig. \ref{fig:pipi1} must be evaluated and with  all
the four pion vertices evaluated on
   shell (i.e. taking $p^2$=$m^2$ for all the meson lines in the
expression
   of the amplitude). This subtle cancellation, which also makes
the work simpler
   technically, was first observed in ref. \cite{chanfrayole}
when checking that the
   results cannot depend on the arbitrary coefficient that one
has in chiral
   theories at the level of three pion fields in the expansion of
the $U$ function
   defined after eq. (\ref{Fi}).
   
    The interplay found here for the pion pole term and the three
meson contact term
  has been investigated in other nuclear problems before, where
also interesting
  cancellations were found. In ref. \cite{toniulf} it was shown
that the real
  part of the kaon self-energy in the nuclear medium tied to the
scattering of
  the kaons with the virtual pion nuclear cloud was zero, because
of exact cancellations
  between the four meson terms and the contact three meson baryon
terms. This
  solved a puzzle at the time where large uncertainties were tied
to the off
  shell part of the $K \pi$ amplitude \cite{JK92,juancarmen}. The
  phenomenology also demanded that this real part be close to
zero
  \cite{juancarmen}. Similarly, in ref. \cite{17} the pion
self-energy tied to the
  interaction of the pions with the virtual pion cloud was found
to be very small, with
  partial cancellations which became complete in the chiral
limit. 
  
\begin{figure}[ht!]
\vspace*{1cm}
\centerline{
\includegraphics[width=0.55\textwidth,angle=0]{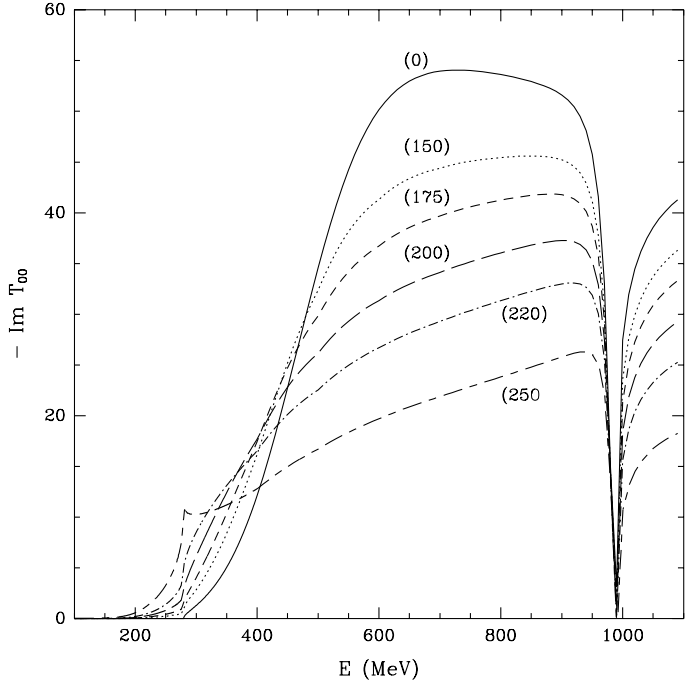}
}
\caption{Im $T_{22}$  for $\pi \pi \rightarrow \pi \pi $
scattering in $J=I=0$
 $(T_{00}$ in the figure) in the
nuclear medium for different values of $k_F$ versus the CM energy
of the 
pion pair. The labels correspond to the values of $k_F$ in MeV. }
\label{fig:pipi4}
\end{figure}

    The results obtained for the imaginary part of the $\pi \pi$
amplitude are
    shown in Fig. \ref{fig:pipi4} for different values of the
Fermi momentum. One can observe
    a depletion of the strength in the region of 600-900 MeV, but
more
    interesting, in connection with the experiments reporting an
enhancement of
    the invariant mass of the two pions close to threshold, is
the strength
    found in the figure in that region.

    The results shown in Fig. \ref{fig:pipi4} are very similar to
those of ref. \cite{Aouissat} where
    minimal chiral constraints were imposed in their models which
eliminated the
   peaks  below threshold found in earlier works.
   
   The situation is nevertheless still puzzling. Indeed, by
   using a model of ref. \cite{Rapp} for the $\pi \pi$
interaction in the medium and a
   model for the $(\pi,\pi \pi)$ reaction from ref. \cite{19}, a
spectrum of
   invariant masses similar to the experimental one could be
reproduced in 
   ref. \cite{rappnew}. A more
   detailed work was carried in ref. \cite{manolo} improving on
the approximations
   done in ref. \cite{rappnew} and it was found that the
enhancement of the invariant
   mass found was narrowly tied to the approximations done, and
when more
   accurate calculations were done there was no much sign of  an
enhanced
   invariant mass in the two pion mass distribution. One of the
reasons for
   the apparent lack of enhancement is that there are large
cancellations of
   pieces of the amplitude  for the case of
   $\pi^+ \pi^-$ production  which is not
   the case for $\pi^+ \pi^+$ production. This latter process
exhibits a peak 
   at low invariant masses mostly due to phase space reasons.
Should there be 
   a renormalization of some terms in the nucleus which
   would alter the cancellation found in free space, then the
results would look
   more like those of the $\pi^+ \pi^+$ production and the
experimental
   observation could be understood. Further work is necessary to
understand 
better the
   process before we can see in this reaction a precursor of the
chiral
   symmetry restoration, which is one of the appealing
possibilities suggested
   so far \cite{hatsuda}.
   
\section{The {\boldmath $K^-$} nucleus interaction}

The properties of the kaons and antikaons in the nuclear medium
have been the object of numerous investigations since the
possibility of the existence of a kaon
condensed phase in dense nuclear matter was pointed out
\cite{KN86}.
If the $K^-$ meson develops sufficient attraction in dense matter
it could
be energetically more favorable,
after a certain critical density, to
neutralize the positive charge with antikaons rather than with
electrons. A
condensed  kaon phase would then start to develop, changing
drastically the
properties of dense neutron star matter
\cite{BKRT92,TPL94,EKP95,FMMT96,TY98,LLB97,GS98}. In fact, the
enhancement
of the $K^-$ yield in Ni+Ni collisions measured recently by the
KaoS
collaboration at
GSI \cite{kaos97} can be explained by assuming the $K^-$ meson to
feel a strong
attraction in the medium \cite{cassing97,LKF94,LK96,Li97},
although
alternative
mechanisms, such as the production of antikaons via $\Sigma$
hyperons, have
also been suggested \cite{SKE99}. 
Kaonic atom data, a compilation of which is given in ref.
\cite{FGB94},
also favor an attractive $K^-$ nucleus interaction.

The theoretical investigations that go beyond pure phenomenology
\cite{sibir98}
have mainly followed two different
strategies. One
line of approach is that of the mean field models, built within
the
framework of chiral Lagrangians
\cite{Li97,CHL95,CHL96b,Mao99}, based on the relativistic Walecka
model
extended to incorporate strangeness in the form of hyperons or
kaons 
\cite{Sch97} or using explicitly
quark degrees of freedom \cite{Tsushi98}.
The other type of approach aims at obtaining
the in-medium
$\bar{K}N$ interaction microscopically by incorporating the
medium
modifications in 
a $\bar{K} N$ amplitude that reproduces the
low energy scattering data 
and generates the $\Lambda(1405)$
resonance dynamically
\cite{alberg76,star87,Koch94,WKW96,Waas97,Lutz98,RO99}.
For instance, Pauli blocking on the intermediate nucleon
states of the Lippmann-Schwinger or, alternatively, the
Bethe-Salpeter 
equation makes the
${\bar K}N$ interaction density dependent and this, in turn,
modifies the $K^-$
properties from those in free space. These medium modifications
were already included long time ago \cite{alberg76} 
in the context of
Brueckner-type many body theory 
using a separable $\bar{K}N$ interaction to obtain the
kaon-nucleus optical potential for kaonic atoms. The more recent
theoretical works \cite{Koch94,WKW96,Waas97,Lutz98,RO99} take the
$\bar{K}N$
interaction from the chiral Lagrangian.
The blocking of intermediate states shifts the resonance to
higher energy 
and this changes the
${\bar K}N$ interaction at threshold from being repulsive in free
space to
being attractive in the medium.
A recent self-consistent calculation
of the  $K^-$ self-energy 
\cite{Lutz98} has shown that the position of the resonance
remains unchanged, due to
a compensation of the repulsive Pauli blocking effects with the
attraction
felt by the $K^-$ meson, in qualitatively agreement with was was
noted
in ref. \cite{star87} using a constant mean field potential for
the $\bar{K}$.

Additional medium effects have been considered in a recent work
\cite{RO99}, including the self-energy of the
pions in the $\pi \Lambda$, $\pi \Sigma$ intermediate
states, which couple
strongly to the ${\bar K} N$ state, as well as the dressing of
the baryons
($N,\Lambda,\Sigma$) through  density-dependent mean-field
binding
potentials. 
The starting point is the chiral model of ref. \cite{AO},
described in
{\it section 3.3.2}, which reproduces the 
$\bar{K}N$ low energy scattering observables. 
The medium effects on the $\bar{K} N$ interaction are
incorporated
replacing the free meson and baryon propagators in the
meson-baryon loop
of Eq. (\ref{eq:loop}) by in-medium ones.
For the nucleon, a mean-field propagator
\begin{equation}
A(\sqrt{s}-q^0,-\vec{q},\rho)=
\frac{1-n(\vec{q}_{\rm
lab})}{\sqrt{s}-q^0-E_l(-\vec{q}\,)+i\epsilon} +
\frac{n(\vec{q}_{\rm
lab})}{\sqrt{s}-q^0-E_l(-\vec{q}\,)-i\epsilon}
\label{eq:nucleon}
\end{equation}
is taken, where $n(\vec{q}_{\rm lab})$ is the occupation
probability of a
nucleon of
momentum $\vec{q}_{\rm lab}$ in the lab frame. For the hyperons
($\Lambda$
and
$\Sigma$), the occupation probability is simply zero. The single
particle
energy,
$E_l(-\vec{q}\,)$, now contains a mean-field potential of
the type $U_0 \rho/\rho_0$, with $\rho_0=0.17$ fm$^{-3}$ being
the normal
nuclear matter density. For the nucleon, a reasonable depth value
is
$U_0^N=-70$ MeV, as suggested by numerous calculations of the
nucleon 
potential in nuclear matter. For the $\Lambda$ hyperon, it is
reasonable
to take $U_0^\Lambda=-30$ MeV, as implied by the extrapolation to
very
heavy systems of the experimental
$\Lambda$ single
particle energies in $\Lambda$ hypernuclei \cite{moto90}. For the
$\Sigma$ hyperon, there is no conclusive information on the
potential. Early phenomenological analyses \cite{Batty78}
and calculations
\cite{oset90}  
found the $\Sigma$ atom data to be compatible with $U_0^\Sigma
\sim -30$
MeV, but more recent analysis do not exclude a repulsive
potential in the 
nuclear interior \cite{Batty94}.
In the work of ref. \cite{RO99} the potential used is
$U^\Sigma=-30\rho/\rho_0$ MeV, as commonly accepted for low
densities, but
the effects of using a repulsive depth of 30 MeV are also
explored.
The meson propagator for the $\bar{K}$ and $\pi$ mesons is
replaced by the
dressed one 
\begin{equation}
D_l(q^0,\vec{q},\rho) = \frac{1}{(q^0)^2-{\vec q\,}^2 - m_l^2 -
\Pi_l(q^0,\vec{q},\rho)} =
\int_0^\infty d\omega \, 2\omega\,
\frac{S_l(\omega,\vec{q},\rho)}{(q^0)^2-\omega^2 + i\epsilon} \ ,
\label{eq:meson}
\end{equation}
where $\Pi_l(q^0,\vec{q},\rho)$ is the meson self-energy. 
The second equality in Eq. (\ref{eq:meson}) is the Lehmann
representation of
the meson propagator and
$S_l(\omega,\vec{q},\rho)=-{\rm Im}
D_l(\omega,\vec{q},\rho)/\pi$
is the meson spectral density which, in the case on undressed
mesons,
reduces to
$\delta(\omega-\omega_l(\vec{q}\,))/2\omega_l(\vec{q}\,)$.
With these modifications the loop
integral becomes
\begin{eqnarray}
G_l(P^0,\vec{P},\rho) &=&
\int_{\mid\vec{q}\,\mid < q_{\rm max} } \frac{d^3 q}{(2 \pi)^3}
\frac{M_l}{E_l (-\vec{q}\,)}
\int_0^\infty d\omega \,
 S_l(\omega,{\vec q},\rho) \nonumber \\
&\times & \left\{
\frac{1-n(\vec{q}_{\rm lab})}{\sqrt{s}- \omega
- E_l (-\vec{q}\,)
+ i \epsilon} +
\frac{n(\vec{q}_{\rm lab})}
{\sqrt{s} + \omega - E_l(-\vec{q}\,) - i \epsilon } \right\} \ ,
\label{eq:gmed}
\end{eqnarray}
where $(P^0,\vec{P})$ is the total four-momentum in the lab frame
and
$s=(P^0)^2-\vec{P}^2$.

The in-medium $\bar{K}N$ interaction, $T_{\rm
eff}(P^0,\vec{P},\rho)$,
is then obtained by solving the coupled-channel Bethe-Salpeter
equation
using the dressed
meson-baryon loop of Eq.~(\ref{eq:gmed}). The S-wave $\bar{K}$
self-energy 
(${\bar K}=K^-$ or $\bar{K}^0$)
is determined by summing the in-medium $\bar{K}N$ interaction
over
the nucleons in the Fermi sea
\begin{equation}
\Pi^s_{\bar{K}}(q^0,{\vec q},\rho)=2\sum_{N=n,p}\int
\frac{d^3p}{(2\pi)^3}
n(\vec{p}\,) \,   T_{\rm eff}^{\bar{K}
N}(q^0+E(\vec{p}\,),\vec{q}+\vec{p},\rho) \ .
\label{eq:selfka}
\end{equation}
Note that a self-consistent approach is required since one
calculates the ${\bar K}$ self-energy from the effective
interaction $T_{\rm eff}$ which uses ${\bar K}$ propagators which
themselves include the self-energy being calculated.
A P-wave contribution to the ${\bar K}$
self-energy coming from the coupling of the ${\bar K}$ meson to
hyperon-hole excitations is also included and the expression can
be found 
in ref. \cite{RO99}.

\begin{figure}[ht!] 
\begin{center}
\includegraphics[width=0.55\linewidth]{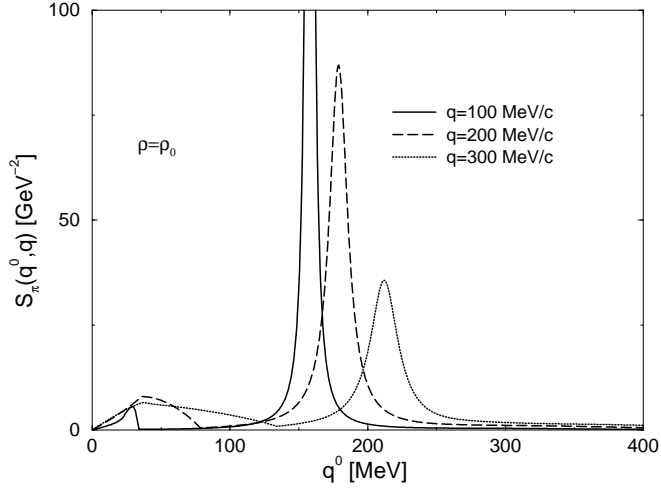}
\end{center}
\caption{Pion spectral density at $\rho=\rho_0$ for several
momenta}
\label{fig:specpi}
\end{figure}

The pion self-energy is built from a model that contains the
effect of one-
and two-nucleon absorption and is
conveniently
modified to include the effect of nuclear short-range
correlations (see ref. \cite{ramos94} for details).
To assess the importance of dressing the pions we show
in Fig. \ref{fig:specpi} the spectral
density of the $\pi$ meson in nuclear matter at density
$\rho=\rho_0$ for several
momenta.  The strength is distributed over a wide range of
energies and,
as the pion momentum increases, the
position of the peak is
increasingly lowered from the corresponding one in free space as
a
consequence of the attractive pion-nuclear potential.
Note that, to the left of the peaks, there appears the typical
structure
of the $1p1h$ excitations which give rise to $1p1h\Lambda$ and
$1p1h\Sigma$ components in the effective ${\bar K}N$ interacion.

\begin{figure}[ht!]
\begin{minipage}{.45\linewidth}
\begin{center}
\includegraphics[width=0.9\linewidth]{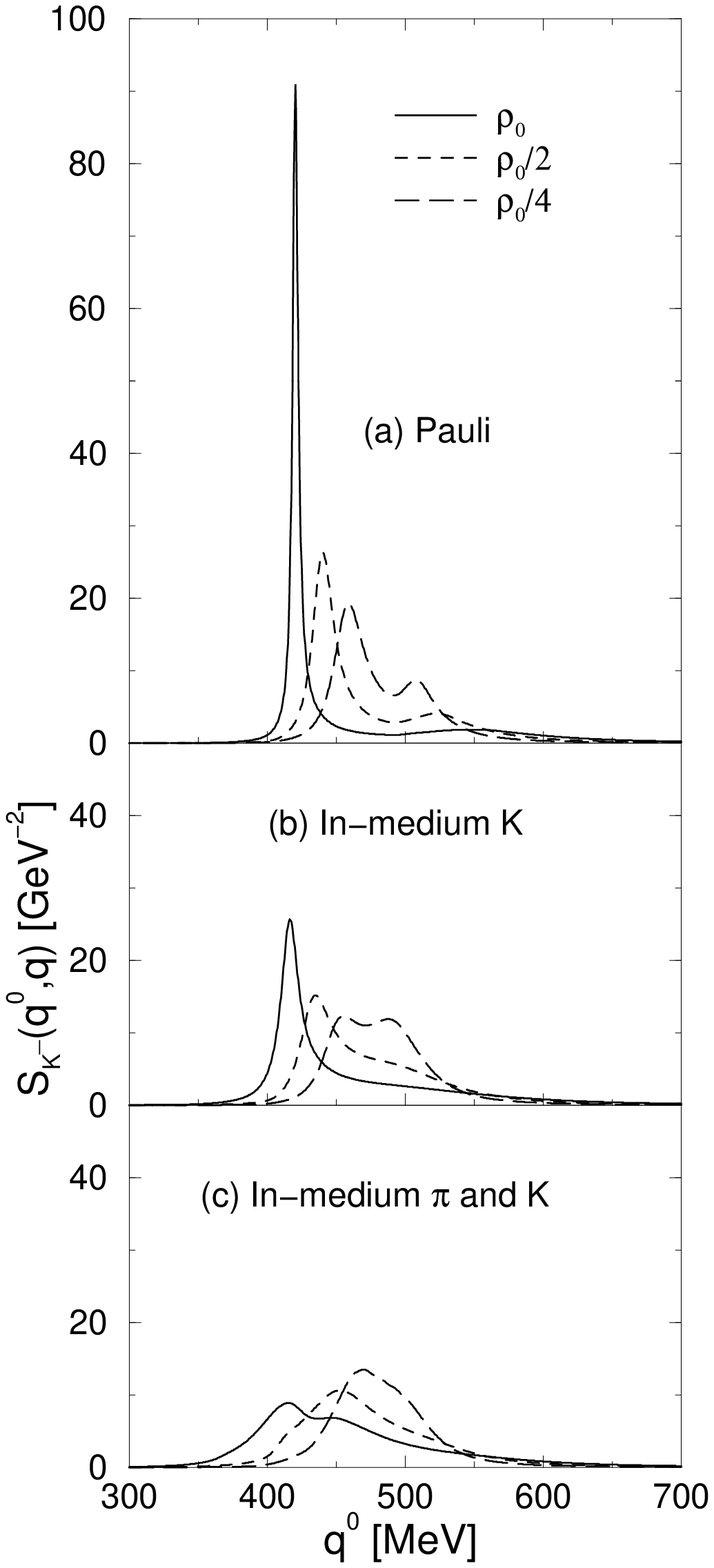}
\end{center}
\caption{$K^-$ spectral density for zero momentum}
\label{fig:kspec}
\end{minipage}
\hfill
\begin{minipage}{0.45\linewidth}
\begin{center}
\includegraphics[width=0.9\linewidth]{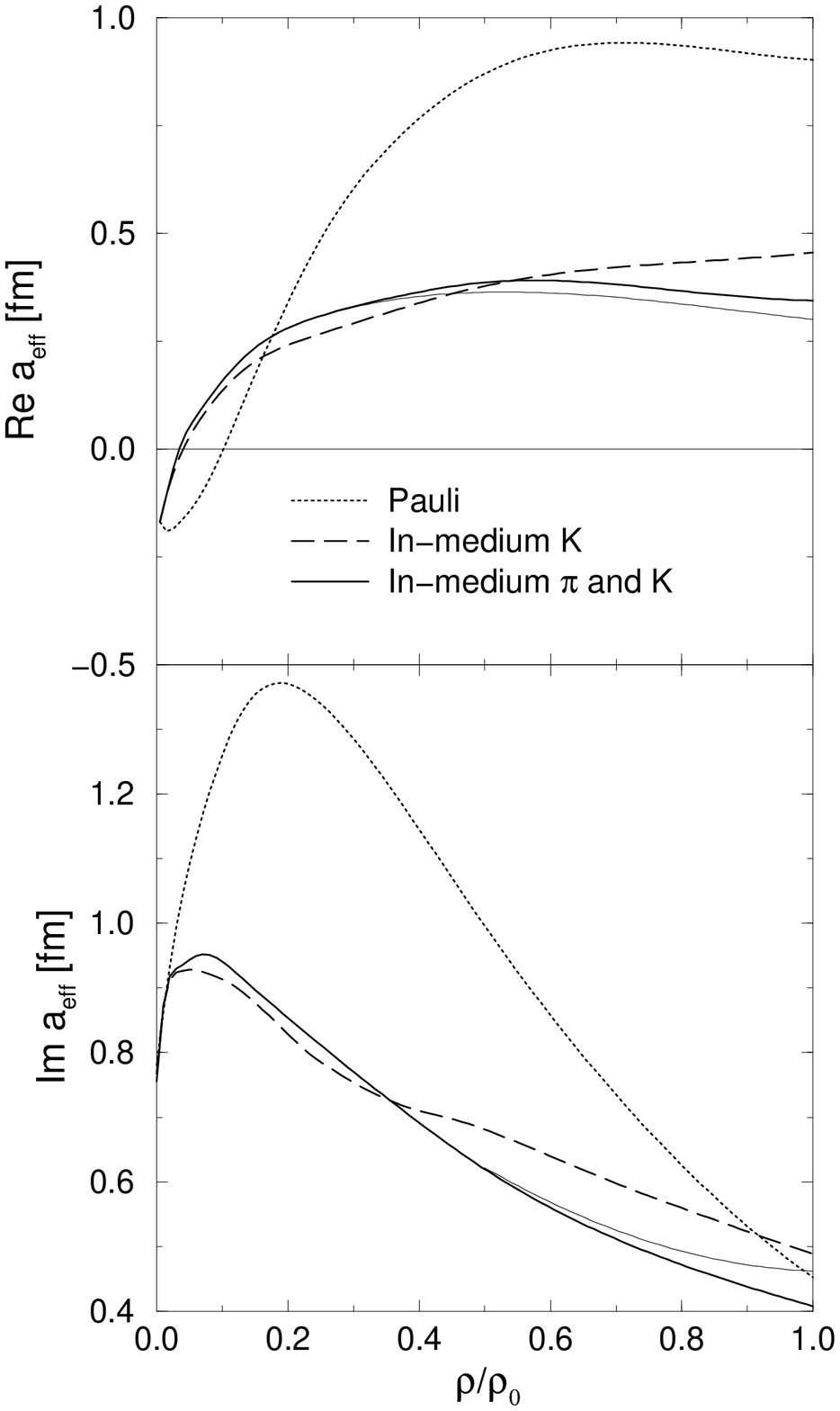}
\end{center}
\caption{$K^-N$
scattering length
as a function of density}
\label{fig:scatlen}
\end{minipage}
\end{figure}

The spectral function of a $K^-$ meson of zero momentum is
shown in Fig. \ref{fig:kspec} for various densities: $\rho_0$, $\rho_0/2$
and
$\rho_0/4$.
The results in the upper panel include
only Pauli blocking effects, i.e. the nucleons propagate as in
Eq.
(\ref{eq:nucleon}) but the mesons behave as in free space. 
At $\rho_0/4$ one clearly sees two excitation modes. The left one
corresponds to the
$K^-$ pole branch, appearing
at an energy smaller than the kaon mass, $m_K$, due to the
attractive medium effects. The peak on the right corresponds to
the
$\Lambda(1405)$-hole excitation mode,
located above $m_K$ because of the shifting of
the $\Lambda(1405)$ resonance to energies above the $K^-p$
threshold.
As density increases, the $K^-$ feels an enhanced
attraction while
the $\Lambda(1405)$-hole peak moves to higher energies and
loses
strength, a reflection of the tendency of the $\Lambda(1405)$
to dissolve in the dense nuclear medium. These features were
already
observed in ref. \cite{Waas97}.
The (self-consistent) incorporation of the ${\bar K}$ propagator
in
the Bethe-Salpeter equation softens the effective interaction,
$T_{\rm eff}$, which becomes
more spread out in energies. The resulting $K^-$ spectral
function (middle
panel in Fig.~\ref{fig:kspec}) shows the displacement of the resonance
to lower energies
because, as noted by Lutz \cite{Lutz98}, the attraction felt by
the ${\bar
K}$ meson lowers the threshold for the
${\bar K}N$ states that had been increased by the Pauli blocking
on the
nucleons. 
This has a compensatory effect
and the resonance moves backwards, slightly below its free
space value.
The $K^-$ pole peak appears at similar or slightly
smaller energies, but its width is larger, due to the
strength of the intermediate ${\bar K}N$ states being distributed
over a wider region of energies.
Therefore the $K^-$ pole and the $\Lambda(1405)$-hole branches
merge into one another and can hardly be distinguished.
Finally, when the pion is dressed according to the spectral
function
shown in Fig. \ref{fig:specpi} the effective interaction
$T_{\rm eff}$ becomes even smoother. The resulting 
$K^-$ spectral function is shown in the bottom panel in
Fig.~\ref{fig:kspec}. 
As seen by the long-dashed line, even at
very small densities one no longer distinguishes the
$\Lambda(1405)$-hole
peak from the $K^-$ pole one.
As density increases the attraction
felt by the $K^-$ is more moderate and the $K^-$ pole peak
appears at
higher energies than in the other two approaches.
However, more strength is found at very
low energies, especially at $\rho_0$,  due to the coupling of
the $K^-$ to the
$1p1h$ and $2p2h$ components of the pionic strength. It is
precisely the
opening of the
$\pi\Sigma$ channel, on top of the already opened $(1p1h)\Sigma$
and
$(2p2h)\Sigma$ ones, the reason for the cusp structure which
appears
slightly above 400 MeV.

The isospin averaged in-medium scattering length, defined as
\begin{equation}
a_{\rm eff}(\rho)= -\frac{1}{4\pi} \frac{M}{m_K + M }
\frac{\Pi_{\bar{K}}(m_K,\vec{q}=0,\rho)}{\rho} \ ,
\end{equation}
is shown in Fig.~\ref{fig:scatlen} as a function of the nuclear density
$\rho$.
The change of ${\rm Re}\,
a_{\rm eff}$ from negative to positive values indicates the
transition from a repulsive interaction in free space to an
attractive one in the medium. As shown by the dotted
line, this transition happens at a density of about $\rho\sim
0.1\rho_0$ when only Pauli effects are considered, in agreement
with what was found in ref. \cite{WKW96}. However, this
transition occurs at even lower densities ($\rho \sim 0.04
\rho_0$) when one considers the self-energy
of the mesons in the description, whether one dresses only the
$\bar{K}$
meson (dashed line) or both the $\bar{K}$ and $\pi$ mesons (solid
line).
The deviations from the approach including only Pauli
blocking or those dressing the mesons are quite appreciable over
a wide range of densities. The thin solid lines show the results
obtained
with a repulsive $\Sigma$ potential of the type
$U^\Sigma=U^\Sigma_0 \rho/\rho_0$, with $U_0^\Sigma=30$ MeV.
The deviations from the thick solid line, obtained for an
attractive
potential depth of $U_0^\Sigma=-30$ MeV, are smaller than 10\%
and only show
up at the higher densities.

\begin{figure}[ht!] 
\begin{center}
\includegraphics[width=0.6\linewidth]{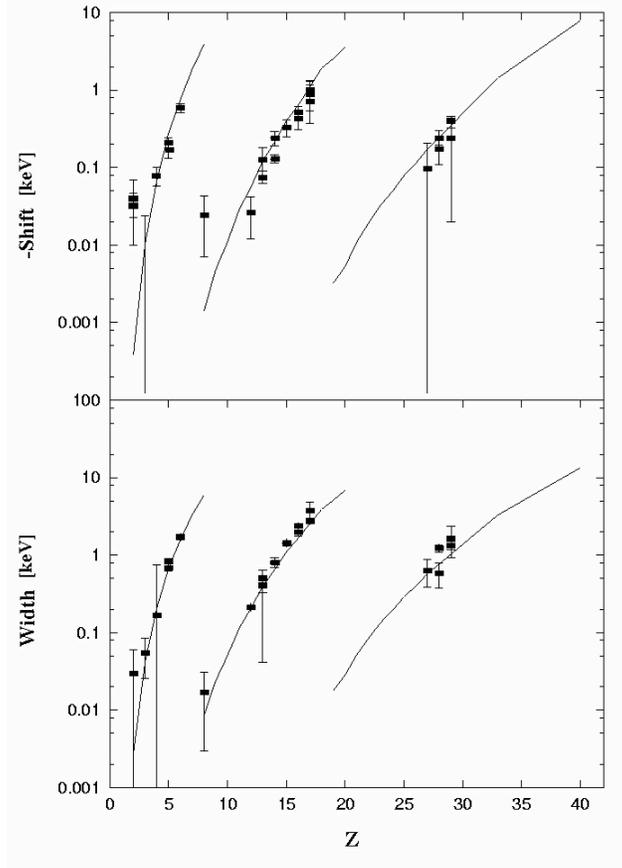}
\end{center}
\caption{Energy shifts and widths of kaonic atom states 
as predicted by
the model of ref. \cite{oku99}. The experimental data are taken from
the compilation given in ref. \cite{BGF97}.}
\label{fig:katom}
\end{figure}

The implications on kaonic atoms of the scattering length
displayed in Fig. \ref{fig:scatlen} or, equivalently,
the $K^-$ optical potential, $V_{\rm
opt}(\rho)=\Pi_{\bar{K}}(m_K,\vec{q}=0,\rho)/2
m_K$, have been analyzed 
in the framework of a local density approximation, where
the nuclear matter density $\rho$ is replaced
by the density profile $\rho(r)$ of the particular nucleus
\cite{oku99}. As can be seen from Fig. \ref{fig:katom},
both the energy shifts
and widths of kaonic atom states agree well with the
bulk of experimental data.

The model reported here gives a $K^-$ nuclear potential depth of
$-44$ MeV at $\rho=\rho_0$. This is
about half the attraction of that obtained with other
recent theories and approximation schemes
\cite{WKW96,Waas97,Li97,Mao99,Sch97,Tsushi98}, which give rise to
potential depths at the center of the nucleus
in between $-140$ and $-75$ MeV, and also lies very far from
the depth of around $-200$ MeV obtained
from a best fit to $K^-$ atomic data with a phenomenological
potential that includes an additional
non-linear density
dependent term \cite{FGB94}. On the other hand,
the early Brueckner-type calculations of ref.
\cite{alberg76} also obtained a shallow $K^-$-nucleus potential,
of
the order of $-40$ MeV at the center of $^{12}$C, and predicted
reasonably well the $K^-$ atomic data available at that time and
recent self-consistent calculations \cite{SKE99} find a moderate
attraction
of $-32$ MeV.
Acceptable fits to kaonic atom data have also been obtained using
charge
densities and
a phenomenological $T_{\rm eff} \rho$ type potential
with a depth of the
order of $-50$ MeV in the nuclear interior\cite{batty81}. But,
when
matter densities are used instead, the fit gives a potential
depth of $-80$
MeV \cite{FGB94}. A comparison of kaonic atom results obtained with various
$K^-$-nucleus potentials can be found in ref. \cite{baca00}.
A hybrid model, combining a relativistic
mean field
approach in the nuclear interior and a phenomenological density
dependent potential at the surface that is fitted to $K^-$ atomic
data, also favors a strongly attractive $K^-$ potential of depth
$-180$ MeV \cite{FGMC99}.

In summary, although all models predict attraction for the
$K^-$-nucleus
potential, there are still large discrepancies for the precise
value of
its depth, which has important
implications for the occurrence of kaon condensation. It is then
necessary
to gather more data that could help in disentangling the
properties of
the $\bar{K}$ in the medium. Apart from the valuable information
that can
be extracted from the production of $K^-$
in heavy-ion collisions, one could also measure deeply bound
kaonic states,
which have been predicted to be narrow \cite{oku99,baca00,FG99a} and
could be measured
in $(K^-,\gamma)$ \cite{oku99} or $(K^-,p)$ reactions
\cite{FG99b,Kishi99}.

%% file: conclu.tex
\chapter{Conclusions}
After a short review of the basic concepts of chiral symmetry
and of several chiral Lagrangians we have discussed various
nonperturbative methods to deal with the meson-meson and meson-baryon
interactions which allow one to extend the region of applicability of the
theory to higher energies than in $\chi PT$, where the low lying mesonic and
baryonic resonances appear. The common ground of all these methods was
the exact implementation of unitarity in coupled channels. The constraints
imposed by unitarity allow one to extract information contained in the
chiral Lagrangians which is not accessible with the standard $\chi PT$
expansion.

One of the procedures followed was the Inverse Amplitude Method,
which relies upon the expansion of the inverse of the scattering matrix
and  gives
rise to an expansion in powers of $p^2$ with a larger convergence
radius than $\chi PT$. In that method one could extend the predictions
for meson-meson interactions up to about 1.2 GeV, and all mesonic resonances
up to this energy were well reproduced, as well as phase shifts and
inelasticities.

A second method relied upon the use of the N/D method and the hypothesis of
resonance saturation. In this case, the use of the information contained in
the
lowest order chiral Lagrangian, together with chiral loops and the explicit
exchange of some resonances, which are genuine QCD states in the sense that
they
would remain in the large $N_c$ limit, allow also a good description of the
meson-meson data up to about 1.5 GeV. This second method is particularly
rewarding
for it allows one to dig into the nature of the mesonic resonances and
separate
those which are preexisting QCD resonances, in the limit of large $N_c$, from 
others which qualify as dynamical meson-meson resonances coming from
the multiple scattering of the mesons. In this way, it was stablished that the
low lying
scalar resonances, the $\sigma$, $\kappa$, $a_0(980)$ and to large extent
the
$f_0(980)$, are generated dynamically from multiple scattering from the
lowest order
chiral Lagrangian. On the contrary, a singlet contribution to the $f_0(980)$
and a
scalar octet around 1.35 GeV would be the lightest preexisting scalar
states. This
latter method also allows one to understand why in the case of the scalar
sector
a succesful reproduction of the data can be obtained simply by means of
the lowest
order chiral Lagrangian and the Bethe-Salpeter equation, together
with a suitable cut off, or regularizing scale.

In the meson-baryon problem, applications were only done in the
scalar sector taking advantage of the simplification of the Bethe-Salpeter
equation, which was found to be a suitable approach much as in the case
of the meson-meson scalar sector. In this case, low lying resonances like
the
$\Lambda(1405)$ or the $N(1535)$ were generated within that approach, and a
good reproduction of the low energy scattering data was found, particularly
in the
case of the $K^- N$ interaction and coupled channels.

 Applications to problems of initial and final state interaction
have also been shown. Since the energy region of applicability of the
reported
methods is much larger than the one of $\chi PT$, one could tackle many new
problems formerly inaccessible with plain $\chi PT$ theory.

 We have also shown how these chiral approaches to the meson-baryon
and meson-meson interactions have repercussions in nuclear physics and
provide a new perspective into problems which have been rather controversial
up to now, like the $\pi \pi$ scattering in a nuclear medium or the $K^-$
nucleus interaction. Several reactions which can bring new light into these
problems have also been reviewed and the implementation of the experiments
is already planned in some laboratories.

 The methods exposed here open new possibilities to face
a large number of problems, of which we have only given a few examples.
Extension of the methods to higher energies by incorporating channels
with more than two particles and application to other physical domains
remain as challenges for the future.

\vspace*{3cm}

{\large\bf Acknowledgements}

\vskip 0.5 cm

This work has been partially supported by CICYT contracts
PB95-1249, PB96-0753 (Spain) and by the Eurodaphne project, EEC-TMR 
contract FMRX-CT98-0169.

We would like to thank our colleagues
H.C.~Chiang,
F.~Guerrero,
S.~Hi\-ren\-zaki,
A.~Ho\-sa\-ka,
T.S.H.~Lee,
E.~Marco,
Ulf-G.~Meissner,
J.C.~Nacher,
M.~Oka,
Y.~Okumura,
A.~Parre\~no,
J.R.~Pel\'aez,
A.~Pich,
H.~Toki and
M.J.~Vicente Vacas
for many stimulating discussions and for their collaboration in 
some of the works reported here.

%% file: paper.bbl
\begin{thebibliography}{299}
\bi{Pich} A. Pich, Rep. Prog. Phys. {{58}} (1995) 563.
\vs
\bi{Meissner}U. G. Meissner, Rep. Prog. Phys. {{56}} (1993)
903; V. Bernard,
N. Kaiser and U. G. Meissner, Int. J. Mod. Phys. {{E4}} (1995)
193.
\vs
\bi{Eckerep}G. Ecker, Prog. Part. Nucl. Phys. {{35}} (1995) 1.
\vs
\bi{1961TR} J. Goldstone, \NC{19} (1961) 154.
\vs
\bi{xpt} J. Gasser and H. Leutwyler, \AN{158} (1984) 142; J.
Gasser and H. 
Leutwyler, \NP{B250} (1985) 465, 517, 539.
\vs
\bi{Mann1962TP} M. Gell-Mann, \PR{125} (1962) 1067.
\vs
\bi{Okubo1962TP} S. Okubo, Prog. Theor. Phys. {{27}} (1962)
949.
\vs
\bi{Wess1971TR} J. Wess and B. Zumino, \PR{163} (1967) 1727;
J. Wess and B.
Zumino, \PL{B37} (1971) 95.
\vs
\bi{Witten1983TR} E. Witten, \AN{128} (1980) 363; E. Witten,
\NP{B223} (1983) 422.

\vs
\bi{Georgi} A. Manohar and H. Georgi, \NP{B234} (1984) 189.
\vs
\bi{EPR} G. Ecker, J. Gasser, A. Pich and E. de Rafael,
\NP{B321} (1989) 311.
\vs
\bi{GT} M. L. Goldberger and S. B. Treiman, Phys. Rev. {{110}}
(1958) 1178. 
\vs
\bi{KR} N. M. Kroll and M. A. Ruderman, Phys. Rev. {{93}}
(1954) 233.
\vs
\bi{Weinberg} S. Weinberg, Physica A {{96}} (1979) 327.
\vs
\bi{appe} T. Applequist and C. Bernard, \PR{D22} (1980) 200;
A. C. Longhitano,
\NP{B188} (1981) 118.
\vs
\bi{solid} H. Leutwyler, \PR{D49} (1994) 3033; H. Leutwyler,
Helv. Phys. Acta
{{70}} (1997) 275; S. C. Zhang, Science {{275}} (1997) 1089; C.
P. Burgess
and A. Lutken, \PR{B57} (1998) 8642.
\vs
\bi{Steele} J. V. Steele, H. Yamagishi and I. Zahed, \NP{A615}
(1997) 305; M. R.
Pennington and J. Portol\'es, \PL{B344} (1995) 399.
\vs
\bi{Longi} A. C. Longhitano, \NP{B188} (1981) 118; \PR{D22}
(1980) 1166.
\vs
\bi{IAM} T. N. Truong, \PRL{61} (1988) 2526; \PRL{67} (1991)
2260; A.
Dobado, M. J. Herrero and T. N. Truong, \PL{B235} (1990) 134; A.
Dobado and J.
R. Pel\'aez, \PR{D47} (1993) 4883; \PR{D56} (1997) 3057.
\vs
\bi{LHCres}A. Dobado, M. J. Herrero and T. N. Truong,
\PL{B235} (1990) 129; A.
Dobado, M. J. Herrero and J. Terr\'on, \ZP{C50} (1991) 205;
\ZP{{C50}}
(1991) 465; J. R. Pel\'aez, \PR{D55} (1997) 4193.
\vs
\bi{Weinis} J. Weinstein and N. Isgur, \PRL{48} (1982) 659; J.
Weinstein and N. 
Isgur, \PR{D27} (1983) 588; \PR{D41} (1990) 2236.
\vs
\bi{Jansen} G. Jansen, B. C. Pearce, K. Holinde and J. Speth,
\PR{D52} (1995)
2690.
\vs
\bi{npa} J. A. Oller and E. Oset, \NP{A620} (1997) 438;
{\it erratum} \NP{A652}
(1999) 407.
\vs
\bi{Tornqvist} N. A. Tornqvist, \PRL{49} (1982) 624; M. Roos
and N. A.
Tornqvist, \PRL{76} (1996) 1575.
\vs
\bi{OOnd} J. A. Oller and E. Oset, \PR{D60} (1999) 074023.
\vs
\bi{prl} J. A. Oller, E. Oset and J. R. Pel\'aez, \PRL{80}
(1998) 3452.
\vs
\bibitem{Martinspearman}A. D. Martin and T. D. Spearman,
Elementary Particle
Theory, John Willey, 1970, p 187.
\vs
\bi{NPB} F. Guerrero and J. A. Oller, \NP{B537} (1999) 459.
\vs
\bi{UMNR} V. Bernard, N. Kaiser and U. G. Meissner, \NP{B357}
(1991) 129.
\vs
\bi{Kaminski} R. Kaminski, L. Lesniak and K. Rybicki, \ZP{C74}
(1997) 79.
\vs
\bi{Hyams} B. Hyams et al., \NP{B64} (1973) 134.
\vs
\bi{Estabrooks} P. Estabrooks et al., AIP Conf. Proc. 13
(1973) 37.
\vs
\bi{Grayer} G. Grayer et al., Proc. 3rd Philadelphia Conf. on
Experimental Meson
Spectroscopy, Philadelphia, 1972 (AIP, NY, 1972), p 5.
\vs
\bi{Protopopescu} S. D. Protopopescu and M. Alson-Granjost,
\PR{D7} (1973) 1279.
\vs
\bi{Frogatt} C. D. Frogatt and J. L. Petersen, \NP{B129}
(1977) 113.
\vs
\bi{Cohen} D. Cohen et al., \PR{D22} (1980) 2595.
\vs
\bi{Martin} A. D. Martin and E. N. Ozmuth, \NP{B158} (1979)
520.
\vs
\bi{Ochs} W. Ochs, University of Munich thesis, 1974.
\vs
\bi{Rosselet} L. Rosselet et al., \PR{D15} (1977) 574.
\vs
\bi{Schenk} A. Schenk, \NP{B363} (1991) 97.
\vs
\bi{PRD} J. A. Oller, E. Oset and J. R. Pel\'aez, \PR{D59}
(1999) 74001; {\it erratum}
\PR{D60} (1999) 099906.
\vs
\bi{kaiser} N. Kaiser, Eur. Phys. J. {{A3}} (1998) 307.
\vs
\bi{Grayer2} G. Grayer et al., in Experimental Meson
Spectroscopy, edited by A.
H. Rosenfeld and K. W. Lai, AIP Conf. Proc. 8 (AIP, New York,
1972) p 117.
\vs
\bi{Grayer3} G. Grayer et al., Paper No.768 Contributed to
the 16th Int. Conf.
on High Energy Physics, Batavia, 1972.
\vs
\bi{Grayer4} G. Grayer et al., \NP{B75} (1974) 189.
\vs
\bi{Manner} W. Manner, Contribution to the 4th Int. Conf. on
Experimental Meson
Spectroscopy, Boston, Massachusetts, USA, April 1974 and CERN
preprint.
\vs
\bi{Linden} S. J. Lindenbaum and R. S. Longacre, \PL{B274}
(1992) 492.
\vs
\bi{Estabrooks3} P. Estabrooks et al., \NP{B133} (1978) 490.
\vs
\bi{Linglin} D. Linglin et al., \NP{B57} (1973) 64.
\vs
\bi{Amsterdam} Amsterdam, CERN, Nijmegen-Oxford Collaboration,
J. B. Gay et 
al., \PL{B63} (1976) 220.
\vs
\bi{Mercer} R. Mercer et al., \NP{B32} (1971) 381.
\vs
\bi{Bingham} H. H. Bingham et al., \NP{B41} (1972) 1.
\vs
\bi{Baker} S. L. Baker et al., \NP{B99} (1975) 211.
\vs
\bi{Aston} D. Aston et al., \NP{B296} (1988) 493.
\vs
\bi{newPRD} J. A. Oller, E. Oset, J. R. Pel\'aez, \textit{The
$\phi \rightarrow
\pi^+ \pi^-$ within a chiral unitary approach}, submitted to
Phys. Rev. D, hep-ph/9911297.
\vs
\bi{UMR}V. Bernard, N. Kaiser and U. G. Meissner, \NP{B364}
(1991) 283.
\vs
\bi{FP} F. Guerrero and A. Pich, \PL{B412} (1997) 382.
\vs
\bi{NewP} M. F. L. Golterman and S. Peris, hep-ph/9908252.
\vs
\bi{Montanet} For general reviews about the issue: L.
Montanet, Rep. Prog. Phys.
{{46}} (1983) 337; F.E. Close, Rep. Prog. Phys. {{51}} (1988) 833;
M. R.
Pennington, Nucl. Phys. {{B21}} (Proc. Suppl.) (1991) 37.
\vs
\bi{Morgan} D. Morgan, \PL{B51} (1974) 71.
\vs
\bi{Jaffe} R. L. Jaffe, \PR{D15} (1977) 267;  R. L. Jaffe, \PR{D15}
(1977) 281.
\vs
\bi{Achasov} N. N. Achasov, S. A. Debyanin and G. N.
Shestakov, \ZP{C22} (1984)
53; N. N. Achasov, S. A. Devyanin and G. N. Shestakov, \PL{B96}
(1980) 168; N.
N. Achasov, S. A. Devyanin and G. N. Shestakov, Phys. Scripta
{{27}} (1983)
330.
\vs
\bi{5deMP93} R. L. Jaffe, \PL{41} (1975) 267; D. Robson,
\NP{B130} (1977) 328.
\vs
\bi{6deMP93} T. Barnes, in Proceedings of the Fourth Workshop
on Polarized Targets Materials
and Techniques, Bad Honned, Germany, 1984, edited by W. Meyer
(Bonn Univeristy,
Bonn, 1984); J.F. Donoghue, in Hadron Spectroscopy-1985,
Proceedings of the
International Conference, College Park, Maryland, edited by S.
Oneda, AIP Conf.
Proc. No. 132 (AIP, New York, 1985), p 460; also, Close
\cite{Montanet}.
\vs
\bi{17nd} G. F. Chew and S. Mandelstam, \PR{119} (1960) 467.
\vs
\bi{nieves}J. Nieves and E. Ruiz Arriola, \PL{B455} (1999)
30; hep-ph/9907469.
\vs
\bi{CDD}L. Castillejo, R. H. Dalitz and F. J. Dyson, \PR{101}
(1956) 453.
\vs
\bi{Witten}G. 't Hooft, \NP{B72} (1974) 461; E. Witten,
\NP{B160} (1979) 57.
\vs
\bi{Bjorken}J. D. Bjorken, \PRL{4} (1960) 473.
\vs
\bi{Jamin} M. Jamin, J. A. Oller and A. Pich, forthcoming.
\vs
\bi{ohiggs} J. A. Oller, \PL{B}, in print. hep-ph/9908493.
\vs
\bi{KSFR} K. Kawarabayashi and M. Suzuki, \PRL{16} (1996) 255;
Rizuddin and
Fayyazuddin, \PR{147} (1966) 1071.
\vs
\bi{PDG} C. Caso et al., The European Physical Journal {{C3}}
(1998) 1.
\vs
\bi{EP2} G. Ecker, J. Gasser, H. Leutwyler, A. Pich and E. de
Rafael, 
\PL{B223} (1989) 425.
\vs
\bi{43}A. Etkin et al., \PR{D28} (1982) 1786.
\vs
\bi{44}W. Wetzel et al., \NP{B115} (1976) 208; V. A.
Polychronakos et al., 
\PR{D19} (1979) 1317; G. Costa et al., \NP{B175} (1980) 402.
\vs
\bi{Arm}T. A. Armstrong et al., \ZP{C52} (1991) 389.
\vs
\bi{Chew2}G. F. Chew and S. C. Frautschi, \PR{124} (1961) 264.
\vs
\bi{AO} E. Oset and A. Ramos,  \NP{A635} (1998) 99.
\vs
\bi{NK} N. Kaiser, P. B. Siegel and W. Weise, \NP{A594} (1995)
325.
\vs
\bibitem{BKM97} V. Bernard, N. Kaiser and U.G. Meissner, Nucl.
Phys. A615
(1997) 483. 
\vs
\bibitem{FMS98} N. Fettes, U.G. Meissner and S. Steininger,
Nucl. Phys. A640
(1998) 199.
\vs
\bibitem{Moj98} M. Mojzis, Eur. Phys. J C2 (1998) 181.
\vs
\bibitem{DP97} A. Datta and S. Pakvasa, Phys. Rev. D56 (1997)
4322.
\vs
\bibitem{ET98} P.J. Ellis and H.B. Tang, Phys. Rev C57 (1998)
3356.
\vs
\bibitem{Lee94} C. H. Lee, H. Jung, D. P. Min and M. Rho,
Phys. 
Lett.
B326 (1994) 14.
\vs
\bibitem{Bro93} G. E. Brown, C.-H. Lee, M. Rho and V.
Thorsson,
Nucl. Phys.
A567 (1994) 937.
\vs
\bibitem{dalitz59} R.H. Dalitz and S.F. Tuan, Phys. Rev. Lett. 5 (1959)
425; R.H. Dalitz and S.F. Tuan, Ann. Phys. (NY) 10 (1960) 307.
\vs
\bibitem{isgur78}
N. Isgur and G. Karl, Phys. Rev. D18 (1978) 4187.
\vs
\bibitem{arima94}
M. Arima, S: Matsui and K. Shimizu, Phys. Rev. C49 (1994) 2831.
\vs
\bibitem{veit84}
E.A. Veit, B.K. Jennings, R.C. Barret and A.W. Thomas, Phys. Lett. B137
(1984) 415;
E.A. Veit, B.K. Jennings, A.W. Thomas and R.C. Barret, Phys. Rev. D31
(1985) 1033.
\vs
\bibitem{schat95}
C.L. Schat, N.N. Scoccola and C. Gobbi, Nucl. Phys. A585 (1995) 627.
\vs
\bibitem{CHL96a} C.-H. Lee, D. P. Min and M. Rho, Nucl. Phys.
A602
(1996) 334.
\vs
\bibitem{To71} D. N. Tovee et al., Nucl. Phys. B33 (1971) 493.
\vs
\bibitem{No78} R. J. Nowak et al., Nucl. Phys. B139 (1978) 61.
\vs
\bibitem{Th73} D. W. Thomas, A. Engler, H. E. Fisk and R. W. 
Kraemer, Nucl. Phys. B56 (1973) 15.
\vs
\bibitem{Hem85}  R. J. Hemingway, Nucl. Phys. B253\
(1985) 742.
\vs
\bibitem{dal91} R.H. Dalitz and A. Deloff, J. Phys. G17 (1991)
289.
\vs
\bibitem{Miw97} M. Iwasaki et al., Phys. Rev. Lett. 78 (1997)
3067.
\vs
\bibitem{Adm81} A. D. Martin, Nucl. Phys. B179 (1981) 33.
\vs
\bibitem{Hump} W.E. Humphrey and R.R. Ross, Phys. Rev. 127
(1962)
1305.
\vs
\bibitem{Sakitt} M. Sakitt et al., Phys. Rev. 139 (1965) 719.
\vs
\bibitem{Kim} J. K. Kim, Phys. Rev. Lett. 21 (1965) 29; J. K.
Kim,
Columbia University Report, Nevis 149 (1966).
\vs
\bibitem{Kittel} W. Kittel, G. Otter and I. Wacek, Phys. Lett.
21
(1966) 349.
\vs
\bibitem{Cibo} J. Ciborowski et al., J. Phys. G8 (1982) 13.
\vs
\bibitem{Evans} D. Evans et al., J. Phys. G9 (1983) 885.
\vs
\bibitem{brmart} B. R. Martin, Nucl. Phys. B94 (1975) 413.
\vs
\bibitem{Do82}See C. Dover and G. Walker, Phys. Reports 89
(1982)
1 for a compilation of data.
\vs
\bibitem{Kai95b} N. Kaiser, P.B. Siegel and W. Weise, Phys.
Lett, B362
(1995) 23.
\vs
\bibitem{KWW97} N. Kaiser, T. Waas and W. Weise, Nucl.
Phys. A612 (1997) 297.
\vs
\bibitem{Kai} J. Caro Ramon, N. Kaiser, S. Wetzel and W. Weise,
Nucl. Phys. A , in print. nucl-th/9912053.
\vs
\bibitem{pelaez99} A. G\'omez Nicola and J.R. Pel\'aez,
hep-ph/9912512.
\vs
\bibitem{nieves00}
J. Nieves and E. Ruiz-Arriola, hep-ph/0001013.
\vs
\bibitem{ulfoller} 
U.G. Meissner and J.A. Oller, Nucl. Phys. A, in print. nucl-th/9912026.
\vs
\bi{10}  M. R. Pennington, The Second DAPHNE Physics Handbook
(1995) Vol. II,
p 531.
\vs
\bi{Feindt} M. Feindt and J. Harjes, \NP{B21} (Proc. Suppl.)
(1991) 61.
\vs
\bi{gamanpa} J. A. Oller and E. Oset, \NP{A629} (1998) 739.
\vs
\bi{Po} P. Ko, \PR{D41} (1990) 1531; J. F. Donoghue and B. R.
Holstein, \PR{D48}
(1993) 137; J. F. Donoghue, B. R. Holstein and D. Wyler, \PR{D47}
(1993) 2089.
\vs
\bi{xpt001} J. Bijnens and F. Cornet, \NP{B296} (1988) 557.
\vs
\bi{xpt002} J. F. Donoghue, B. K. Holstein and Y. C. Lin,
\PR{D37} (1988) 2423.
\vs
\bi{26} D. Morgan and M. R. Pennington, \ZP{C37} (1988) 431.
\vs
\bi{5} H. Albrecht et al., \ZP{C48} (1989) 183.
\vs
\bi{hep-ph-BM} M. Boglione and M. R. Pennington, Eur. Phys. J.
{{C9}} (1999) 
11.
\vs
\bi{PLB} J. A. Oller, \PL{B426} (1998) 7.
\vs
\bi{Emarco} E. Marco, S. Hirenzaki, E. Oset and H. Toki,
\PL{B}, in print, hep-ph/9903217.

\vs
\bibitem{10C} I. Dunietz, J. Hauser and J. L. Rosner, \PR{D35}
(1987) 2166.
\vs
\bibitem{1C} J. Lucio and J. Pestiau, \PR{D42} (1990) 3253;
J. Lucio and J. Pestiau, \PR{D43} 
(1991) 2447.
\vs
\bibitem{2C} S. Nussinov and T. N. Truong, \PRL{63} (1989)
1349.
\vs
\bibitem{3C} S. Nussinov and T. N. Truong, \PRL{63} (1989)
2003.
\vs
\bibitem{5C} N. N. Achasov and V. N. Ivachenko, \NP{B315}
(1989) 465. 
\vs
\bi{muli} T. L. Yhuang, M. L. Yan, X. J. Wang, hep-ph/9907233.
\vs
\bibitem{4C} N. Paver and Riazuddin, \PL{B246} (1990) 240.
\vs
\bibitem{C} F. E. Close, N. Isgur, S. Kumano, \NP{B389} (1993)
513.
\vs
\bi{Bramon} A. Bramon, A. Grau and G. Pancheri, \PL{B289}
(1992) 97.

\vs
\bibitem{Novo} M.N. Achasov et al., Phys. Lett. B440 (1998)
442.
\vs
\bibitem{CMDpi0pi0} R.R. Akhmetshin et al., Phys. Lett. B462 (1999)
380.
\vs
\bibitem{pi0eta} M.N. Achasov et al., Phys. Lett. B438 (1998)
441.
\vs
\bibitem{CMDpi+pi-} R.R. Akhmetshin et al., Phys. Lett.B462 (1999) 371.
\vs
\bibitem{Huber} K. Huber and H. Neufeld, Phys. Lett. B357
(1995) 221.
\vs
\bibitem{jaojr}
 E. Oset, S. Hirenzaki, E. Marco, J.A. Oller, J.R. Pel\'aez and H. Toki,
Contribution to the Workshop on physics and detectors for DAPHNE,
Frascati, November 1999.
\vs
\bi{Wa1} K. M. Watson, \PR{95} (1955) 228.
\vs
\bi{Mu1}N. I. Muskhelishivili, Singular Integral Equations
(Noordhoof,
Groningen, 1953).
\vs
\bi{Om1} R. Omn\`es, \NC{8} (1958) 316.
\vs
\bi{ulfff} J. Gasser and U. G. Meissner, Nucl. Phys. B357
(1991) 90.
\vs
\bi{Bar1} L. M. Barkov, \NP{B256} (1985) 365.
\vs
\bi{F2} F. Guerrero, \PR{D57} (1998) 4136.
\vs
\bi{T1} See the Phys. Rev. Lett. papers by Truong in
\cite{IAM}.
\vs
\bi{T2} L. Beldjoudi and T. N. Truong, hep-ph/9403348.
\vs
\bi{H1} T. Hannah, \PR{D55} (1997) 5613.




\vs
\bibitem{ugemeson} E. Marco, E. Oset and H. Toki, Phys. Rev.
C60 (1999) 015202.
\vs
\bibitem{Adam} C. R. Ji, R. Kaminsky, L. Lesniak, A.
Szczpaniak and R.
 Williams, Phys. Rev. C58 (1998) 1205.
\vs
\bibitem{Fries} C.D. Fries et al., Nucl. Phys. B143 (1978)
408.
\vs
\bibitem{Titov} A. I. Titov, Y. Oh, S. N. Yang and T. Mori,
Phys. Rev. C58 (1998) 2429; A. I. Titov, T.S.H. Lee and H. Toki,
nucl-th/9812074.
\vs
\bibitem{Tejedor} J.A. G\'omez Tejedor and E. Oset, Nucl.
Phys. A571
 (1994) 667;  J.A. G\'omez Tejedor and E. Oset, Nucl.
Phys. A600 (1996) 413.
\vs
\bibitem{Ochi} K. Ochi, M. Hirata and T. Takaki, Phys. Rev. C56 (1997) 1472.
\vs
\bibitem{WT76} W. Weise and L. Tauscher, Phys. Lett. B64
(1976) 424.
\vs
\bibitem{CT77} S.R. Cotanch and F. Tabakin, Phys. Rev. C15
(1977) 1379.
\vs
\bibitem{T7778} M. Thies, Phys. Lett. B70 (1977) 401; 
Nucl. Phys. A298 (1978) 344.
\vs
\bibitem{D80} A. Deloff, Phys. Rev. C21 (1980) 1516. 
\vs
\bibitem{RT80} A. S. Rosenthal and F. Tabakin,
 Phys. Rev. C22 (1980) 711. 
\vs
\bibitem{CE92} C. M. Chen and D. J. Ernst, Phys. Rev. C45
 (1992) 2019. 
\vs
\bibitem{JEC95}
M. J. Jiang, D. J. Ernst and C. M. Chen, Phys. Rev. C51 (1995)
857.
\vs
\bibitem{JK92}
M. F. Jiang and D. S. Koltun, Phys. Rev. C46 (1992) 2462.
\vs
\bibitem{MHT94}
M. Mizoguchi, S. Hirenzaki and H. Toki, Nucl. Phys. A567 (1994)
893.
\vs
\bibitem{FGB94} E. Friedman, A. Gal, C.J. Batty, Nucl. Phys.
A579 (1994) 518.
\vs
\bibitem{BGF97}
C. J. Batty, E. Friedman and A. Gal, Phys. Reports 287 (1997)
385.
\vs
\bibitem{alberg76} M. Alberg, E.M. Henley and L. Wilets, Ann.
Phys.  96 (1976) 43.
\vs
\bibitem{star87} L.R. Staronski and S. Wycech, 
J. Phys. G: Nucl. Phys. {13} (1987) 1361.
\vs
\bibitem{Koch94} V. Koch, Phys. Lett. B337 (1994) 7. 

\vs
\bibitem{WKW96} T. Waas, N. Kaiser and W. Weise, Phys. Lett. B365 (1996) 12.

\vs
\bibitem{Waas97} T. Waas and W. Weise, Nucl. Phys. A625 (1997)
287.



\vs
\bibitem{Lutz98} M. Lutz, Phys. Lett. B426 (1998) 12; M. Lutz,
nucl-th/9802033.
\vs
\bibitem{RO99} A. Ramos and E. Oset, Nucl. Phys. A, in print,
nucl-th/9906016.

\vs
\bibitem{Schuck}
 P. Schuck, W. N\"orenberg and G. Chanfray, Z. Phys. A330 (1988)
119.
\vs
\bibitem{Aouissat}
 Z. Aouissat, R. Rapp, G Chanfray, P Schuck and J. Wambach,
 Nucl. Phys. A581 (1995) 471.
\vs
\bibitem{Soyeur}
 M. Soyeur, G.E. Brown and M. Rho, Nucl. Phys. A556 (1993) 355.
\vs
\bibitem{Hermann} M. Hermann, B.L. Friman and W. N\"orenberg,
Nucl. Phys. A560 (1993) 411.
\vs
\bibitem{manolonew} M.J. Vicente-Vacas and E. Oset, forthcoming.


\vs
\bibitem{jones}
M. Jones, R.H. Dalitz and R.R. Horgan, Nucl. Phys. B129 (1977)
45.
\vs
\bibitem{darewych}
J.D. Darewych, R. Koniuk and N. Isgur, Phys. Rev. D{32} (1985)
1765.
\vs
\bibitem{veit}
E.A. Veit, B.K. Jennings, A.W. Thomas and R.C. Barrett,
Phys. Rev. D{31} (1985) 1033.
\vs
\bibitem{zhong}
Y.S. Zhong, Q.W. Thomas, B.K. Jennings and R.C. Barrett,
Phys. Rev. D{38} (1988) 837.
\vs
\bibitem{kaxiras}
E. Kaxiras, E.J. Moniz and M. Soyeur, Phys. Rev. D{32} (1985)
695.
\vs
\bibitem{he}
G. He and R.H. Landau, Phys. Rev. C{48} (1993) 3047.
\vs
\bibitem{arima}
A. Arima, S. Matsui and K. Shimizu, Phys. Rev. C{49} (1994) 2831.
\vs
\bibitem{workman}
R.L. Workman and Harold W. Fearing, Phys. Rev. D{37} (1988) 3117;
J. Lowe, Nuovo Cimento, A{102} (1989) 16, and other
references therein.
\vs
\bibitem{saghai}
Peter B. Siegel and Bijan Saghai, Phys. Rev. C{52} (1995) 392.
\vs
\bibitem{LOOR98} T.-S.H. Lee, J.A. Oller, E. Oset and A.
Ramos, Nucl.
Phys. A643 (1998) 402.
\vs
\bibitem{gammadat}
D.A. Whitehouse et al., Phys. Rev. Lett. {63} (1989) 1352.


\vs
\bibitem{nacherphot} J.C. Nacher, E. Oset, H. Toki and A.
Ramos, Phys. Lett.
B455 (1999) 55.
\vs
\bibitem{nacherrad} J.C. Nacher, E. Oset, H. Toki and A.
Ramos, 
Phys. Lett. B461 (1999) 299.
\vs
\bibitem{peterson} R. J. Peterson, private communication.
\vs
%

\bibitem{2} N. Grion et al., Phys. Rev. Lett. 59 (1987) 1080.
\vs
\bibitem{3} F. Bonutti et al., Phys. Rev. Lett. 77 (1996) 603.
\vs
\bibitem{JL86} J.A. Johnstone and T.S. H. Lee, Phys. Rev. C34
(1986) 243.
\vs
\bibitem{6} J. W. Durso, A. D. Jackson and B. J. Verwest,
Nucl. Phys. A345 (1980) 471.
\vs
\bibitem{4} G. Chanfray, Z. Aouissat, P. Schuck and W.
N\"{o}renberg, Phys. Lett. B256 (1991) 
325.

\vs
\bibitem{7} Z. Aouissat, G. Chanfray and P. Schuck, Mod. Phys.
Lett. A8 (1993) 1379.
\vs
\bibitem{9} D. Lohse, J.W. Durso, K. Holinde and J. Speth,
Phys. Lett.
B234 (1989) 235; Nucl. Phys. A516 (1990) 513.
\vs
\bibitem{Rapp} R. Rapp, J.W. Durso and J. Wambach, Nucl. Phys.
A596
(1996) 436.
\vs
\bibitem{chiang} H.C. Chiang, E. Oset and M.J. Vicente-Vacas,
Nucl. Phys.
A644 (1998) 77.
\vs
\bibitem{wolferic} T.E.O. Ericson and W. Weise , Pions and
Nuclei, Clarendon
Press, 1988.
\vs
\bibitem{15} S. Weinberg, Phys. Rev. Lett. 17 (1966) 616;
Phys. Lett.
18 (1967) 188.
\vs
\bibitem{16} M.G. Olsson and L. Turner, Phys. Rev. Lett. 20
(1968) 1127.
\vs
\bibitem{19} E. Oset and M.J. Vicente Vacas, Nucl. Phys. A446
(1985)
584.
\vs
\bibitem{19a} O. Jaekel, M. Dillig, C. A. Z. Vasconcelos,
Nucl. Phys. A541 (1992) 673.
\vs
\bibitem{20} V. Bernard, N. Kaiser and U.G. Meissner, Nucl.
Phys.
B457 (1995) 147; V. Bernard, N. Kaiser and U.G. Meissner, Nucl.
Phys. A619
(1997) 261.
\vs
\bibitem{21} T.S. Jensen and A.F. Miranda, Phys. Rev. C55
(1997) 1039;
Nucl. Phys. A541 (1992)673.
\vs
\bibitem{toniulf} U.G. Meissner, E. Oset and A. Pich, Phys.
Lett. B353 (1995) 161.
\vs
\bibitem{chanfrayole} G. Chanfray and D. Davesne, Nucl. Phys.
A646 (1999) 125.
\vs
\bibitem{juancarmen} C. Garc\'\i a-Recio, J. Nieves and E.
Oset, Phys. Rev. C51
(1995) 237.
\vs
\bibitem{17} E. Oset, C. Garc\'{\i}a-Recio and J. Nieves,
Nucl. Phys.
A584 (1995) 653.
\vs
\bibitem{rappnew} R. Rapp et al., Phys. Rev. C59 (1999) R1237.
\vs
\bibitem{manolo} M.J. Vicente-Vacas and E. Oset, Phys. Rev. C60 (1999)
064621.
\vs
\bibitem{hatsuda} T. Hatsuda, T. Kunihiro and H. Shimizu,
Phys. Rev. Lett.
82 (1999) 2840; T. Kunihiro, hep-ph/9905262.
\vs
\bibitem{KN86} D.B. Kaplan and A.E. Nelson, Phys. Lett.
B175 (1986) 57.
\vs
\bibitem{BKRT92}
G.E. Brown, K. Kubodera, M. Rho and V. Thorsson, Phys. Lett. B291
(1992)
355.
\vs
\bibitem{TPL94}
V. Thorsson, M. Prakash and J. Lattimer, Nucl. Phys. A572 (1994)
693.
\vs
\bibitem{EKP95}
P.J. Ellis, R. Knorren and M. Prakash, Phys. Lett. B349 (1995)
11.
\vs
\bibitem{FMMT96}
H. Fuji, T. Maruyama, T. Muto and T. Tatsumi, Nucl. Phys. A597
(1996) 645.
\vs
\bibitem{TY98}
T. Tatsumi and M. Yasuhira, Phys. Lett. B441 (1998) 9; T. Tatsumi and M.
Yasuhira, Nucl.
Phys. A653 (1999) 133.
\vs
\bibitem{LLB97}
G.Q. Li, C.-H. Lee and G.E. Brown, Phys. Rev. Lett. 79 (1997)
5214.
\vs
\bibitem{GS98}
N.K. Glendenning and J. Schaffner-Bielich, Phys. Rev. Lett. 81
(1998) 4564.
\vs
\bibitem{kaos97} R. Barth et al., Phys. Rev. Lett. 78 (1997)
4007; F. Laue et al., Phys. Rev. Lett. 82 (1999) 1640.
\vs
\bibitem{cassing97} W. Cassing, E.L. Bratkovskaya, U. Mosel,
S. Teis and A.
Sibirtsev, Nucl. Phys.  A614 (1997) 415; E.L. Bratkovskaya,
W. Cassing and
U. Mosel, Nucl. Phys.  A622 (1997) 593.
\vs
\bibitem{LKF94}
G.Q. Li, C.N. Ko and X.S. Fang, Phys. Lett. B329 (1994) 149.
\vs
\bibitem{LK96}
G.Q. Li and C.N. Ko, Phys. Rev. C54 (1996) 2159.
\vs
\bibitem{Li97} G.Q. Li, C.-H. Lee and G.E. Brown, Nucl. Phys.
A625 (1997) 372.
\vs
\bibitem{SKE99} J. Schaffner-Bielich, V. Koch and M.
Effenberg, Nucl. Phys.
A, in print, nucl-th/9907095.
\vs
\bibitem{sibir98} A. Sibirtsev and W. Cassing, Nucl. Phys.
A641 (1998) 476.
\vs
\bibitem{CHL95} C.-H. Lee, G.E. Brown and M. Rho, Phys. Lett.
B335
(1994) 266; C.-H. Lee, G.E. Brown, D.P. Min and M. Rho, Nucl.
Phys.  A585
(1995) 401.
\vs
\bibitem{CHL96b} C.-H. Lee, Phys. Reports 275 (1996) 255
\vs
\bibitem{Mao99} G. Mao, P. Papazoglou, S. Hofmann, S. Schramm,
H. St\"ocker
and W. Greiner, Phys. Rev. C59 (1999) 3381.
\vs
\bibitem{Sch97} J. Schaffner and I.N. Mishustin, Phys. Rev.
C53 (1996) 1416; J. Schaffner-Bielich, I.N. Mishustin and
J. Bondorf, Nucl. Phys. A625 (1997) 325.
\vs
\bibitem{Tsushi98} K. Tsushima, K. Saito, A.W. Thomas and S.V.
Wright, Phys.  Lett. B429 (1998) 239.

\vs
\bibitem{moto90} H. Band\={o}, T. Motoba and J. \v{Z}ofka,
Int.
J. Mod.  Phys. A5 (1990) 4021.
\vs
\bibitem{Batty78} C.J. Batty et al, Phys. Lett. 74B (1978) 27,
C.J. Batty et al, Phys. Lett. 87B (1979) 324.
\vs
\bibitem{oset90} E. Oset, P. Fern\'andez de C\'ordoba,
L.L. Salcedo and R. Brockmann, Phys. Reports 188 (1990) 79.
\vs
\bibitem{Batty94}
C.J. Batty, E. Friedman and A. Gal, Phys. Lett. B335 (1994) 273.
\vs
\bibitem{ramos94} A. Ramos, E. Oset and L.L. Salcedo, Phys.
Rev.
C50 (1994) 2314.
\vs
\bibitem{oku99} S. Hirenzaki, Y. Okumura, H. Toki, E. Oset and
A.
Ramos, Phys. Rev. C, in print and in 
proceedings of $XV^{th}$ Particles and Nuclei
International
Conference, Eds. G. F\"aldt, B. H\"oistad and S. Kullander,
Nucl. Phys. A, in print.
\vs
\bibitem{batty81} C.J. Batty, Nucl. Phys. A372 (1981) 418.
\vs
\bibitem{baca00}
A. Baca, C. Garc\'{\i}a-Recio and J. Nieves, Nucl. Phys. A, in press,
nucl-th/0001060.
\vs
\bibitem{FGMC99} E. Friedman, A. Gal, J. Mare\v{s} and A.
Ciepl\'y,
Phys. Rev. C60 (1999) 024314.
\vs
\bibitem{FG99a} E. Friedman and A. Gal, Phys. Lett. B459
(1999) 43.
\vs
\bibitem{FG99b} E. Friedman and A. Gal, Nucl. Phys. A658 (1999) 345.
\vs
\bibitem{Kishi99} T. Kishimoto, Phys. Rev. Lett. 83 (1999) 4701.

\end{thebibliography}
